\documentclass[12pt]{iopart}

\usepackage{hyperref, multirow, color, bbold, tabularx}
\usepackage[dvipsnames]{xcolor}
\usepackage{orcidlink}

\newcolumntype{C}{>{\centering\arraybackslash}X}

\newcommand{\blue}[1]{{#1}}
\newcommand{\eqref}[1]{(\ref{#1})}

\usepackage{ulem}   %for \sout

\usepackage{url,hyperref,lineno,microtype}
\usepackage[onehalfspacing]{setspace}

\newcommand{\CornellPhysics}{Department of Physics, Cornell University, Ithaca,
  NY, 14853, USA}
\newcommand{\Cornell}{Cornell Center for Astrophysics and Planetary
    Science, Cornell University, Ithaca, New York 14853, USA}
\newcommand{\Caltech}{Theoretical Astrophysics 350-17, California
    Institute of Technology, Pasadena, CA 91125, USA}
\newcommand{\Fullerton}{Nicholas and Lee Begovich Center for
    Gravitational-Wave Physics and Astronomy, California State University
    Fullerton, Fullerton, CA 92834, USA}

\begin{document}

\title[A positivity-preserving adaptive-order finite-difference scheme for
GRMHD]{A positivity-preserving adaptive-order finite-difference scheme for GRMHD}

\author{
  Nils Deppe\,$^{1,2,3}$\orcidlink{0000-0003-4557-4115},
  Lawrence E.~Kidder\,$^{2}$\orcidlink{0000-0001-5392-7342},
  Saul A.~Teukolsky\,$^{2,3}$\orcidlink{0000-0001-9765-4526},
  Marceline S.~Bonilla\,$^4$\orcidlink{0000-0003-4502-528X},
  Fran\c{c}ois H\'{e}bert\,$^{3}$\orcidlink{0000-0001-9009-6955},
  Yoonsoo Kim\,$^{3}$\orcidlink{0000-0002-4305-6026},
  Mark A.~Scheel\,$^{3}$\orcidlink{0000-0001-6656-9134},
  William Throwe\,$^{2}$\orcidlink{0000-0001-5059-4378},
  and Nils L.~Vu\,$^{3}$\orcidlink{0000-0002-5767-3949}}
\address{$^1$\CornellPhysics}
\address{$^2$\Cornell}
\address{$^3$\Caltech}
\address{$^4$\Fullerton}
\ead{nd357@cornell.edu}

\begin{abstract}
We present an adaptive-order positivity-preserving conservative
finite-difference scheme that allows a high-order solution away from shocks
and discontinuities while guaranteeing positivity and robustness at
discontinuities. This is achieved by monitoring the relative power in the
highest mode of the reconstructed polynomial and reducing the order when the
polynomial series no longer converges. Our approach is similar to the
multidimensional optimal order detection (MOOD) strategy, but differs in
several ways. The approach is \textit{a priori} and so does not require
retaking a time step. It can also readily be combined with
positivity-preserving flux limiters that have gained significant traction in
computational astrophysics and numerical relativity. This combination
ultimately guarantees a physical solution both during reconstruction and time
stepping. We demonstrate the capabilities of the method using a standard suite
of very challenging 1d, 2d, and 3d general relativistic magnetohydrodynamics
test problems.
\end{abstract}


\submitto{\CQG}

\section{Introduction}\label{sec:introduction}

Godunov's theorem\cite{Godunov1959} tells us that in the numerical
solution of conservation laws, high-order linear schemes
cannot prevent oscillations from appearing, be they physical or
unphysical.  Nonlinear hybridization has been used to circumvent
Godunov's theorem by reconstructing a high-order polynomial using
nonlinear combinations of the variables. All modern methods such as
essentially non-oscillitary (ENO)\cite{chakravarthy1986essentially,
  harten1986some, Harten1987231, Shu1988439, Shu198932}, weighted ENO
(WENO)\cite{JIANG1996202, LIU1994200}, and weighted compact nonlinear schemes
(WCNS)\cite{DENG200022, Nonomura2007, ZHANG20087294} use nonlinear hybridization
to achieve higher-than-first-order accuracy. WENO and WCNS schemes are commonly
used to construct high-order finite-difference (FD) schemes. One major challenge
in using high-order FD schemes with complicated equations is maintaining
physical realizability of the solution, such as positivity of the density and
pressure. We develop a new adaptive order WENO-type method that guarantees
physical realizability during the reconstruction by reconstructing the primitive
variables as opposed to the conserved or characteristic variables.
We expect such methods to be particularly effective for
binary neutron star merger simulations, both magnetized and unmagnetized, where 
achieving high-order convergence is attracting increasingly more
efforts~\cite{Radice:2013hxh, Radice:2013xpa, Most:2019kfe, Raithel:2022san,
  Cipolletta:2020kgq}.

Existing WENO and WCNS schemes work by combining several low-order stencils into
a high-order stencil that interpolates the solution within a FD cell. How much
each low-order stencil contributes to the final result is determined by use of
an oscillation or smoothness indicator. The more oscillatory a stencil is, the
less it contributes to the overall reconstruction polynomial. FD schemes fall
into two broad categories: flux vector splitting (FVS) and flux difference
splitting (FDS). FVS schemes write the numerical flux as a linear combination of
a right- and left-moving part, then reconstruct the numerical fluxes to the cell
boundaries. FDS schemes reconstruct the conserved, primitive, or characteristic
variables to the cell boundaries, and then use a numerical flux at the cell
boundary. The advantage of FDS over FVS is that more numerical fluxes can be
used with them since the numerical flux need not be split (see,
e.g.~\cite{Toro2009} for a helpful review).

Reconstructing the characteristic variables yields very good results with FDS
schemes (e.g.~\cite{Nonomura20138}). However, for increasingly complex physical
systems computing the characteristic variables becomes analytically intractable
and needs to be done numerically.
Unfortunately, computing the
characteristic variables numerically is often very expensive for these systems
and so does not present a realistic alternative.
An important example in practice is the equations of general
relativistic magnetohydrodynamics (GRMHD).

Another issue that arises when
reconstructing the conserved or characteristic variables is that the
reconstructed solution may not be physically realizable. For example, the
density of the reconstructed state may be negative (see,
e.g.~\cite{doi:10.2514/6.2017-0845, 2016ComAC...3....1Z,
  IVAN2014830}). Reconstructing the primitive variables permits straightforward
guarding
against such unphysical reconstructed states. Flux limiters have been used to
maintain a physically realizable solution~\cite{HU2013169} when using FVS
schemes, even in numerical relativity~\cite{Radice:2013xpa}. However, when a
characteristic decomposition is not available, the dissipative Rusanov or local
Lax-Friedrichs numerical flux needs to be used in the FVS scheme. Another
strategy for maintaining positivity when reconstructing the conserved variables
is the flattener algorithm~\cite{BALSARA20127504}, which was originally
proposed for finite-volume methods. The flattener smoothly
interpolates between a first-order and high-order reconstruction in a way that
the reconstructed polynomial is physically realizable. The main disadvantage of
the flattener compared to the new method presented here is that the flattener
does not provide a way of determining what order FD derivative should be used.

Adaptive-order WENO schemes have recently been presented that combine one or
more high-order stencils with robust low-order stencils\cite{BALSARA2016780,
  SUN201681, Semplice2020, Guercilena:2016fdl, Doulis:2022vkx}. The high-order
stencil is used if it is not too oscillatory, while the low-order standard WENO
stencils are used if the high-order stencil is inadmissible. A similar approach
called multidimensional optimal-order detection (MOOD)\cite{CLAIN20114028,
  DIOT201243, doi:10.1002/fld.3804} has also been presented where unlimited
reconstruction is generally used, but the order is decreased \textit{a
  posteriori} and discretely. That is, after a time step is taken,
the result is
accepted only if the solution
passes numerical and physical admissibility conditions;
otherwise the time step is redone using a lower order scheme.

We address the difficulty of achieving high-order accuracy while maintaining
physical realizability and robustness by:
\begin{itemize}
\item introducing a new oscillation indicator that directly measures the amount
  of power in the high-order polynomials of the reconstruction,
\item weighting the reconstructed stencils by physical realizability instead of
  just numerical admissibility,
\item adapting the order of the FD derivative according to the order of the
  reconstruction to avoid differentiating across discontinuities,
\item providing two general implementations, one using weighting of the different
  stencils and one discretely selecting the highest order admissible stencil.
\end{itemize}
We present the new oscillation indicator in \S\ref{sec:oi}, describe how to
include physical realizability into the scheme in \S\ref{sec:pp strategy}, and
discuss the FD derivative adaptation in \S\ref{sec:fd order}.  In
\S\ref{sec:weighted ao} we discuss how to make the discrete stencil selection
process described in \S\ref{sec:oi} continuous by combining the new oscillation
indicator with physical realizability conditions and using a sigmoid-like
function for blending the stencils. This method of including the physical
realizability conditions into the stencil weights can readily be incorporated
into other WENO and WCNS schemes. Finally, in \S\ref{sec:numerical results} we
show results from a large variety of standard and difficult test problems in 1d,
2d, and 3d GRMHD. \blue{In \ref{sec:dg-fd} we briefly comment on how
  the scheme presented here can be used in a discontinuous Galerkin-finite
  difference hybrid method as described in~\cite{Deppe:2021ada}.}

\section{Method}\label{sec:method}

We first present the new \textit{a priori} adaptive order scheme in 1d,
including a description of the positivity-preserving strategy. We then use the
equations of ideal general relativistic magnetohydrodynamics (GRMHD) as a concrete
example.

We are interested in solving general conservation laws of the form
\begin{equation}
  \label{eq:conservation law}
  \partial_t u+\partial_i F^i(u)=S(u),
\end{equation}
where $u$ is the vector of conserved variables, Latin indices later in the
alphabet (such as $i$) denote spatial indices, $F^i$ is the flux vector, and
$S$ is the source  vector. Here and throughout we implicitly sum over repeated
indices.  We denote the primitive variables as $p$. Since we are ultimately
interested in systems where the characteristic variables are either not known or
are very expensive to compute, we seek to develop a robust, positivity-preserving
non-oscillatory scheme where the primitive variables are reconstructed to cell
interfaces.

Conservative finite-difference schemes evolve the cell-center values, but
require the cell-face values (the midpoints along each axis) for solving the
Riemann problem and for computing the finite-difference derivatives of the
flux. That is, the semi-discrete form of Eq.~\ref{eq:conservation law} is
\begin{equation}
  \label{eq:sd conservation law}
  \partial_t u_{\hat{\imath}}+
  \frac{G^i_{\hat{\imath}+1/2} - G^i_{\hat{\imath}-1/2}}{\Delta
  x^i}=S_{\hat{\imath}},
\end{equation}
where hatted indices denote grid points/cells, and $G^i$ is the numerical flux
given by (approximately) solving the Riemann problem at the cell interface. The
FD derivative in Eq.~\ref{eq:sd conservation law} is of second order, but we
show how this can be extended to higher order in \S\ref{sec:fd order}. In
\S\ref{sec:fd order} we also describe our new method of adjusting the order of the
finite-difference derivative based on the local smoothness of the solution.

WENO and WCNS schemes use nonlinear reconstruction and are very robust when
applied to the characteristic variables. When applied to the primitive variables
they can lead to staircasing, where
small-scale oscillations
are present in smooth regions~\cite{SURESH199783}.
We opt for a different approach to obtain a
high-order non-oscillatory scheme that somewhat resembles multidimensional
optimal-order detection (MOOD)\cite{CLAIN20114028, DIOT201243,
  doi:10.1002/fld.3804}. We combine several new ingredients:
first, we present a
new method of obtaining a high-order polynomial based on spectral elements;
second we present a new \textit{a priori} detection algorithm for determining
whether or not the reconstructed stencil is acceptable; third we present a
method to combine our adaptive-order scheme with a positivity-preserving
strategy; and finally we present a method of adjusting the FD order to increase
stability without the accuracy of the scheme deteriorating in smooth regions. We
will refer to the method we present as a positivity-preserving adaptive-order
(PPAO\footnote{Pronounced ``pow''.}) scheme where we postfix with hyphens the
orders used. For example, we denote a PPAO scheme with the order decrementing
from fifth to third to first as PPAO5-3-1.
In \S\ref{sec:reconstruction} we carry out
a detailed derivation of the fifth-order scheme. We then
provide a table of the
coefficients for seventh- and ninth-order schemes.

\subsection{High-order reconstruction}\label{sec:reconstruction}

We construct a fifth-order scheme using a polynomial of degree four from the
solution
$\{p_{\hat{\jmath}-2}, p_{\hat{\jmath}-1}, p_{\hat{\jmath}}, p_{\hat{\jmath}+1},
p_{\hat{\jmath}+2}\}$, where $\hat{\jmath}$ indexes the cell. We do this by
setting up a spectral element on the interval
$[x_{\hat{\jmath}-5/2}, x_{\hat{\jmath}+5/2}]$ with Legendre basis
functions. The interval is remapped to the logical coordinate $\xi\in[-1,1]$
where the Legendre basis functions are defined. It is important to note that
this differs from the more common approach of remapping the interval
$[x_{\hat{\jmath}-1/2}, x_{\hat{\jmath}+1/2}]$ to $\xi\in[-1,1]$. Defining the
basis functions over the larger interval is key to detecting oscillations over
the region the polynomials are fit. This is one crucial way our method differs
from previous literature and is what makes it very robust at
detecting non-smooth solutions. We denote the $n^{\mathrm{th}}$ degree
Legendre polynomials by $P_n(\xi)$. Within the spectral element we expand the
solution as
\begin{equation}
  \label{eq:spectral solution expansion}
  p_{\hat{\jmath}}(\xi)=\sum_{n=0}^{N}c_{\hat{\jmath},n} P_n(\xi),
\end{equation}
where $c_{\hat{\jmath},n}$ are the modal coefficients for the expansion about the
$\hat{\jmath}^{\mathrm{th}}$ cell. The coefficients $c_{\hat{\jmath},n}$ are computed by solving the
linear system
\begin{equation}
  \label{eq:AO5 coefficients}
  \hspace{-4em}
  \left( \matrix{
    P_0(\xi_{\hat{\jmath}-2}) & P_1(\xi_{\hat{\jmath}-2}) &
    P_2(\xi_{\hat{\jmath}-2}) & P_3(\xi_{\hat{\jmath}-2}) &
    P_4(\xi_{\hat{\jmath}-2})  \cr
    P_0(\xi_{\hat{\jmath}-1}) & P_1(\xi_{\hat{\jmath}-1}) &
    P_2(\xi_{\hat{\jmath}-1}) & P_3(\xi_{\hat{\jmath}-1}) &
    P_4(\xi_{\hat{\jmath}-1}) \cr
    P_0(\xi_{\hat{\jmath}})   & P_1(\xi_{\hat{\jmath}})   &
    P_2(\xi_{\hat{\jmath}})   & P_3(\xi_{\hat{\jmath}})   &
    P_4(\xi_{\hat{\jmath}}) \cr
    P_0(\xi_{\hat{\jmath}+1}) & P_1(\xi_{\hat{\jmath}+1}) &
    P_2(\xi_{\hat{\jmath}+1}) & P_3(\xi_{\hat{\jmath}+1}) &
    P_4(\xi_{\hat{\jmath}+1}) \cr
    P_0(\xi_{\hat{\jmath}+2}) & P_1(\xi_{\hat{\jmath}+2}) &
    P_2(\xi_{\hat{\jmath}+2}) & P_3(\xi_{\hat{\jmath}+2}) &
    P_4(\xi_{\hat{\jmath}+2}) \cr
  }
  \right)
  \left(\matrix{
    c_0 \cr c_1 \cr c_2 \cr c_3 \cr c_4 \cr
  }\right) =
\left(\matrix{
    p_{\hat{\jmath}-2} \cr p_{\hat{\jmath}-1} \cr p_{\hat{\jmath}} \cr
    p_{\hat{\jmath}+1} \cr p_{\hat{\jmath}+2} \cr
  }\right),
\end{equation}
where we have used the notation $\xi(x_{\hat{\jmath}})$ to represent the
function that maps the $x$ coordinates into the $\xi$ coordinates. Assuming
uniform spacing in $\xi$ we find that
\begin{equation}
  \label{eq:AO5 matrix}
  \hspace{-6em}
  \left(\matrix{
    c_0 \cr c_1 \cr c_2 \cr c_3 \cr c_4
  }\right) =
  \left(\matrix{
    275/1152 &   25/288 &  67/192 &   25/288 &  275/1152 \cr
    -55/96 &    -5/48 &       0 &     5/48 &     55/96 \cr
    1525/2016 & -475/504 & 125/336 & -475/504 & 1525/2016 \cr
    -25/48 &    25/24 &       0 &   -25/24 &     25/48 \cr
    125/336 & -125/84 & 125/56 & -125/84 & 125/336
  }\right)
  \left(\matrix{
      p_{\hat{\jmath}-2} \cr
      p_{\hat{\jmath}-1} \cr
      p_{\hat{\jmath}} \cr
      p_{\hat{\jmath}+1} \cr
      p_{\hat{\jmath}+2} \cr
    }\right).
\end{equation}
In practice we never need to perform the matrix multiplication in
Eq.~\eqref{eq:AO5 matrix} since we can precompute the values
$p_{\hat{\jmath}-1/2}$ and $p_{\hat{\jmath}+1/2}$ analytically in terms of
$\{p_{\hat{\jmath}-2},p_{\hat{\jmath}-1},p_{\hat{\jmath}}, p_{\hat{\jmath}+1},
p_{\hat{\jmath}+2}\}$. The reconstructed face values are
\begin{equation}
  \label{eq:AO5 boundary values}
  \begin{array}{ll}
    p_{\hat{\jmath}-1/2} &= - \frac{5}{128}p_{\hat{\jmath}-2} +
    \frac{15}{32}p_{\hat{\jmath}-1} +
    \frac{45}{64}p_{\hat{\jmath}} -\frac{5}{32}p_{\hat{\jmath}+1} +
    \frac{3}{128}p_{\hat{\jmath}+2}, \\
    p_{\hat{\jmath}+1/2} &= \frac{3}{128}p_{\hat{\jmath}-2} -
    \frac{5}{32}p_{\hat{\jmath}-1} + \frac{45}{64}p_{\hat{\jmath}} +
    \frac{15}{32}p_{\hat{\jmath}+1} - \frac{5}{128}p_{\hat{\jmath}+2}.
  \end{array}
\end{equation}
\blue{Here $p_{\hat{\jmath}-1/2}$ is the \textit{right} state of the
  $\hat{\jmath}-1/2$ interface and $p_{\hat{\jmath}+1/2}$ is the \textit{left}
  state of the $\hat{\jmath}+1/2$ interface. By using the same stencil but
  shifted by one cell to the left or right, the left state on $\hat{\jmath}-1/2$
  and the right state on $\hat{\jmath}+1/2$ are computed.}
We show coefficients for evaluating polynomials of degree 2, 4, 6, and 8 at the
cell interfaces in table~\ref{tab:reconstruction coefficients}.

\begin{table}
  \caption{Coefficients for reconstructing the polynomials at the cell-face
    values at different orders.}
  \label{tab:reconstruction coefficients}
  \begin{indented}
    \lineup
  \item[]\hspace{-5em}\begin{tabular}{@{}ccccccccccc}
    \br
    Degree & Face & $\hat{\jmath}-4$ & $\hat{\jmath}-3$ & $\hat{\jmath}-2$ & $\hat{\jmath}-1$ & $\hat{\jmath}$ & $\hat{\jmath}+1$ & $\hat{\jmath}+2$ & $\hat{\jmath}+3$
    & $\hat{\jmath}+4$ \\ \hline
    \multirow{2}{*}{2} & $\hat{\jmath}-1/2$ & & & & $\frac{3}{8}$ & $\frac{3}{4}$ & $-\frac{1}{8}$ & & \\
          & $\hat{\jmath}+1/2$ & & & & $-\frac{1}{8}$ & $\frac{3}{4}$ & $\frac{3}{8}$ & & \\
    \multirow{2}{*}{4} & $\hat{\jmath}-1/2$ & & & $-\frac{5}{128}$ & $\frac{15}{32}$
          & $\frac{45}{64}$ & $-\frac{5}{32}$ & $\frac{3}{128}$ & & \\
          & $\hat{\jmath}+1/2$ & & & $\frac{3}{128}$ & $-\frac{5}{32}$ & $\frac{45}{64}$
          & $\frac{15}{32}$ & $-\frac{5}{128}$ & & \\
    \multirow{2}{*}{6} & $\hat{\jmath}-1/2$ & & $\frac{7}{1024}$ & $-\frac{35}{512}$ & $\frac{525}{1024}$
          & $\frac{175}{256}$ & $-\frac{175}{1024}$ & $\frac{21}{512}$
          & $-\frac{5}{1024}$ & \\
          & $\hat{\jmath}+1/2$ & & $-\frac{5}{1024}$ & $\frac{21}{512}$ & $-\frac{175}{1024}$ & $\frac{175}{256}$
          & $\frac{525}{1024}$ & $-\frac{35}{512}$ & $\frac{7}{1024}$ & \\
    \multirow{2}{*}{8} & $\hat{\jmath}-1/2$ &  $-\frac{45}{32768}$ & $\frac{63}{4096}$ & $-\frac{735}{8192}$
          & $\frac{2205}{4096}$ & $\frac{11025}{16384}$ & $-\frac{735}{4096}$
          & $\frac{441}{8192}$ & $-\frac{45}{4096}$ & $\frac{35}{32768}$ \\
          & $\hat{\jmath}+1/2$ & $\frac{35}{32768}$ & $-\frac{45}{4096}$ & $\frac{441}{8291}$ & $-\frac{735}{4096}$
          & $\frac{11025}{16384}$ & $\frac{2205}{4096}$ & $-\frac{735}{8192}$ & $\frac{63}{4096}$
          & $-\frac{45}{32768}$ \\
           \br
         \end{tabular}
       \end{indented}
\end{table}

\subsection{Oscillation detection}\label{sec:oi}

With a polynomial for reconstructing
the face values in hand, we need to determine how oscillatory the polynomial is
and whether or not to use it. For a spectral expansion such as the one given in
Eq.~\ref{eq:spectral solution expansion}, higher-degree basis functions are more
oscillatory and so a large coefficient $c_N$ for the highest-degree basis
function means the solution is quite oscillatory. We make this precise by using
the troubled cell indicator presented in~\cite{PerssonTci} for the discontinuous
Galerkin method. We define $\hat{p}_j$ as
\begin{equation}
  \label{eq:u hat}
  \hat{p}_{\hat{\jmath}}(\xi)=c_{\hat{\jmath},N} P_N(\xi),
\end{equation}
i.e.~only the highest-degree term in Eq.~\ref{eq:spectral solution expansion}.
Then we consider a polynomial admissible if
\begin{equation}
  N^{2\alpha_N}\int_{-1}^{1}\hat{p}_{\hat{\jmath}}^2 d\xi
  \le\int_{-1}^{1}p_{\hat{\jmath}}^2d\xi,
\end{equation}
where $\alpha_N=4$ is generally a good choice since this is when the Legendre
basis stops converging~\cite{doi:10.1137/1.9781611970425}. Specifically, this
guarantees that the coefficients decay at least as $1/N^2$. We can rewrite the
integrals as
\begin{equation}
  \label{eq:sensor}
  N^{2\alpha_N} \frac{c_N^2}{2N+1} \le \sum_{n=0}^{N}\frac{c_n^2}{2n + 1}.
\end{equation}
This effectively requires that the total power in the highest mode of the
Legendre basis expansion is a sufficiently small fraction of the total
power. In practice, our PPAO method drops to the next lowest order if
Eq.~\ref{eq:sensor} is violated. If the lower-order polynomial
violates Eq.~\ref{eq:sensor}, the order is dropped again. This is repeated until
a second-order scheme, such as monotonized central~\cite{VANLEER1977276}, is
used.

We can rewrite the integrals in terms of the primitives directly. For
convenience we define
\begin{eqnarray}
  \bar{\kappa}_N &= \frac{c_N^2}{2N+1}, \\
  \kappa_N &= \sum_{n=0}^{N}\frac{c_n^2}{2n + 1}.
\end{eqnarray}
Then we write the oscillation indicator at third order as
\begin{eqnarray}
  \label{eq:bar kappa_3}
  \bar{\kappa}_2 &= \frac{2}{5}\left(\frac{3}{4}p_{\hat{\jmath}+1} - \frac{3}{2}p_{\hat{\jmath}} +
                   \frac{3}{4}p_{\hat{\jmath}-1}\right)^{2}, \\
  \label{eq:kappa_3}
  \kappa_2 &=  \left(\frac{3}{8}p_{\hat{\jmath}+1} + \frac{1}{4}p_{\hat{\jmath}} +
             \frac{3}{8}p_{\hat{\jmath}-1}\right)
             \left(\frac{31}{20}p_{\hat{\jmath}+1} - \frac{1}{10}p_{\hat{\jmath}} +
             \frac{11}{20}p_{\hat{\jmath}-1}\right).
\end{eqnarray}
The fifth-order terms are given by:
\begin{eqnarray}
  \label{eq:bar kappa_5}
  \bar{\kappa}_4&=
  \frac{2}{9}\left(\frac{125}{336}p_{\hat{\jmath}-2} - \frac{125}{84}p_{\hat{\jmath}-1} +
  \frac{125}{56}p_{\hat{\jmath}} - \frac{125}{84}p_{\hat{\jmath}+1} + \frac{125}{336}p_{\hat{\jmath}+2}
  \right)^{2}, \\
  \label{eq:kappa_5}
  \kappa_4&=\left(\frac{4015}{12096}p_{\hat{\jmath}-2} - \frac{925}{3024}p_{\hat{\jmath}-1}
            + \frac{2707}{2016}p_{\hat{\jmath}} - \frac{2305}{3024}p_{\hat{\jmath}+1}
            + \frac{16855}{12096}p_{\hat{\jmath}+2}\right) \nonumber \\
          &\times\left(\frac{275}{1152}p_{\hat{\jmath}-2} + \frac{25}{288}p_{\hat{\jmath}-1}
            + \frac{67}{192}p_{\hat{\jmath}} + \frac{25}{288}p_{\hat{\jmath}+1}
            + \frac{275}{1152}p_{\hat{\jmath}+2}\right).
\end{eqnarray}
The seventh-order terms are given by:
\begin{eqnarray}
  \label{eq:bar kappa_7}
  \bar{\kappa}_6
  &=\frac{2}{13}\left(\frac{16807}{95040}p_{\hat{\jmath}-3} -
    \frac{16807}{15840}p_{\hat{\jmath}-2} + \frac{16807}{6336}p_{\hat{\jmath}-1}
    - \frac{16807}{4752}p_{\hat{\jmath}} \right.\nonumber \\
  &\left.+ \frac{16807}{6336}p_{\hat{\jmath}+1} - \frac{16807}{15840}p_{\hat{\jmath}+2} +
    \frac{16807}{95040}p_{\hat{\jmath}+3}\right)^{2}, \\
  \label{eq:kappa_7}
  \kappa_6
  &=\left(\frac{52087}{285120}p_{\hat{\jmath}-3} -
    \frac{2989}{190080}p_{\hat{\jmath}-2}+ \frac{43757}{95040}p_{\hat{\jmath}-1} -
    \frac{72629}{285120}p_{\hat{\jmath}}\right.\nonumber \\
  &\left.+\frac{43757}{95040}p_{\hat{\jmath}+1} - \frac{2989}{190080}p_{\hat{\jmath}+2} +
    \frac{52087}{285120}p_{\hat{\jmath}+3}\right)\times \nonumber \\
  &\left(\frac{748741}{2965248}p_{\hat{\jmath}-3} - \frac{559769}{1235520}p_{\hat{\jmath}-2}
    + \frac{9241603}{4942080}p_{\hat{\jmath}-1}\right.\nonumber \\
  &\left.- \frac{145879}{57915}p_{\hat{\jmath}} + \frac{2928709}{988416}p_{\hat{\jmath}+1} -
    \frac{1769887}{1235520}p_{\hat{\jmath}+2} + \frac{1777657}{1347840}p_{\hat{\jmath}+3}\right).
\end{eqnarray}
The ninth-order terms are given by:
\begin{eqnarray}
  \label{eq:bar kappa_9}
  \bar{\kappa}_8
  &=\frac{2}{17}\left(
    \frac{531441}{6406400}p_{\hat{\jmath}-4} - \frac{531441}{800800}p_{\hat{\jmath}-3}+
    \frac{531441}{228800}p_{\hat{\jmath}-2} -
    \frac{531441}{114400}p_{\hat{\jmath}-1}
    \right.\nonumber \\
  &\left.+\frac{531441}{91520}p_{\hat{\jmath}}
    - \frac{531441}{114400}p_{\hat{\jmath}+1} +
    \frac{531441}{228800}p_{\hat{\jmath}+2} -
    \frac{531441}{800800}p_{\hat{\jmath}+3}
    \right.\nonumber \\
  &\left.+
    \frac{531441}{6406400}p_{\hat{\jmath}+4}\right)^{2},\\
  \label{eq:kappa_9}
  \kappa_8
  &=\left(\frac{62705408397}{243955712000}p_{\hat{\jmath}-4} -
    \frac{25040347967}{30494464000}p_{\hat{\jmath}-3} +
    \frac{26778669537}{8712704000}p_{\hat{\jmath}-2} \right.\nonumber\\
  &- \frac{23833259907}{4356352000}p_{\hat{\jmath}-1} +
    \frac{25763924917}{3485081600}p_{\hat{\jmath}} -
    \frac{28082791047}{4356352000}p_{\hat{\jmath}+1} \nonumber \\
  &\left.+ \frac{39963815817}{8712704000}p_{\hat{\jmath}+2} -
    \frac{54884086987}{30494464000}p_{\hat{\jmath}+3} +
    \frac{27239637687}{22177792000}p_{\hat{\jmath}+4}\right) \times \nonumber \\
  &\left(\frac{9600579}{63078400}p_{\hat{\jmath}-4} -
    \frac{673539}{7884800}p_{\hat{\jmath}-3} +
    \frac{9532053}{15769600}p_{\hat{\jmath}-2} \right.\nonumber\\
  &- \frac{508383}{716800}p_{\hat{\jmath}-1} +
    \frac{1357457}{1261568}p_{\hat{\jmath}} -
    \frac{508383}{716800}p_{\hat{\jmath}+1} \nonumber\\
  &\left. + \frac{9532053}{15769600}p_{\hat{\jmath}+2}
    -\frac{673539}{7884800}p_{\hat{\jmath}+3} +
    \frac{9600579}{63078400}p_{\hat{\jmath}+4}\right).
\end{eqnarray}

\subsection{Positivity-preserving strategy}\label{sec:pp strategy}

In many physical systems there are requirements on certain variables, for
example the density and pressure must remain positive.  A few different
strategies for maintaining positivity with WENO-type schemes have been
\blue{presented for Newtonian hydrodynamics\cite{Zhang2011a, ZHANG20108918,
    ZHANG20122245, BALSARA20127504, HU2013169} and also for ideal
  non-relativistic MHD\cite{2015SJSC...37A1825C, 2018SJSC...40B1302W}}. For more
complicated physical systems, such as general relativistic magnetohydrodynamics,
generalizations of these strategies are non-trivial or not possible. We seek to
ensure that the reconstructed polynomials~\eqref{eq:AO5 boundary values} are
physically realizable in a way that is easily tailored to whatever physical
constraints are present in the system. While negative densities and pressures
could still occur from time integration, this can be avoided by using a
positivity-preserving flux limiter~\cite{HU2013169, Radice:2013xpa}. A very
simple, efficient, and robust method for preserving positivity in the
reconstruction is to check that the cell-face values are positive. If the
reconstructed cell-face values are not positive we consider the polynomial to be
invalid and switch to a lower-order polynomial. This is repeated until the
reconstructed polynomial is positive, which is guaranteed to be true for the
first-order scheme.

It is also possible to guarantee that the reconstructed polynomial is positive
over the entire region used to construct it. For the fifth-order case this can
be done as follows. We can write the \textit{derivative} of the fourth-degree
reconstructed polynomial as
\begin{equation}
  \label{eq:reconstructed fourth order polynomial deriv}
  \partial_\xi p_{\hat{\jmath}}(\xi)=a \xi^3 + b \xi^2 + c \xi + d,
\end{equation}
where
\begin{eqnarray}
  \label{eq:fourth order a}
  a&=\frac{625}{96}\left(p_{\hat{\jmath}-2} - 4 p_{\hat{\jmath}-1} + 6
     p_{\hat{\jmath}} - 4 p_{\hat{\jmath}+1} + p_{\hat{\jmath}+2}\right), \\
  \label{eq:fovrth order b}
  b&=\frac{125}{32}\left(- p_{\hat{\jmath}-2} +
     2p_{\hat{\jmath}-1}-2p_{\hat{\jmath}+1} + p_{\hat{\jmath}+2}\right), \\
  \label{eq:fovrth order c}
  c&=-\frac{25}{48}p_{\hat{\jmath}-2} + \frac{25}{3}p_{\hat{\jmath}-1} -
     \frac{125}{8}p_{\hat{\jmath}} + \frac{25}{3}p_{\hat{\jmath}+1} -
     \frac{25}{48}p_{\hat{\jmath}+2},\\
  \label{eq:fovrth order d}
  d&=\frac{5}{24}p_{\hat{\jmath}-2} - \frac{5}{3}p_{\hat{\jmath}-1} +
     \frac{5}{3}p_{\hat{\jmath}+1} - \frac{5}{24}p_{\hat{\jmath}+2}.
     _S
\end{eqnarray}
At this point one can use standard methods for finding the roots of a cubic to
obtain the $\xi$ coordinates at which the extrema of the reconstructed
polynomial occur. If the extrema occur on the interval $\left[-1,1\right]$ one
must check that the reconstructed polynomial is positive there. This can be done
by using the coefficients computed from Eq.~\eqref{eq:AO5 matrix} to evaluate
the reconstructed polynomial at the extrema and check it is positive.

Rigorously verifying that the entire polynomial is positive is quite expensive,
especially for higher than fifth-order reconstruction. An alternative is to
evaluate the polynomial at several additional predetermined nodes, similar
to~\cite{BALSARA20127504}. For a linear solution, the extrema occur at the end
points so evaluating at the additional nodes $\xi\in\{-1,1\}$ seems
natural. Reasonable additional choices are the cell faces not already required
for the Riemann problem. We provide the stencils for evaluating the polynomial
at the additional cell faces in table~\ref{tab:positivity reconstruction
  coefficients}. We only write the $\hat{\jmath}-N/2$ coefficients, since the
$\hat{\jmath}+N/2$ coefficients are given by reflecting the stencils about the
cell. In practice we have not found evaluating at the additional
nodes to be necessary since the coefficients are already decaying as $\sim1/N^5$
or faster and thus the high frequency components that would lead to large
extrema between the nodes are absent. However, this argument does not in principle prevent the end points from being negative.
Nevertheless, our current implementation checks for positivity
only on the cell-face values,
as described in the first paragraph of this subsection,
and does not check for positivity at any additional points.

\begin{table}[h]
  \caption{Coefficients for reconstructing the polynomials at the additional
    cell-face values for ensuring positivity of the reconstructed polynomial at
    different orders. The coefficients for the faces to the right of the cell
    are obtained by reflecting the coefficients about $\hat{\jmath}$.}
  \label{tab:positivity reconstruction coefficients}
  \begin{indented}
    \lineup
  \item[]\hspace{-7.5em}\begin{tabular}{@{}ccccccccccc}
    \br
    Degree & Face & $\hat{\jmath}-4$ & $\hat{\jmath}-3$ & $\hat{\jmath}-2$
    & $\hat{\jmath}-1$ & $\hat{\jmath}$ & $\hat{\jmath}+1$ & $\hat{\jmath}+2$
    & $\hat{\jmath}+3$ & $\hat{\jmath}+4$ \\ \hline
    2 & $\hat{\jmath}-3/2$ & & & & $\frac{15}{8}$ & -$\frac{5}{4}$ & $-\frac{1}{8}$ & & \\
    4 & $\hat{\jmath}-5/2$ & & & $\frac{315}{128}$ & $-\frac{105}{32}$
          & $\frac{189}{64}$ & $-\frac{45}{32}$ & $\frac{35}{128}$ & & \\
          & $\hat{\jmath}-3/2$ & & & $\frac{35}{128}$ & $\frac{35}{32}$ & $-\frac{35}{64}$
          & $\frac{7}{32}$ & $-\frac{5}{128}$ & & \\
    6 & $\hat{\jmath}-7/2$ & & $\frac{3003}{1024}$ & $-\frac{3003}{512}$ & $\frac{9009}{1024}$
          & $-\frac{2145}{256}$ & $\frac{5005}{1024}$ & $-\frac{819}{512}$
          & $\frac{231}{1024}$ & \\
          & $\hat{\jmath}-5/2$ & & $\frac{231}{1024}$ & $\frac{693}{512}$ & $-\frac{1155}{1024}$ & $\frac{231}{256}$
          & $-\frac{495}{1024}$ & $\frac{77}{512}$ & $-\frac{21}{1024}$ & \\
          & $\hat{\jmath}-3/2$ & & $-\frac{21}{1024}$ & $\frac{189}{512}$ & $\frac{945}{1024}$ & $-\frac{105}{256}$
          & $\frac{189}{1024}$ & $-\frac{27}{512}$ & $\frac{7}{1024}$ & \\
    8 & $\hat{\jmath}-9/2$ &  $\frac{109395}{32768}$ & $-\frac{36465}{4096}$ & $\frac{153153}{8192}$
          & $-\frac{109395}{4096}$ & $\frac{425425}{16384}$ & $-\frac{69615}{4096}$ & $\frac{58905}{8192}$
          & $-\frac{7293}{4096}$ & $\frac{6435}{32768}$ \\
          & $\hat{\jmath}-7/2$ & $\frac{6435}{32768}$ & $\frac{6435}{4096}$ & $-\frac{15015}{8192}$ & $\frac{9009}{4096}$
          & $-\frac{32175}{16384}$ & $\frac{5005}{4096}$ & $-\frac{4095}{8192}$ & $\frac{495}{4096}$
          & $-\frac{429}{32768}$ \\
          & $\hat{\jmath}-5/2$ & $-\frac{429}{32768}$ & $\frac{1287}{4096}$ & $\frac{9009}{8192}$ & $-\frac{3003}{4096}$
          & $\frac{9009}{16384}$ & $-\frac{1287}{4096}$ & $\frac{1001}{8192}$ & $-\frac{117}{4096}$
          & $\frac{99}{32768}$ \\
          & $\hat{\jmath}-3/2$ & $\frac{99}{32768}$ & $-\frac{165}{4096}$ & $\frac{3465}{8192}$ & $\frac{3465}{4096}$
          & $-\frac{5775}{16384}$ & $\frac{693}{4096}$ & $-\frac{495}{8192}$ & $\frac{55}{4096}$
          & $-\frac{45}{32768}$ \\
           \br
         \end{tabular}
       \end{indented}
\end{table}

\subsection{Finite-difference order adaptation}\label{sec:fd order}

The final ingredient in our adaptive-order scheme is changing the order of the
FD derivative used to approximate the flux divergence in Eq.~\ref{eq:sd
  conservation law}. We must first decide how to take the FD derivative,
that is,
what nodes to use. In the ECHO scheme presented in~\cite{DelZanna:2007pk}, a
high-order FD derivative is obtained by directly using a fourth-order stencil
using the interface numerical fluxes
$G_{\hat{\jmath}-3/2}, G_{\hat{\jmath}-1/2}, G_{\hat{\jmath}+1/2},
G_{\hat{\jmath}+3/2}$. Nonomura and Fujii~\cite{Nonomura20138} find that using
cell-centered fluxes in the FD derivative helps stabilize the scheme, making the
conservative FD scheme nearly as robust as FV schemes, while remaining
computationally cheaper. For a sixth-order FD derivative Nonomura and Fujii use
$G_{\hat{\jmath}-3/2}, F_{\hat{\jmath}-1}, G_{\hat{\jmath}-1/2},
G_{\hat{\jmath}+1/2}, F_{\hat{\jmath}+1}, G_{\hat{\jmath}+3/2}$. This idea was
further expanded by Chen, T\'oth, and Gombosi~\cite{CHEN2016604} (CTG), who
use
only the nearest cell-interface numerical fluxes and cell-centered fluxes
otherwise. For a sixth-order FD derivative CTG use
$F_{\hat{\jmath}-2}, F_{\hat{\jmath}-1}, G_{\hat{\jmath}-1/2},
G_{\hat{\jmath}+1/2}, F_{\hat{\jmath}+1}, F_{\hat{\jmath}+2}$. The advantage of
using only cell-centered fluxes to obtain high-order convergence is that this
minimizes reconstruction and communication between different
processors on a distributed system. CTG also write the high-order FD derivatives
as corrections to the numerical flux $G$. This makes applying a
positivity-preserving flux limiter straightforward, unlike if high-order FD
derivative stencils are used.

For the above reasons we use the approach taken by CTG, writing the numerical
flux $G$ as
\begin{equation}
  \label{eq:correction}
  G_{\hat{\jmath}+1/2}=G^{(2)}_{\hat{\jmath}+1/2} - G^{(4)}_{\hat{\jmath}+1/2} +
  G^{(6)}_{\hat{\jmath}+1/2} - G^{(8)}_{\hat{\jmath}+1/2} +
  G^{(10)}_{\hat{\jmath}+1/2},
\end{equation}
where $G^{(2)}_{\hat{\jmath}+1/2}$ is the standard numerical flux obtained by
solving the Riemann problem at the interface and
\begin{eqnarray}
  \label{eq:correction terms}
  G^{(4)}_{\hat{\jmath}+1/2}&=\frac{1}{6}\left(F_{\hat{\jmath}} -2
                              G^{(2)}_{\hat{\jmath}+1/2} +
                         F_{\hat{\jmath}+1}\right), \\
  G^{(6)}_{\hat{\jmath}+1/2}&=\frac{1}{180}\left(F_{\hat{\jmath}-1} - 9
                              F_{\hat{\jmath}} + 16 G^{(2)}_{\hat{\jmath}+1/2}
                         -9 F_{\hat{\jmath}+1} + F_{\hat{\jmath}+2}\right), \\
  G^{(8)}_{\hat{\jmath}+1/2}&=\frac{1}{2100}\left(F_{\hat{\jmath}-2} -
                              \frac{25}{3} F_{\hat{\jmath}-1}
                         + 50 F_{\hat{\jmath}} - \frac{256}{3}
                              G^{(2)}_{\hat{\jmath}+1/2} + 50
                              F_{\hat{\jmath}+1} \right. \nonumber \\
                            &\left.- \frac{25}{3}
                              F_{\hat{\jmath}+2} + F_{\hat{\jmath}+3}\right), \\
  G^{(10)}_{\hat{\jmath}+1/2}&=\frac{1}{17640}
                         \left(F_{\hat{\jmath}-3} - \frac{49}{5}
                               F_{\hat{\jmath}-2} + 49 F_{\hat{\jmath}-1}
                         - 245 F_{\hat{\jmath}} + \frac{2048}{5}
                               G^{(2)}_{\hat{\jmath}+1/2}\right.
                               \nonumber \\
                       &\left.- 245 F_{\hat{\jmath}+1}+ 49 F_{\hat{\jmath}+2} -
                         \frac{49}{5} F_{\hat{\jmath}+3} +
                         F_{\hat{\jmath}+4}\right).
\end{eqnarray}
We now need to decide which correction orders to include. CTG and Nonomura
and Fujii use one order higher than their reconstruction scheme. That is, if
fifth-order reconstruction is used, a sixth-order FD derivative is used, which
would mean we include $G^{(4)}$ and $G^{(6)}$. We store the polynomial order
used for reconstruction in each cell. Then the order we use at
$\hat{\jmath}+1/2$ is given by $\mathcal{O}_{\hat{\jmath}+1/2} =
\min(\mathcal{O}_{\hat{\jmath}},\mathcal{O}_{\hat{\jmath}+1})$, guaranteeing
that we do not differentiate across a discontinuity. For example, if cell
$\hat{\jmath}$ used fifth-order reconstruction and cell $\hat{\jmath}+1$ used
second-order reconstruction, $G_{\hat{\jmath}+1/2}=G^{(2)}_{\hat{\jmath}+1/2}$.

\subsection{Extension to higher dimensions}\label{sec:higher dimensions}

The extension to higher dimensions is straightforward since for a
finite-difference scheme each dimension can be treated separately.

\subsection{Weighted adaptive order schemes}\label{sec:weighted ao}

We can hybridize the multiple stencils into a single WENO-type stencil, replacing
the conditionals by shifted $\tanh$ functions. We denote the weight functions
for hybridization by $\Theta_{(N)}$ where $N$ is the degree of the
scheme. $\Theta_{(N)}$ is one where the stencil should be used and zero where
the stencil is invalid. Denoting the stencil of order $N$ by $S^{(N)}$ we write
the PPAO9-5-3-1 scheme as:
\begin{eqnarray}
  \label{eq:single ppao9-5-3-1}
  p_{\hat{\jmath}\pm1/2}
  &=S^{(8)}_{\hat{\jmath}\pm1/2}\Theta_{(8)} \nonumber \\
  &+\left(1 - \Theta_{(8)}\right)
    \left\{S^{(4)}_{\hat{\jmath}\pm1/2}\Theta_{(4)}\right.
     \nonumber  \\
  &+\left.\left(1 - \Theta_{(4)}\right)
    \left[S^{(2)}_{\hat{\jmath}\pm1/2}\Theta_{(2)}
    + \left(1 - \Theta_{(2)}\right) S^{(0)}_{\hat{\jmath}\pm1/2}
    \right]\right\}.
\end{eqnarray}
The weight functions based on the oscillation indicators are
\begin{equation}
  \label{eq:weight functions}
  \Theta_N(s^{\hat{\jmath}}_N;\alpha_N,\gamma) = \frac{1}{1 + \exp\left(-\gamma
  s^{\hat{\jmath}}_N + \alpha_N \gamma\right)},
\end{equation}
where $\gamma$ controls how quickly the transition occurs and should be chosen
the same for all stencils. For positivity preservation we use the weight
function
\begin{equation}
  \label{eq:positivity weight}
  \Theta_p\left(S^{(N)}_{\hat{\jmath}\pm1/2}; p_{\min}\right) =
  \frac{1}{1 + \exp\!\left[-(100/p_{\min}) S^{(N)}_{\hat{\jmath}\pm1/2} +
  50\right]},
\end{equation}
where $p_{\min}$ is the value above which the stencil should be used. That is,
if we require the stencil $S^{(N)}$ be used when its value at the reconstruction
point is equal to $10^{-13}$ then $p_{\min}=10^{-13}$ should be used. When both
smoothness and positivity need to be enforced, the weight functions can simply be
multiplied together. In this way a general weight function can be obtained that
enforces any number of constraints. Finally, the order of the reconstructed
polynomial is given by
\begin{equation}
  \label{eq:weighted order}
  \mathcal{O}_{\hat{\jmath}}=9\Theta_{(8)} + \left(1 - \Theta_{(8)}\right)
  \left\{5\Theta_{(4)} + \left(1 -\Theta_{(4)}\right)
  \left[3\Theta_{(2)} + \left(1 - \Theta_{(2)}\right)1\right]\right\}.
\end{equation}

\section{Numerical results}\label{sec:numerical results}

We now present a series of numerical tests to demonstrate the capabilities of
our PPAO schemes. Unless stated otherwise, all tests use a third-order
strong-stability-preserving Runge-Kutta scheme (SSP-RK3) for the time
evolution\cite{Shu1988439}, and an HLL Riemann solver~\cite{Harten1983}. We
solve the GRMHD equations in conservative form with divergence cleaning. Unless
stated otherwise, the second-order reconstruction scheme is monotonized
central~\cite{VANLEER1977276}. See~\cite{Deppe:2021bhi} for details of the
implementation in \texttt{SpECTRE} and~\cite{2006ApJ...637..296A, Font:2008fka,
  Baumgarte:2010ndz} for more detailed discussions of the GRMHD system. All
simulations are performed using \texttt{SpECTRE}~\cite{spectrecode} with the
implementation of the algorithms described being publicly available.
All features used to perform the simulations described are available in the
v2023.04.07 release of \texttt{SpECTRE}~\cite{spectrecode}.

\blue{The goal of this section is to get an overview of whether certain choices
  of reconstruction and FD derivative order are significantly better or worse
  than others. When comparing schemes we name them both by the PPAO approach and
  the associated FD derivative approach. PPAO9-5-2-1 means that first unlimited
  ninth-order reconstruction is attempted, if that fails unlimited fifth-order
  reconstruction is attempted, if that fails second-order reconstruction is
  attempted, and if that results in an unphysical solution first-order
  reconstruction is used. By comparison, PPAO5-2-1 means unlimited fifth-order
  reconstruction is attempted, if that fails second-order reconstruction is
  attempted, if that results in an unphysical solution first-order
  reconstruction is used. When using the same FD derivative order independent of
  the reconstruction order we use notation like FD-8 for eighth-order FD
  derivatives, and FD-2 for second-order FD derivatives. When using
  adaptive-order FD derivatives, we note the order associated with each
  reconstruction order. For example, if we use PPAO9-5-2-1 with FD-10-6-2-2
  (denoted PPAO9-5-2-1+FD-10-6-2-2), then for ninth-order reconstruction we use
  tenth-order derivatives, for fifth-order reconstruction we use sixth-order
  derivatives, for second-order reconstruction we use second-order derivatives,
  and for first-order reconstruction we also use second-order derivatives.}

\subsection{1d Smooth Flow}\label{sec:Smooth Flow}

We consider a simple 1d smooth flow problem to verify that the algorithm
converges at the expected order for smooth solutions in the absence of magnetic
fields. A smooth density perturbation is advected across the domain with a
velocity $v^i$. The analytic solution is given by
\begin{eqnarray}
  \rho&=1 + 0.7 \sin[k^i (x^i-v^it)], \\
  v^i&=(0.8,0,0),\\
  k^i&=(1,0,0),\\
  p&=1,\\
  B^i&=(0,0,0),
\end{eqnarray}
and we close the system with an adiabatic equation of state,
\begin{equation}
  \label{eq:ideal fluid eos}
  p = \rho\epsilon\left(\Gamma-1\right),
\end{equation}
where $\Gamma$ is the adiabatic index, which we set to 1.4. We use a domain
given by $[0,2\pi]^3$, and apply periodic boundary conditions in all directions.
The time step size is $\Delta t = 2\pi/ 5120$ so that the spatial discretization
error is larger than the time stepping error for all resolutions that we
use. The final time is chosen to be $2\pi$, and we use a 5th-order
Dormand-Prince time integrator~\cite{DORMAND19861007, Hairer:1993a,
  NumericalRecipes}. The high-order time stepper combined with the small step
size ensure that the spatial errors dominate.

We perform convergence tests at different FD derivative orders and present the
results in table~\ref{tab:Smooth flow errors}. We show both the $L_2$ norm of
the error and the convergence rate. The $L_2$ norm is defined as
\begin{equation}
  \label{eq:L2 norm}
  L_2(u)=\sqrt{\frac{1}{M}\sum_{i=0}^{M-1}u_i^2},
\end{equation}
where $M$ is the total number of grid points and $u_i$ is the value of $u$ at
grid point $i$ and the convergence order is given by
\begin{equation}
  \label{eq:convergence order}
  L_2\;\mathrm{Order} =
  \log_2\left[\frac{L_2(\mathcal{E}_{N_x/2})}{L_2(\mathcal{E}_{N_x})}\right].
\end{equation}
We always use the PPAO9-5-2-1 scheme, but since the solution is smooth
ninth-order reconstruction is used. We observe the expected convergence rate
except for the 8th-order derivative case, where the convergence order is
slightly above the expected rate. This ultimately demonstrates that our PPAO
scheme is able to achieve high-order convergence for smooth solutions in the
absence of magnetic fields.

\begin{table}
  \caption{\label{tab:Smooth flow errors} The errors and local convergence order
    for the smooth flow problem using different FD derivative orders but always
    using a PPAO9-5-2-1 reconstruction. Since the solution is smooth the
    reconstruction always ends up being 9th order. We observe the expected
    convergence rate except for the 8th-order derivative it is slightly
    higher than expected.}
  \begin{indented}
    \lineup
  \item[]\begin{tabular}{@{\extracolsep{\fill}}cccc}
    \br
    Derivative Order & $N_x$ & $L_2(\mathcal{E}(\rho))$ & $L_2$ Order\\
    \hline
    2 & 11 & 2.41440e-2 & \\
    & 22 & 6.04972e-3 & \02.00\\
    & 44 & 1.51327e-3 & \02.00\\
    & 88 & 3.78368e-4 & \02.00\\
    \hline
    4 & 11 & 2.81416e-4 & \\
    & 22 & 1.76480e-5 & \04.00\\
    & 44 & 1.10441e-6 & \04.00\\
    & 88 & 6.90479e-8 & \04.00\\
    \hline
    6 & 11 & 7.40386e-6 & \\
    & 22 & 9.86855e-8 & \06.23\\
    & 44 & 1.53525e-9 & \06.01\\
    & 88 & 2.39498e-11 & \06.00\\
    \hline
    8 & 11 & 3.25675e-6 & \\
    & 22 & 6.79011e-9 & \08.91\\
    & 44 & 1.37058e-11 & \08.95\\
    & 88 & 1.19152e-13 & \06.85\\
    \hline
    10 & 11 & 3.27319e-6 & \\
    & 22 & 6.79768e-9 & \08.91\\
    & 44 & 1.35104e-11 & \08.97\\
    & 88 & 1.15890e-13 & \06.87\\
    \hline
    10-6-2-2 & 11 & 3.27319e-6 & \\
    & 22 & 6.79768e-9 & \08.91\\
    & 44 & 1.35104e-11 & \08.97\\
    & 88 & 1.15890e-13 & \06.87\\
    \hline
    10-4-2-2 & 11 & 3.27319e-6 & \\
    & 22 & 6.79768e-9 & \08.91\\
    & 44 & 1.35104e-11 & \08.97\\
    & 88 & 1.15890e-13 & \06.87\\
           \br
         \end{tabular}
       \end{indented}
\end{table}

\subsection{Alfv\'en Wave}\label{sec:alfven wave}

We now verify that our scheme achieves high-order convergence in the
presence of magnetic fields. To achieve this we evolve an Alfv\'en wave across
the domain~\cite{DelZanna:2007pk}. The magnetic field $\textbf{B}$ in the Alfv\'en wave
solution is the sum of a background static magnetic field $\textbf{B}_0$ and a
transverse time-dependent magnetic field $\textbf{B}_1$.  We define the
auxiliary magnetic velocities $v_{B_0}$ and $v_{B_1}$ as
\begin{equation}
  v^2_{B_0} = \frac{(B_0)^2}{\rho h + B^2}, \;\;\;\;\;
  v^2_{B_1} = \frac{(B_1)^2}{\rho h + B^2}.
\end{equation}
The Alfv\'en speed and fluid speed are then given by:
\begin{equation}
  v^2_A = \frac{v^2_{B_0}}{\frac{1}{2} +
           \sqrt{\frac{1}{4}-v^2_{B_0}v^2_{B_1}}}, \;\;\;\;\;
  v^2_f = \frac{v^2_{B_1}}{\frac{1}{2} +
           \sqrt{\frac{1}{4}-v^2_{B_0}v^2_{B_1}}}.
\end{equation}
We use an ideal fluid equation of state with $\Gamma=1.6$

The circularly polarized wave is best described in a basis aligned with the
initial magnetic fields at the origin. To this end we define the unit vectors
$\hat{\textbf{b}}_0$, $\hat{\textbf{b}}_1$, and $\hat{\textbf{e}}$ as
\begin{equation}
  \hat{\textbf{b}}_0 = \frac{\textbf{B}_0}{B_0},\;\;\;\;\;
  \hat{\textbf{b}}_1 =\left.\frac{\textbf{B}_1}{B_1}
                       \right\vert_{\textbf{x}=\textbf{0},t=0},\;\;\;\;\;
  \hat{\textbf{e}} = \hat{\textbf{b}}_0\times\hat{\textbf{b}}_1.
\end{equation}
The analytic solution is given by
\begin{eqnarray}
  \textbf{v}(\textbf{x},t) &=
                         -v_f\left[\cos(\delta\phi)\hat{\textbf{b}}_1 +
                         \sin(\delta\phi)\hat{\textbf{e}}\right], \\
  \textbf{B}(\textbf{x},t) &= \textbf{B}_0 +
                         B_1\left[\cos(\delta\phi)\hat{\textbf{b}}_1 +
                         \sin(\delta\phi)\hat{\textbf{e}}\right],
\end{eqnarray}
with $\delta\phi$ given by
\begin{equation}
  \delta\phi(\textbf{x},t)=k\left(\textbf{x}\cdot\hat{\textbf{b}}_0-v_A
  t\right),
\end{equation}
where $k$ is the wave number, and $\rho$ and $p$ are spatially constant. We
choose $\rho=1$, $p=1$, $k=\sqrt{3}$, $\textbf{B}_0 = [1,1,1]$, and
$\textbf{B}_1(\textbf{0},0)
=\left[\sqrt{2},-1/\sqrt{2},-1/\sqrt{2}\right]$.

We use an adaptive Dormand-Prince 5~\cite{DORMAND19861007, Hairer:1993a,
  NumericalRecipes} time integrator with an absolute tolerance of $10^{-15}$ and
a relative tolerance of $10^{-13}$, and evolve to a final time of $\pi$. In
table~\ref{tab:alfven wave errors} we present convergence results of $B^i$ at
the final time. We always use the PPAO9-5-2-1 scheme, but since the solution is
smooth, ninth-order reconstruction is used. From table~\ref{tab:alfven wave
  errors} we see that we get high-order convergence for the Alfv\'en wave
problem, though when using tenth-order FD derivatives we still only achieve
eighth-order convergence. We have not yet understood why this is, though given
the time stepper tolerance and the small errors, we are not concerned about
it. In realistic astrophysical simulations one is highly unlikely
to be able to achieve relative errors of $10^{-12}$.

\begin{table}
  \caption{\label{tab:alfven wave errors} The errors and local $L_2$ convergence
    order for the Alfv\'en wave problem using different FD derivative orders but
    always using a PPAO9-5-2-1 reconstruction. We show the error and order for
    each component of the magnetic field separately, and in all cases observe
    high-order convergence. We have not yet understood why the tenth-order
    derivatives only converge at eighth order, though given the time stepper
    tolerance and the small errors, we are not concerned about it.}
  \centering
    \begin{indented}
    \lineup
  \item[]\begin{tabular}{@{\extracolsep{\fill}}cccccccc}
    \br
    $\mathcal{O}(\mathrm{FD})$ & $N_x$ & $L_2(\mathcal{E}(B^x))$ & $\mathcal{O}(B_x)$ &
    $L_2(\mathcal{E}(B^y))$ & $\mathcal{O}(B^y)$ & $L_2(\mathcal{E}(B^z))$ &
    $\mathcal{O}(B^z)$\\
    \hline
    2 & 22 & 1.08968e-2 &  &  1.08968e-2 &  &  1.08968e-2 & \\
    & 44 & 2.72631e-3 & \02.00 &  2.72631e-3 & \02.00 &  2.72631e-3 & \02.00\\
    & 88 & 6.81709e-4 & \02.00 &  6.81709e-4 & \02.00 &  6.81709e-4 & \02.00\\
    \hline
    4 & 22 & 4.43192e-5 &  &  4.43192e-5 &  &  4.43192e-5 & \\
    & 44 & 2.77829e-6 & \04.00 &  2.77829e-6 & \04.00 &  2.77829e-6 & \04.00\\
    & 88 & 1.73745e-7 & \04.00 &  1.73745e-7 & \04.00 &  1.73745e-7 & \04.00\\
    \hline
    6 & 22 & 3.10693e-7 &  &  3.10693e-7 &  &  3.10693e-7 & \\
    & 44 & 4.95066e-9 & \05.97 &  4.95066e-9 & \05.97 &  4.95066e-9 & \05.97\\
    & 88 & 7.89883e-11 & \05.97 &  7.89881e-11 & \05.97 &  7.89882e-11 & \05.97\\
    \hline
    8 & 22 & 1.45926e-7 &  &  1.45926e-7 &  &  1.45926e-7 & \\
    & 44 & 5.66948e-10 & \08.01 &  5.66949e-10 & \08.01 &  5.66949e-10 & \08.01\\
    & 88 & 5.58414e-12 & \06.67 &  5.58425e-12 & \06.67 &  5.58425e-12 & \06.67\\
    \hline
    10 & 22 & 1.49272e-7 &  &  1.49272e-7 &  &  1.49272e-7 & \\
    & 44 & 5.80746e-10 & \08.01 &  5.80746e-10 & \08.01 &  5.80745e-10 & \08.01\\
    & 88 & 5.63870e-12 & \06.69 &  5.63867e-12 & \06.69 &  5.63866e-12 & \06.69\\
    \hline
    10-6-2-2 & 22 & 1.49272e-7 &  &  1.49272e-7 &  &  1.49272e-7 & \\
    & 44 & 5.80746e-10 & \08.01 &  5.80746e-10 & \08.01 &  5.80745e-10 & \08.01\\
    & 88 & 5.63870e-12 & \06.69 &  5.63867e-12 & \06.69 &  5.63866e-12 & \06.69\\
    \hline
    10-4-2-2 & 22 & 1.49272e-7 &  &  1.49272e-7 &  &  1.49272e-7 & \\
    & 44 & 5.80746e-10 & \08.01 &  5.80746e-10 & \08.01 &  5.80745e-10 & \08.01\\
    & 88 & 5.63870e-12 & \06.69 &  5.63867e-12 & \06.69 &  5.63866e-12 & \06.69\\
    \br
         \end{tabular}
       \end{indented}
\end{table}

\subsection{1d Riemann Problems}

Having verified the order of convergence of our PPAO scheme for smooth
solutions, we now turn to testing them in the presence of
discontinuities. One-dimensional Riemann problems are a standard test for any
scheme that must be able to handle shocks. We will focus on the five Riemann
problems (RP-5) of~\cite{Balsara2001} and the fast shock problem
of~\cite{Komissarov1999}. The initial conditions for the Riemann problems are
given in table~\ref{tab:Rp1 conditions}. We use a domain given by
$[-0.5,0.5]\times[-1,1]^2$, with 704 grid points in the $x$-direction and 11
grid points in $y$ and $z$, analytic boundary conditions in the $x$-direction
and periodic in $y$ and $z$, and use a time step size $\Delta
t=5\times10^{-4}$. The initial conditions for the fast shock are given
by~\cite{Komissarov1999}
\begin{eqnarray}
  \rho&=\left\{
        \begin{array}{ll}
          1.0,
          & \mathrm{if} \; x \le 0.0 \\
          25.48, & x > 0.0,
        \end{array}\right. \\
  p&=\left\{
     \begin{array}{ll}
       1.0,
       & \mathrm{if} \; x \le 0.0 \\
       367.5, & x > 0.0,
     \end{array}\right. \\
  B^i&=\left\{
       \begin{array}{ll}
         (20.0, 25.02, 0.0),
         & \mathrm{if} \; x \le 0.0 \\
         (20.0, 49.0, 0.0), & x > 0.0,
       \end{array}\right. \\
  u^i&=\left\{
       \begin{array}{ll}
         (25.0, 0.0, 0.0),
         & \mathrm{if} \; x \le 0.0 \\
         (1.091, 0.392, 0.0), & x > 0.0,
       \end{array}\right. \\
\end{eqnarray}
using a domain $[-0.5,2]\times[-1,1]^2$ with 352 grid points in the
$x$-direction and 11 grid points in $y$ and $z$, analytic boundary conditions in
all directions, and we use a time step size of $\Delta t=10^{-3}$.

\begin{table}
  \caption{\label{tab:Rp1 conditions}The initial conditions for Riemann Problems
    1, 2, 3, 4, and 5 of~\cite{Balsara2001}. The domain is $x\in[-0.5,0.5]$, the
    final time is $t_f=0.4$ for problems 1--4 and $t_f=0.55$ for problem 5. An
    ideal fluid equation of state is used with an adiabatic index of 2 for
    Riemann Problem 1 and $5/3$ for Riemann Problems 2, 3, 4, and 5.}

  \begin{indented}
    \lineup
  \item[]\begin{tabular}{@{\extracolsep{\fill}}lccccc}
    \br
    Problem &  & $\rho$ & $p$ & $v^i$ & $B^i$ \\
    \hline
    RP 1 & $x < 0$ & 1.0\0\0 & \0\0\01.0 & $(\phantom{-}0.0\0\0,\phantom{-}0.0,0.0)$
                                      & $(\00.5,\phantom{-}1.0,\phantom{-}0.0)$ \\
            & $x \ge 0$ & 0.125 & \0\0\00.1 & $(\phantom{-}0.0\0\0,\phantom{-}0.0,0.0)$
                                      & $(\00.5,-1.0,\phantom{-}0.0)$ \\ \hline
    RP 2 & $x < 0$ & 1.0\0\0 & \0\030.0 & $(\phantom{-}0.0\0\0,\phantom{-}0.0,0.0)$
                                      & $(\05.0,\phantom{-}6.0,\phantom{-}6.0)$ \\
            & $x \ge 0$ & 1.0\0\0 & \0\0\01.0 & $(\phantom{-}0.0\0\0,\phantom{-}0.0,0.0)$
                                      & $(\05.0,\phantom{-}0.7,\phantom{-}0.7)$ \\ \hline
    RP 3 & $x < 0$ & 1.0\0\0 & 1000.0 & $(\phantom{-}0.0\0\0,\phantom{-}0.0,0.0)$
                                      & $(10.0,\phantom{-}7.0,\phantom{-}7.0)$ \\
            & $x \ge 0$ & 1.0\0\0 & \0\0\00.1 & $(\phantom{-}0.0\0\0,\phantom{-}0.0,0.0)$
                                      & $(10.0,\phantom{-}0.7,\phantom{-}0.7)$ \\ \hline
    RP 4 & $x < 0$ & 1.0\0\0 & \0\0\00.1 & $(\phantom{-}0.999,\phantom{-}0.0,0.0)$
                                      & $(10.0,\phantom{-}7.0,\phantom{-}7.0)$ \\
            & $x \ge 0$ & 1.0\0\0 & \0\0\00.1 & $(-0.999,\phantom{-}0.0,0.0)$
                                      & $(10.0,-7.0,-7.0)$ \\ \hline
    RP 5 & $x < 0$ & 1.08\0 & \0\0\00.95 & $(\phantom{-}0.4\0\0,\phantom{-}0.3,0.2)$
                                      & $(\02.0,\phantom{-}0.3,\phantom{-}0.3)$ \\
            & $x \ge 0$ & 1.0\0\0 & \0\0\01.0 & $(-0.45\0,-0.2,0.2)$
                                      & $(\02.0,-0.7,\phantom{-}0.5)$ \\
    \br
  \end{tabular}
\end{indented}
\end{table}

\begin{figure}[h]
  \raggedleft
  \begin{minipage}{0.48\columnwidth}
    \centering
    \includegraphics[width=1.0\textwidth]{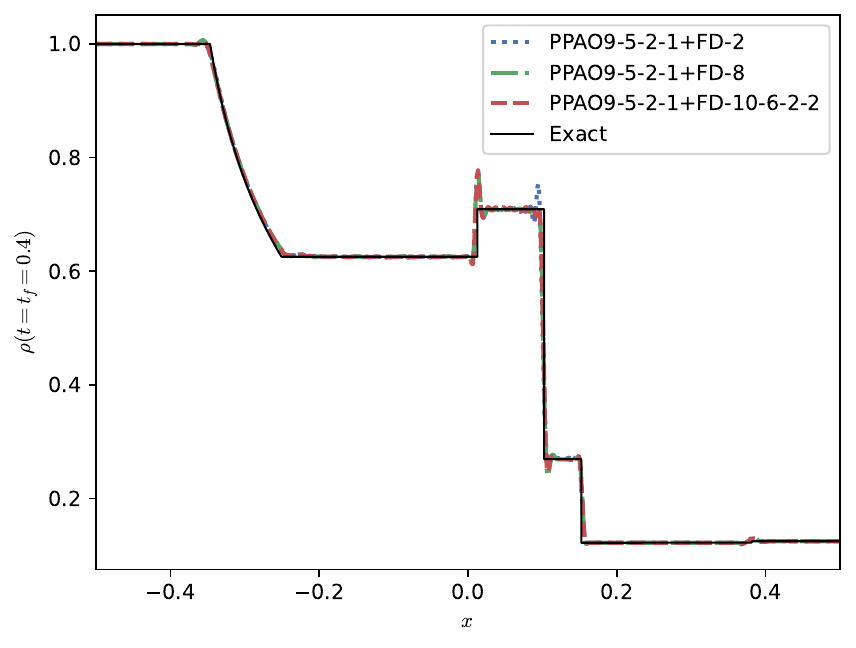}
    \\
    RP1: $\rho$
  \end{minipage}
  \begin{minipage}{0.48\columnwidth}
    \centering
    \includegraphics[width=1.0\textwidth]{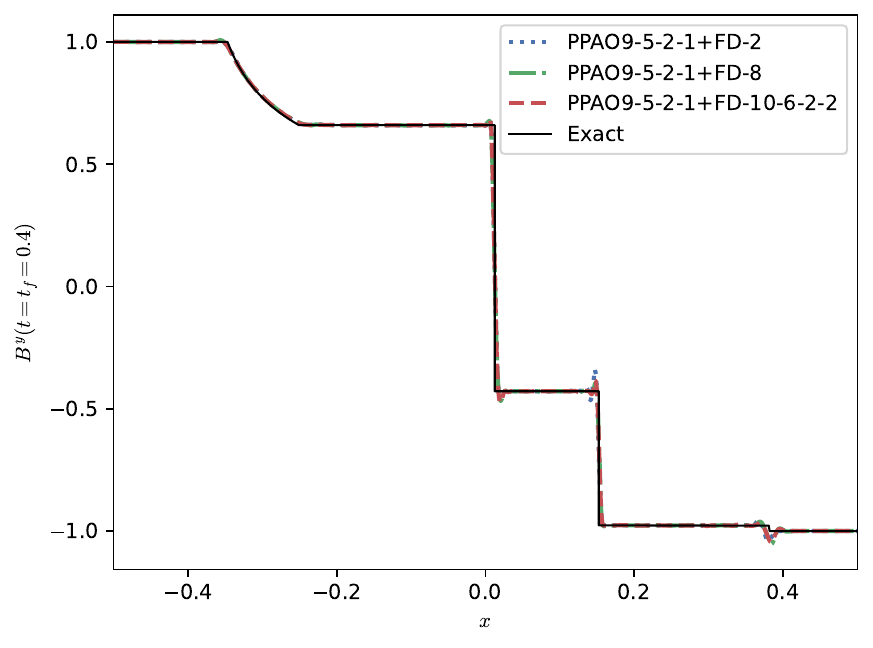}
    \\
    RP1: $B^y$
  \end{minipage}

  \begin{minipage}{0.48\columnwidth}
    \centering
    \includegraphics[width=1.0\textwidth]{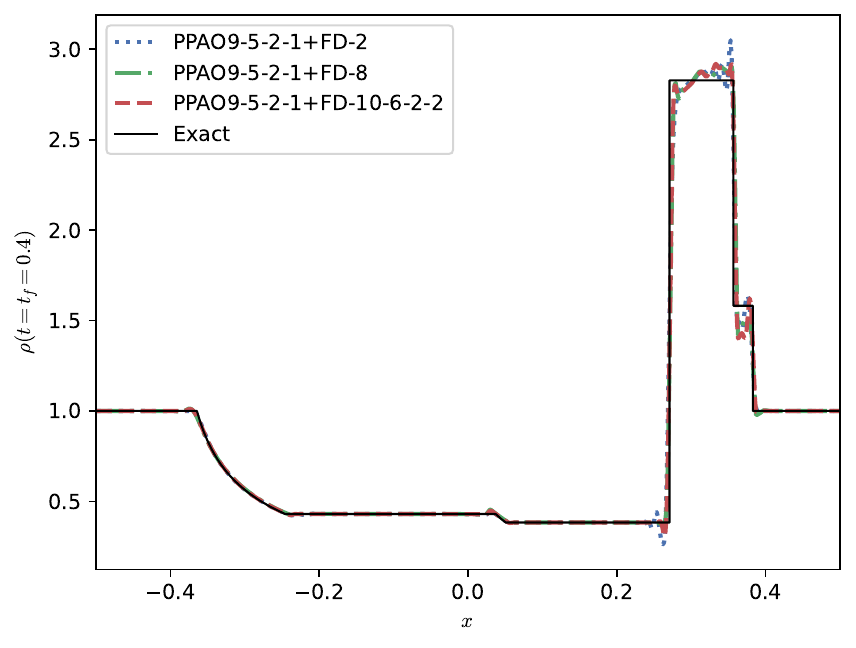}
    \\
    RP2: $\rho$
  \end{minipage}
  \begin{minipage}{0.48\columnwidth}
    \centering
    \includegraphics[width=1.0\textwidth]{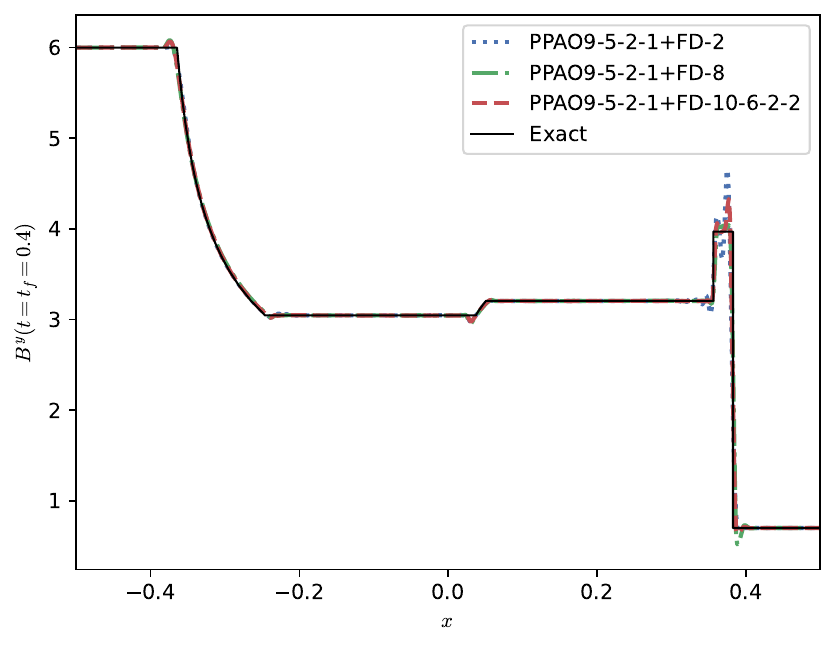}
    \\
    RP2: $B^y$
  \end{minipage}

  \begin{minipage}{0.48\columnwidth}
    \centering
    \includegraphics[width=1.0\textwidth]{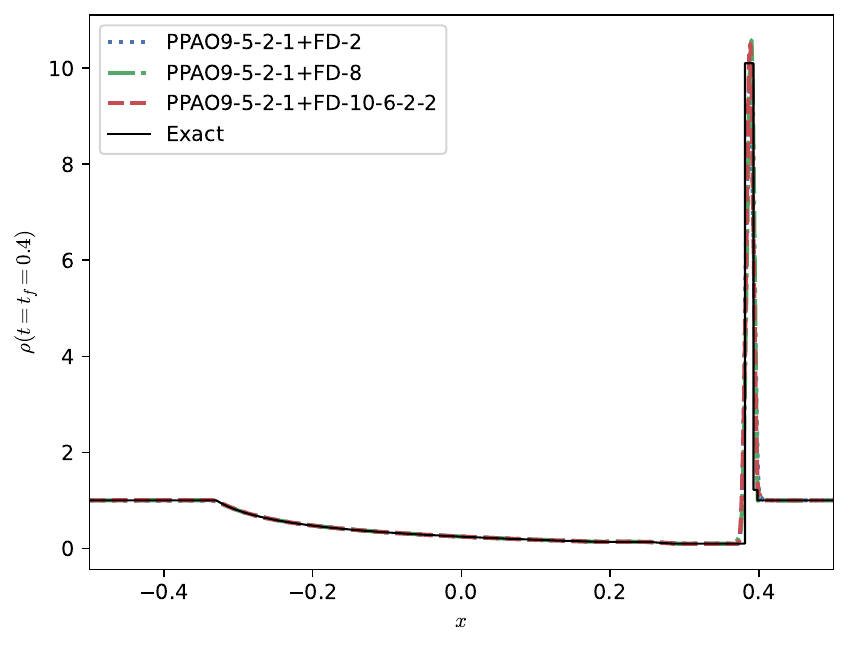}
    \\
    RP3: $\rho$
  \end{minipage}
  \begin{minipage}{0.48\columnwidth}
    \centering
    \includegraphics[width=1.0\textwidth]{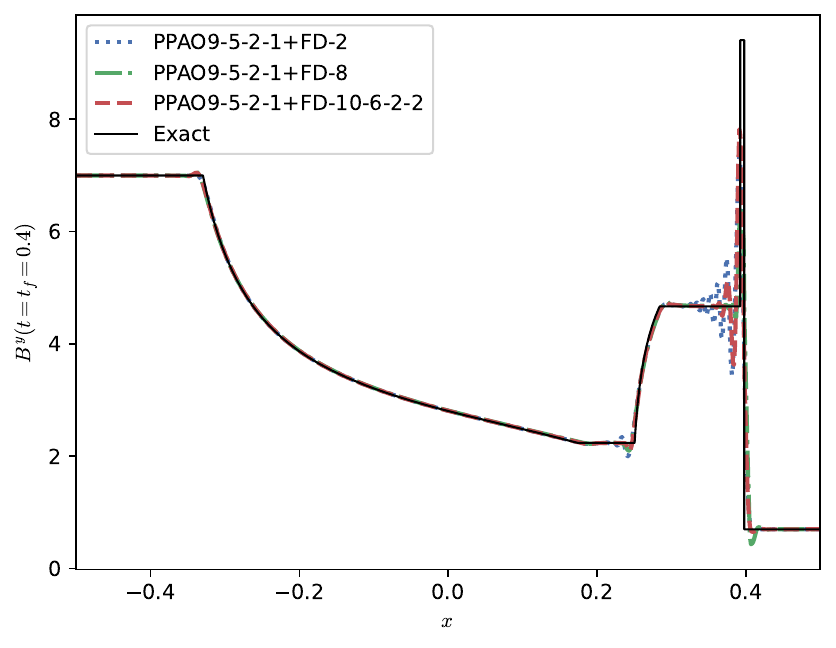}
    \\
    RP3: $B^y$
  \end{minipage}
  \caption{The left panels show the rest mass density $\rho$ at the final for
    Riemann problems 1-3 going from top to bottom, while the right panel shows
    $B^y$ at the final time. In all cases we show the exact solution in solid
    black and our PPAO-9-5-2-1 scheme with different derivative orders. We see
    that for all three Riemann problems the adaptive-order derivatives provide
    the best balance between accurately resolving shocks and minimizing
    oscillations near discontinuities.
  \label{fig:Rp1-3}}
\end{figure}

\begin{figure}[h]
  \raggedleft
  \begin{minipage}{0.48\columnwidth}
    \centering
    \includegraphics[width=1.0\textwidth]{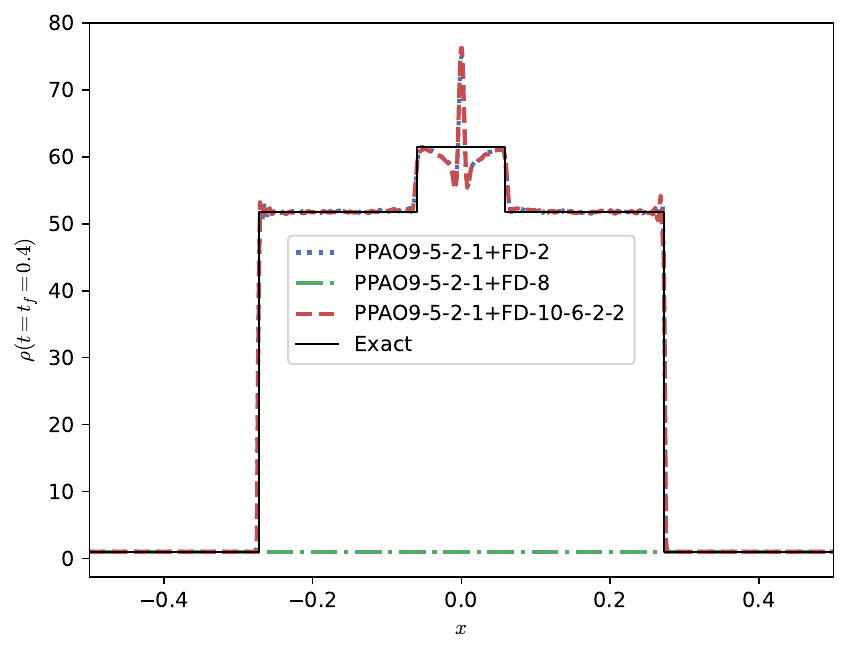}
    \\
    RP4: $\rho$
  \end{minipage}
  \begin{minipage}{0.48\columnwidth}
    \centering
    \includegraphics[width=1.0\textwidth]{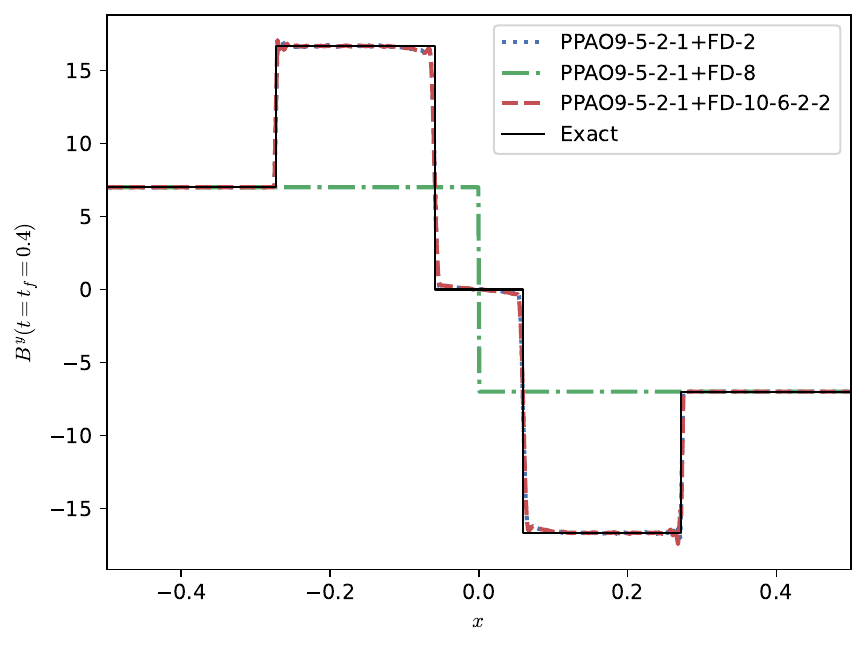}
    \\
    RP4: $B^y$
  \end{minipage}

  \begin{minipage}{0.48\columnwidth}
    \centering
    \includegraphics[width=1.0\textwidth]{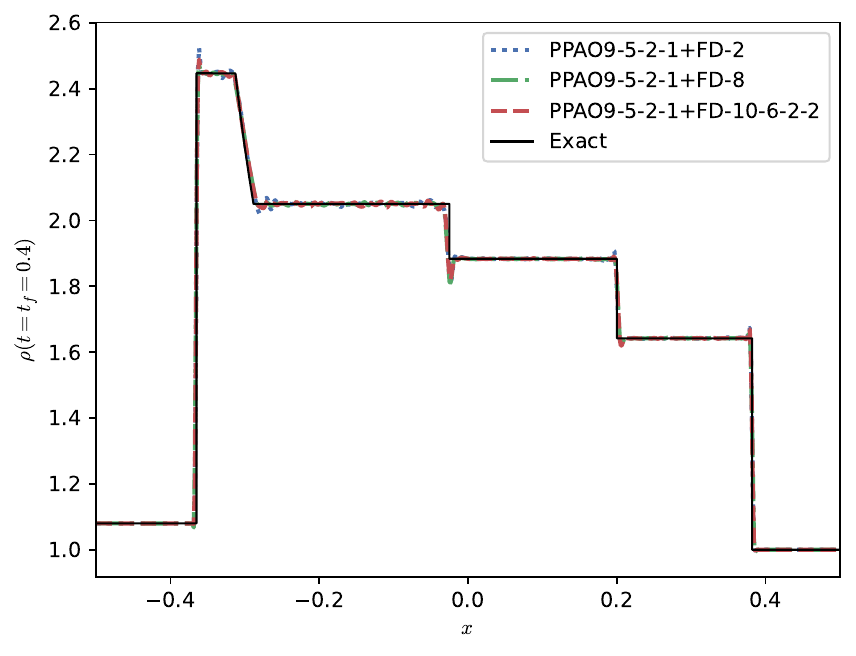}
    \\
    RP5: $\rho$
  \end{minipage}
  \begin{minipage}{0.48\columnwidth}
    \centering
    \includegraphics[width=1.0\textwidth]{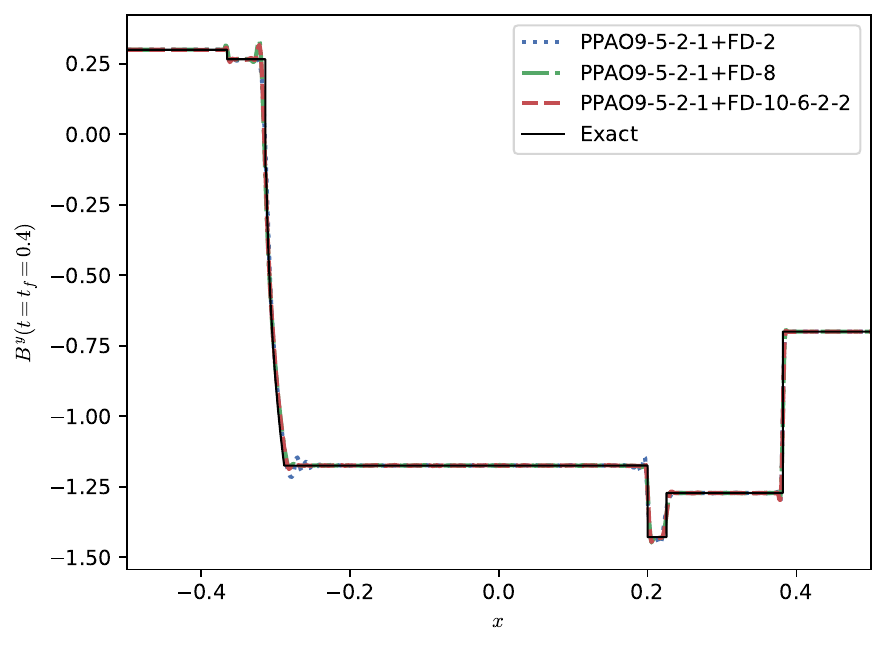}
    \\
    RP5: $B^y$
  \end{minipage}

  \begin{minipage}{0.48\columnwidth}
    \centering
    \includegraphics[width=1.0\textwidth]{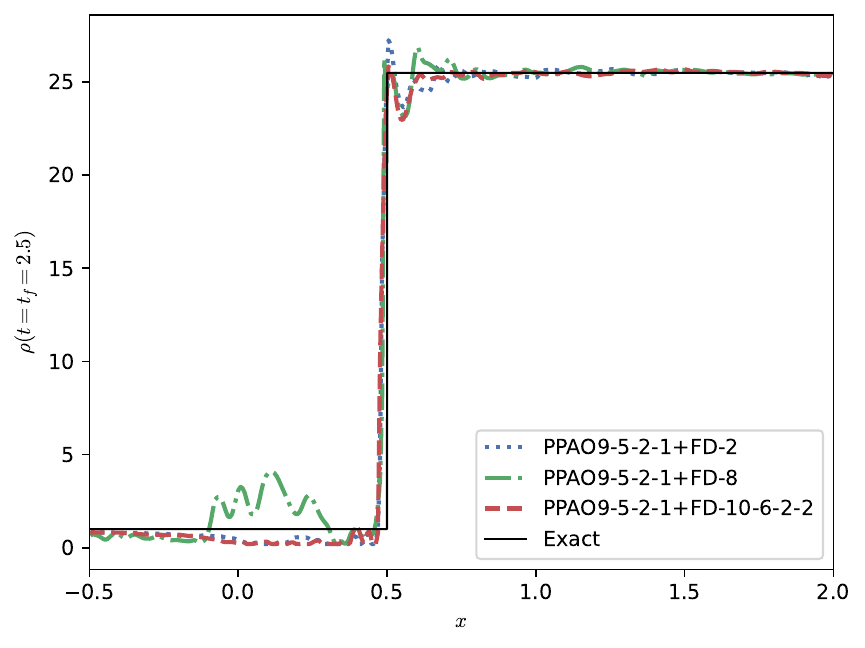}
    \\
    Fast Shock: $\rho$
  \end{minipage}
  \begin{minipage}{0.48\columnwidth}
    \centering
    \includegraphics[width=1.0\textwidth]{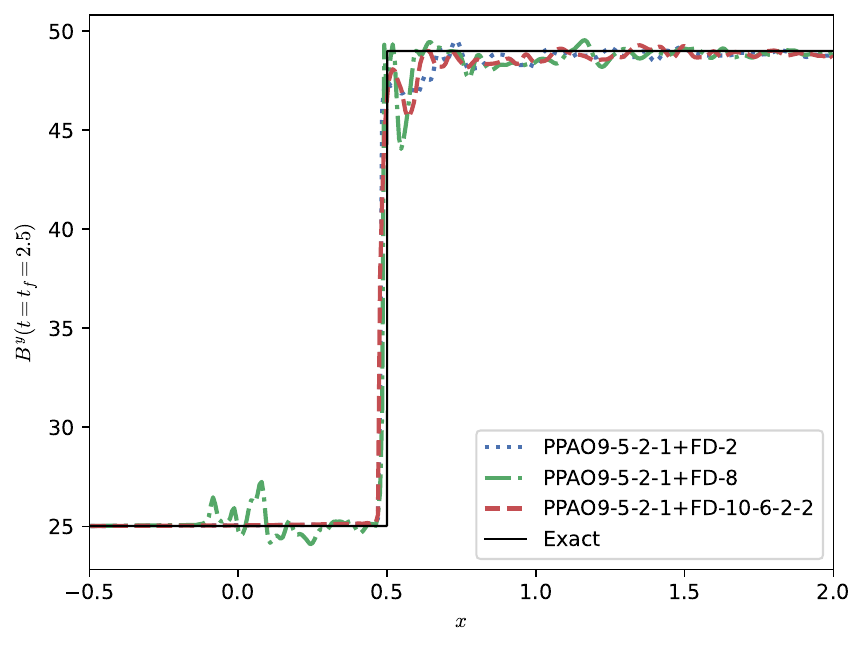}
    \\
    Fast Shock: $B^y$
  \end{minipage}
  \caption{The left panels show the rest mass density $\rho$ at the final for
    Riemann problems 4-5 and the Komissarov fast shock going from top to bottom,
    while the right panel shows $B^y$ at the final time. In all cases we show
    the exact solution in solid black and our PPAO-9-5-2-1 scheme with different
    derivative orders. The FD-8 solver fails for RP4 and does quite poorly for
    the fast shock problem. In all cases the adaptive order scheme is accurate
    and robust.
  \label{fig:Rp4-5 Fast Shock}}
\end{figure}

We plot the rest mass density $\rho$ at the final time in the left panels of
figure~\ref{fig:Rp1-3} and figure~\ref{fig:Rp4-5 Fast Shock}, and the $y$-component
of the magnetic field $B^y$ at the final time in the right panels for the
various Riemann problems and the fast shock problem. The PPAO9-5-2-1
reconstruction scheme is always used and different FD derivative orders are
shown. We find that FD-8 fails for RP4, while both FD-2 and the adaptive
FD-10-6-2-2 are robust. For RP3, we see that the adaptive-order FD derivatives
leads to larger oscillations in $B^y$ around $x=0.35$ than the FD-8 scheme. This
is because the FD-8 scheme is more dissipative than the adaptive approach, and
so there is essentially a low-pass filter applied that smooths out additional
oscillations, be they physical or unphysical. In all cases we see that adapting
the order of the FD
derivative increases accuracy without additional Gibbs phenomena or decreased
shock capturing compared to the second-order scheme. This demonstrates that the
combined PPAO9-5-2-1+FD-10-6-2-2 scheme is able to achieve high-order
convergence in smooth regions while accurately and robustly capturing strong
shocks and discontinuities, at least for 1d test problems. We explore the
scheme's capabilities of handling 2d and 3d problems below.

\subsection{2d Cylindrical Blast Wave}

The cylindrical blast wave~\cite{2005A&A...436..503L,DelZanna:2007pk} is a
standard test problem in which a magnetized fluid obeying a $\Gamma=4/3$ ideal
fluid equation of state starts at rest in a constant magnetic field along the
$x$-axis is evolved. A dense cylinder is surrounded by a lower density fluid
into which the cylinder expands. The presence of a magnetic field causes the
expansion to be non-axially symmetric. The initial density $\rho$ and pressure
$p$ of the fluid are
\begin{eqnarray}
  \rho(r < 0.8) & = 10^{-2}, \\
  \rho(r > 1.0) & = 10^{-4}, \\
  p(r < 0.8) & = 1, \\
  p(r > 1.0) & = 5 \times 10^{-4}.
\end{eqnarray}
In the region $0.8 \leq r \leq 1$, the solution transitions such that the
logarithms of the pressure and density are linear functions of $r$.  The fluid
begins threaded with a magnetic field:
\begin{equation}
  (B^x, B^y, B^z) = (0.1, 0, 0).
\end{equation}
For all simulations we use a time step size $\Delta t=10^{-2}$.

\begin{figure}[h]
  \raggedleft
  \begin{minipage}{0.43\columnwidth}
    \centering
    \includegraphics[width=1.0\textwidth]{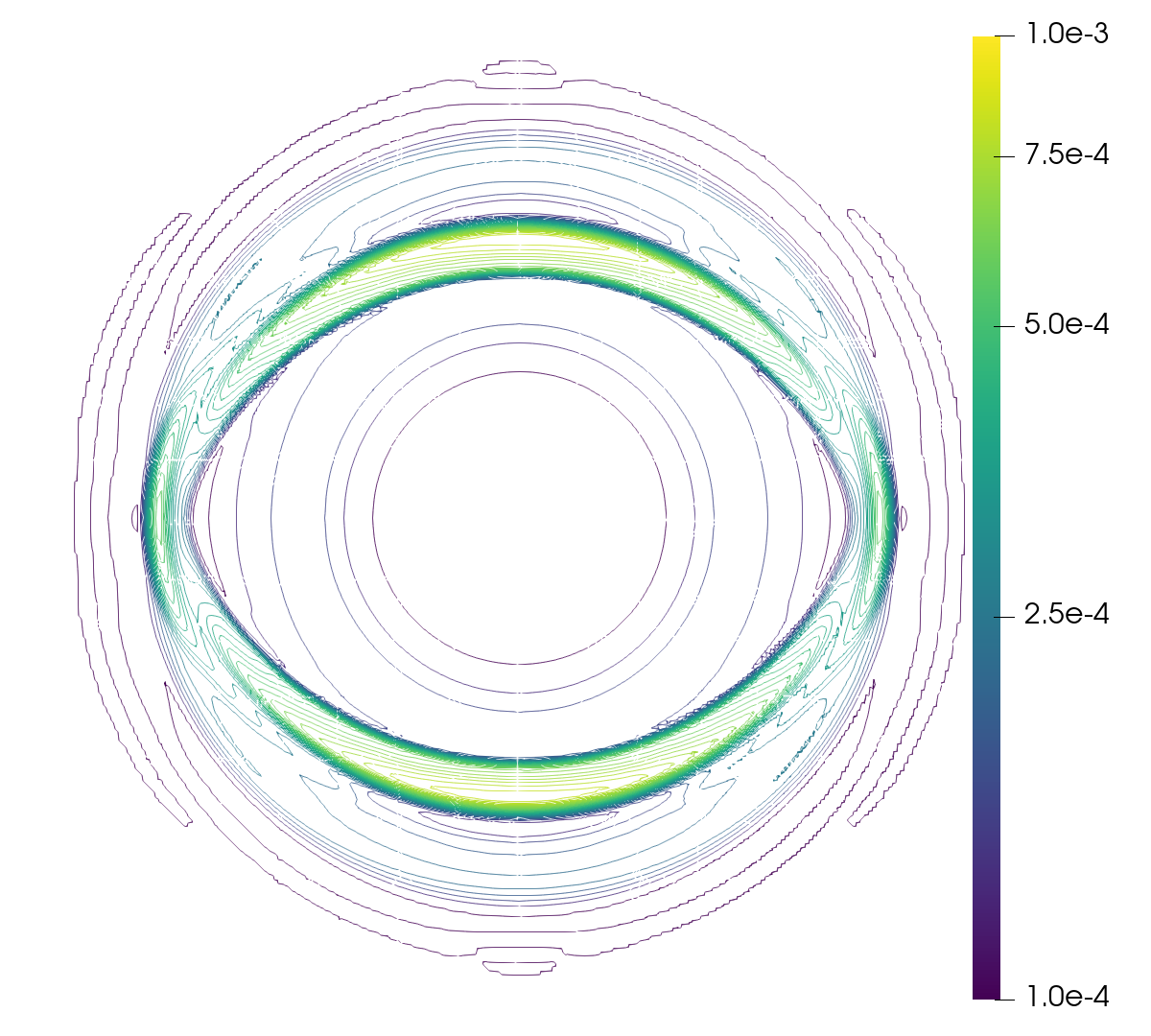}
    \\
    FD-2
  \end{minipage}
  \begin{minipage}{0.43\columnwidth}
    \centering
    \includegraphics[width=1.0\textwidth]{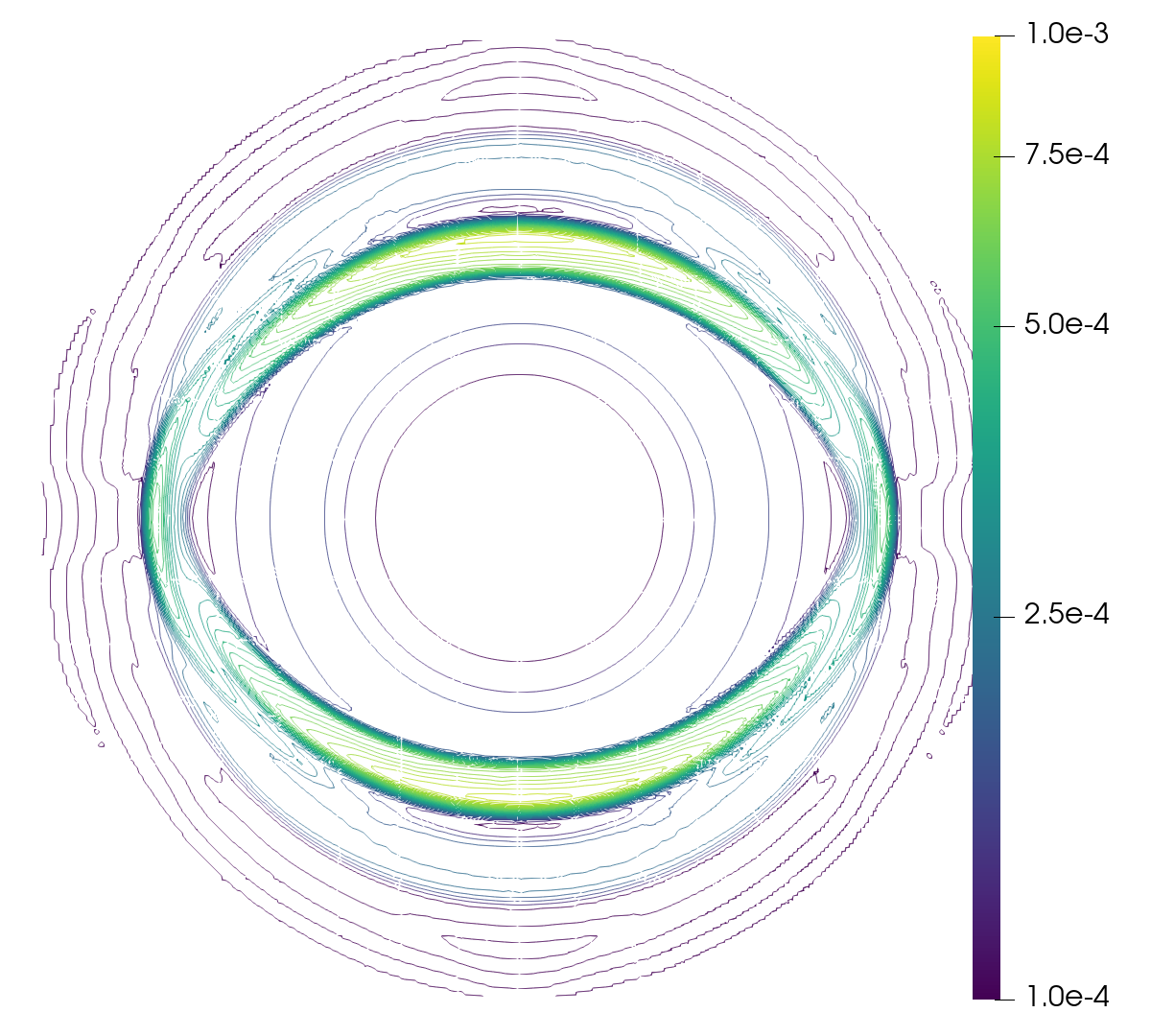}
    \\
    FD-4
  \end{minipage}

  \begin{minipage}{0.43\columnwidth}
    \centering
    \includegraphics[width=1.0\textwidth]{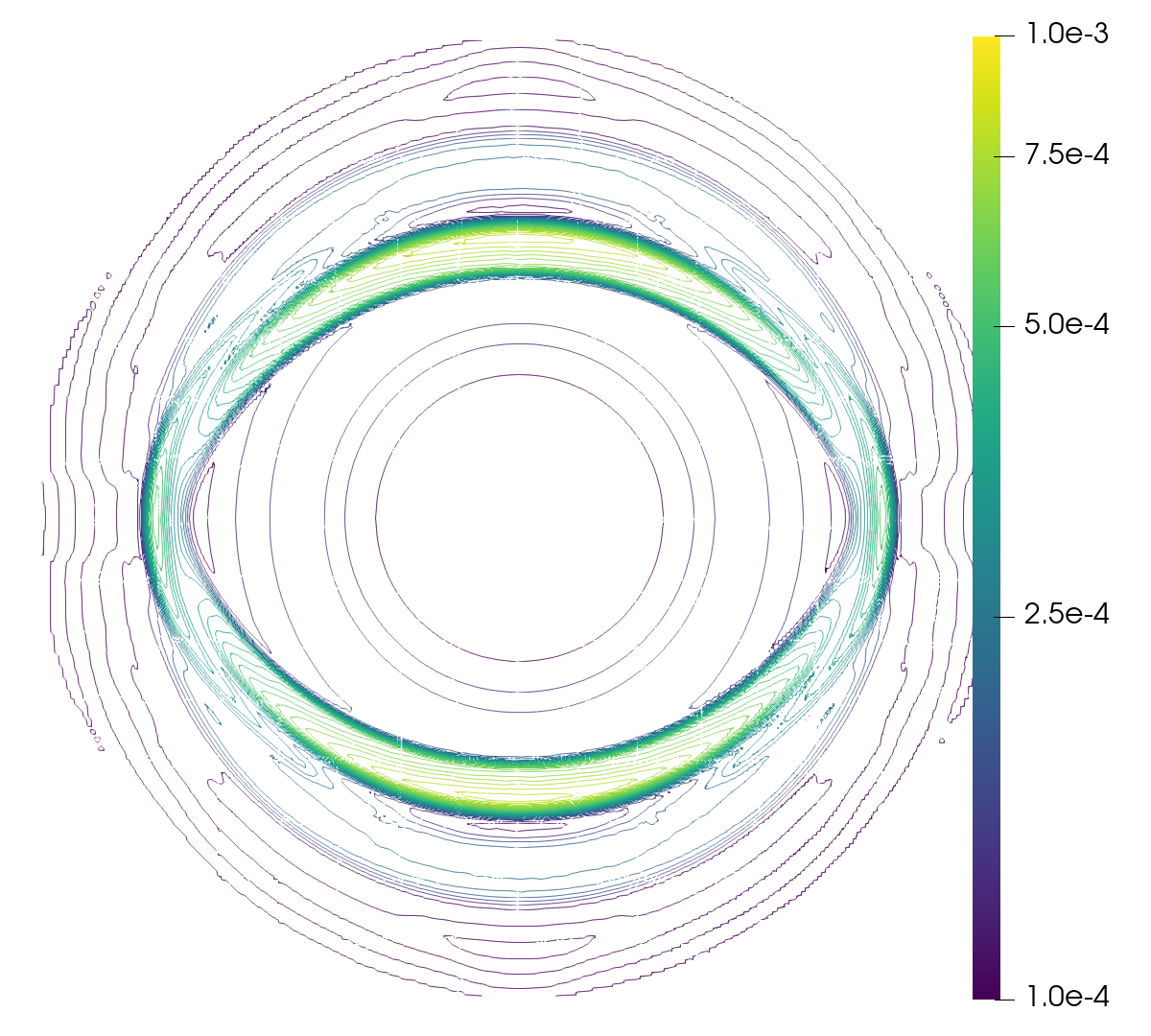}
    \\
    FD-6
  \end{minipage}
  \begin{minipage}{0.43\columnwidth}
    \centering
    \includegraphics[width=1.0\textwidth]{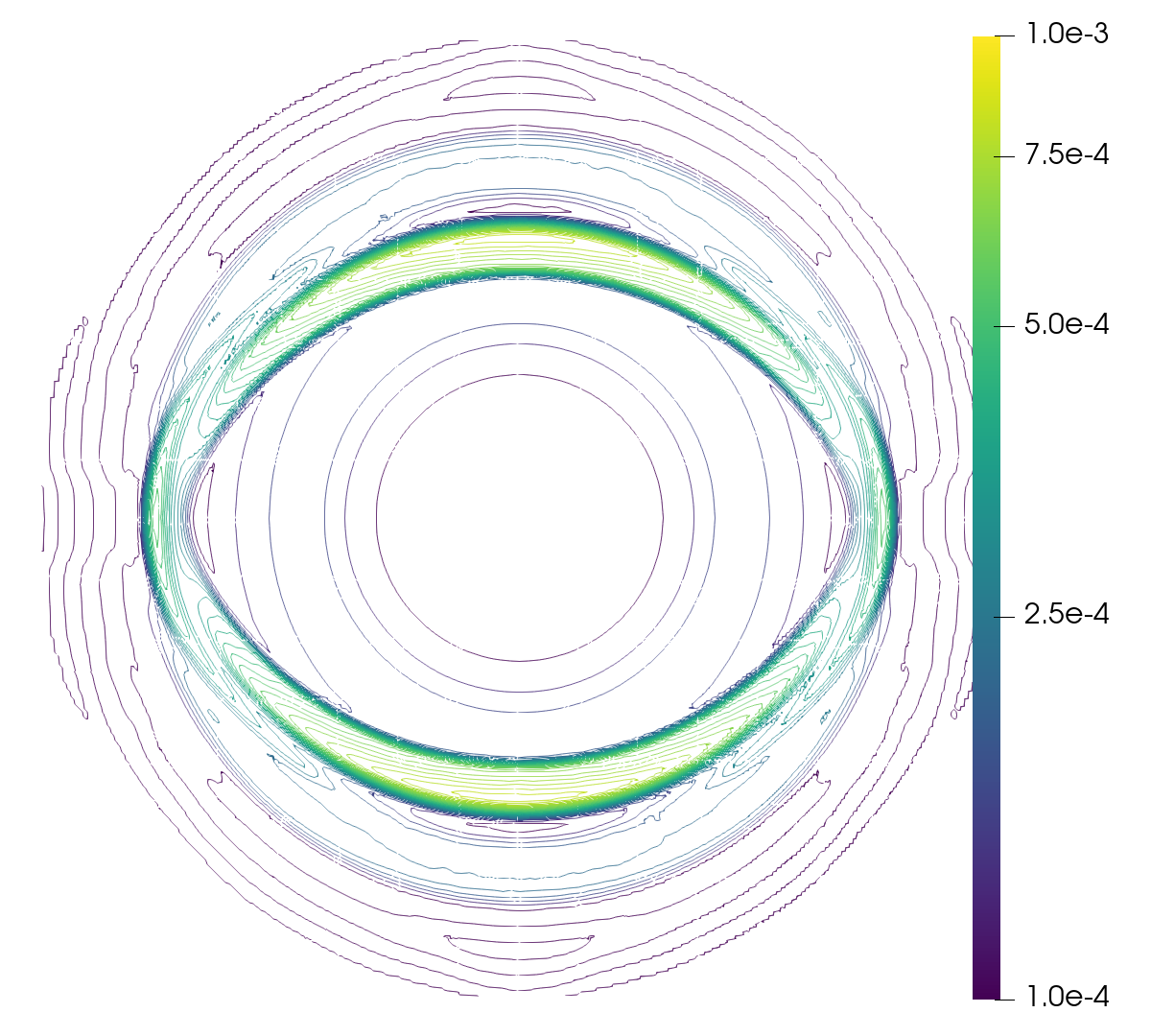}
    \\
    FD-10
  \end{minipage}

  \begin{minipage}{0.43\columnwidth}
    \centering
    \includegraphics[width=1.0\textwidth]{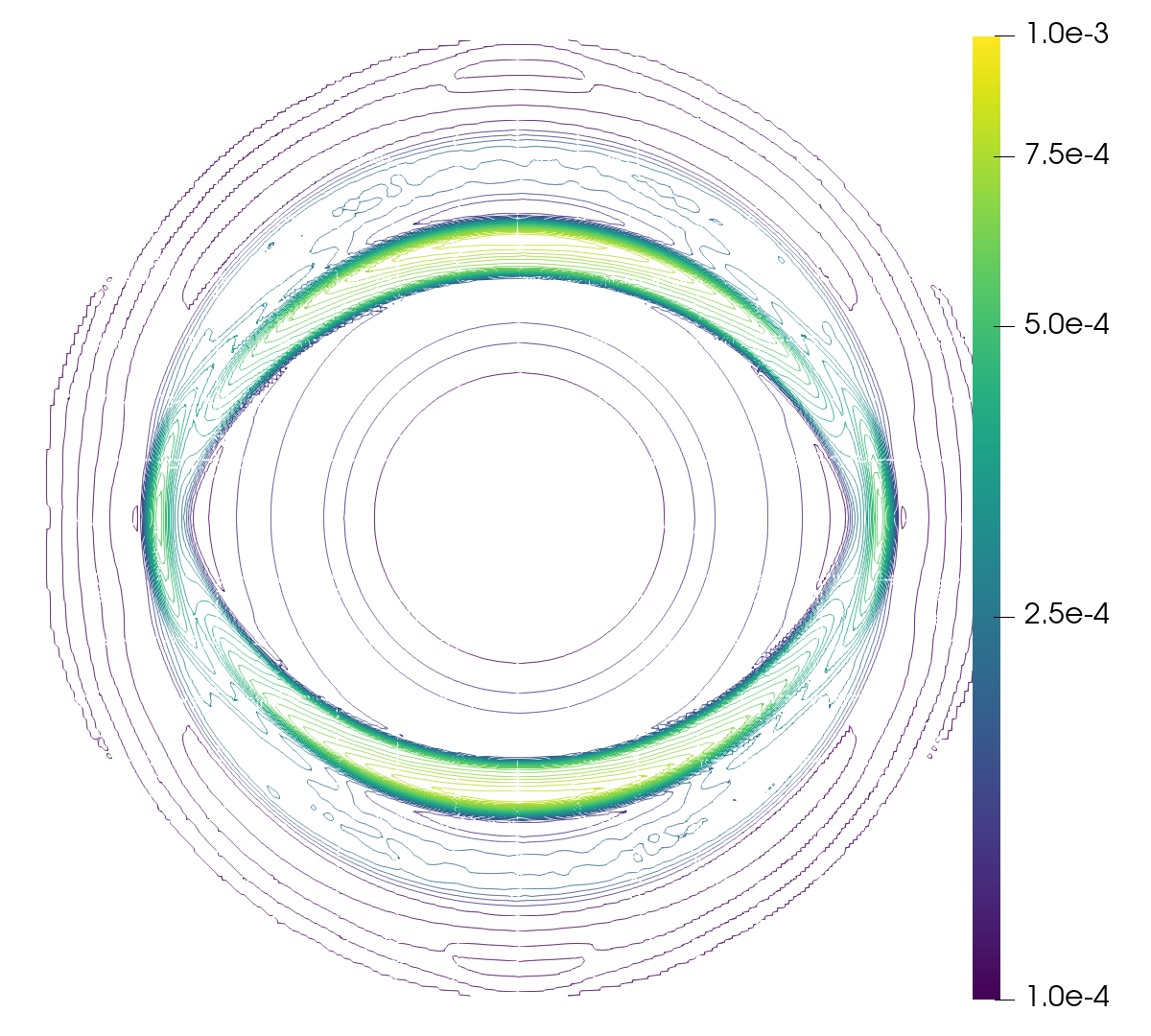}
    \\
    FD-10-6-2-2
  \end{minipage}
  \begin{minipage}{0.43\columnwidth}
    \centering
    \includegraphics[width=1.0\textwidth]{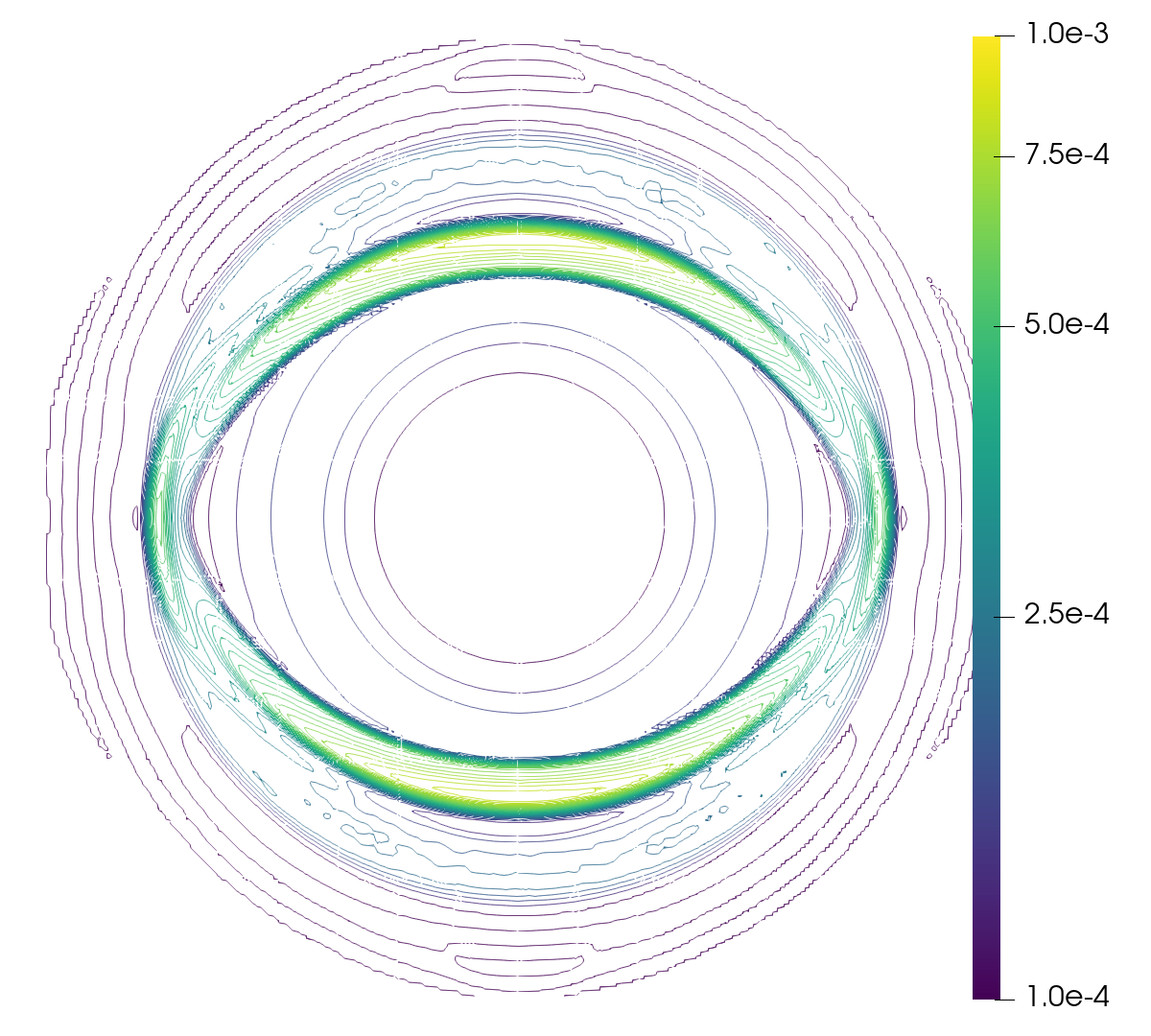}
    \\
    FD-10-4-2-2
  \end{minipage}
  \caption{Cylindrical blast wave $\rho$ at $t=4$ showing the results using
    different FD derivative orders and always using the PPAO9-5-2-1
    reconstruction method. In all cases the scheme is stable, meaning we are
    able to achieve high-order in smooth regions while being robust and stable
    at discontinuities and shocks. In all cases the scheme is stable. There are
    30 contours linearly spaced between $10^{-3}$ and $10^{-4}$.
    \label{fig:BlastWave}}
\end{figure}

\begin{figure}[h]
  \raggedleft
  \begin{minipage}{0.42\columnwidth}
    \centering
    \includegraphics[width=1.0\textwidth]{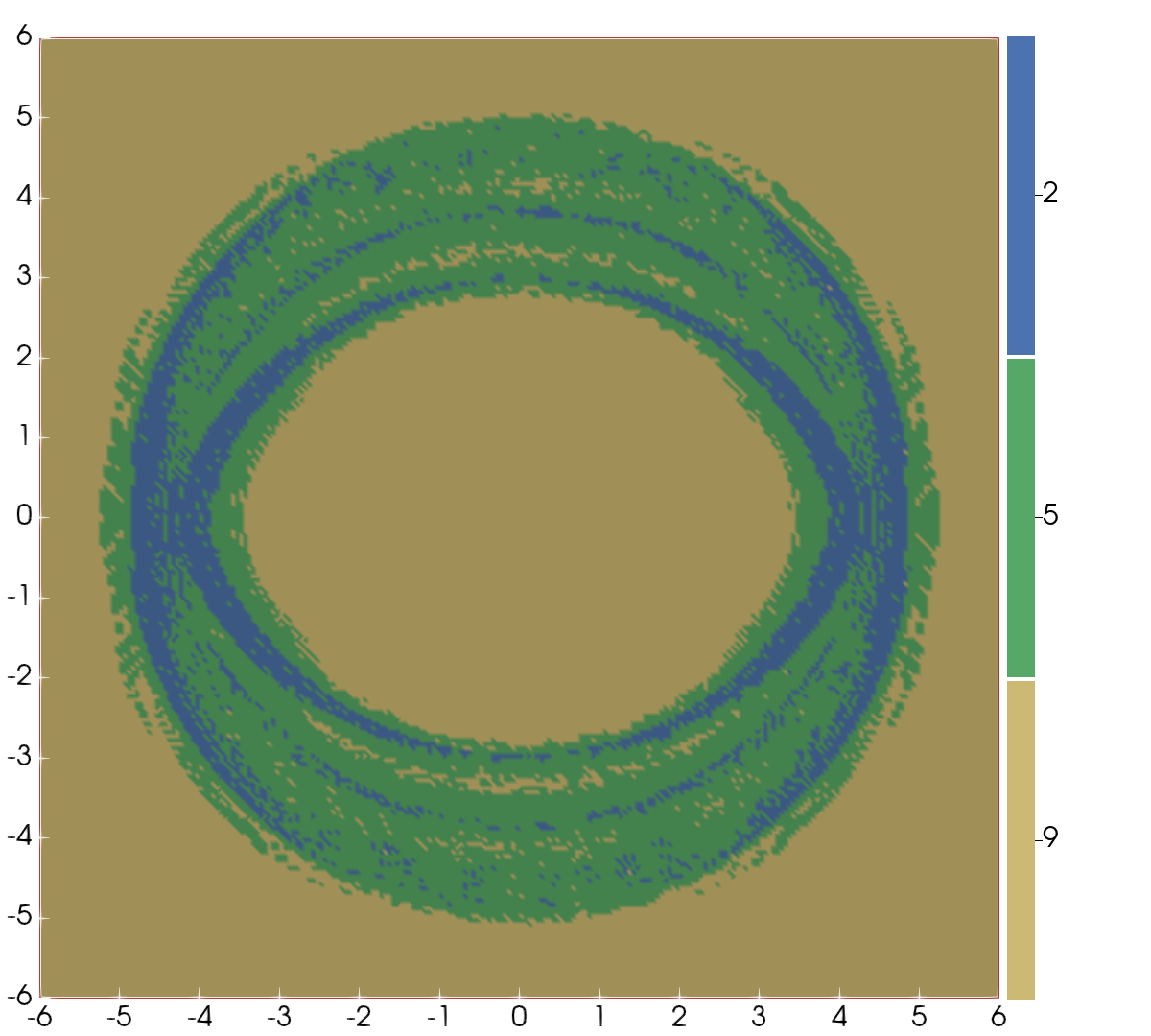}
    \\
    Reconstruction Order in $x$
  \end{minipage}
  \begin{minipage}{0.42\columnwidth}
    \centering
    \includegraphics[width=1.0\textwidth]{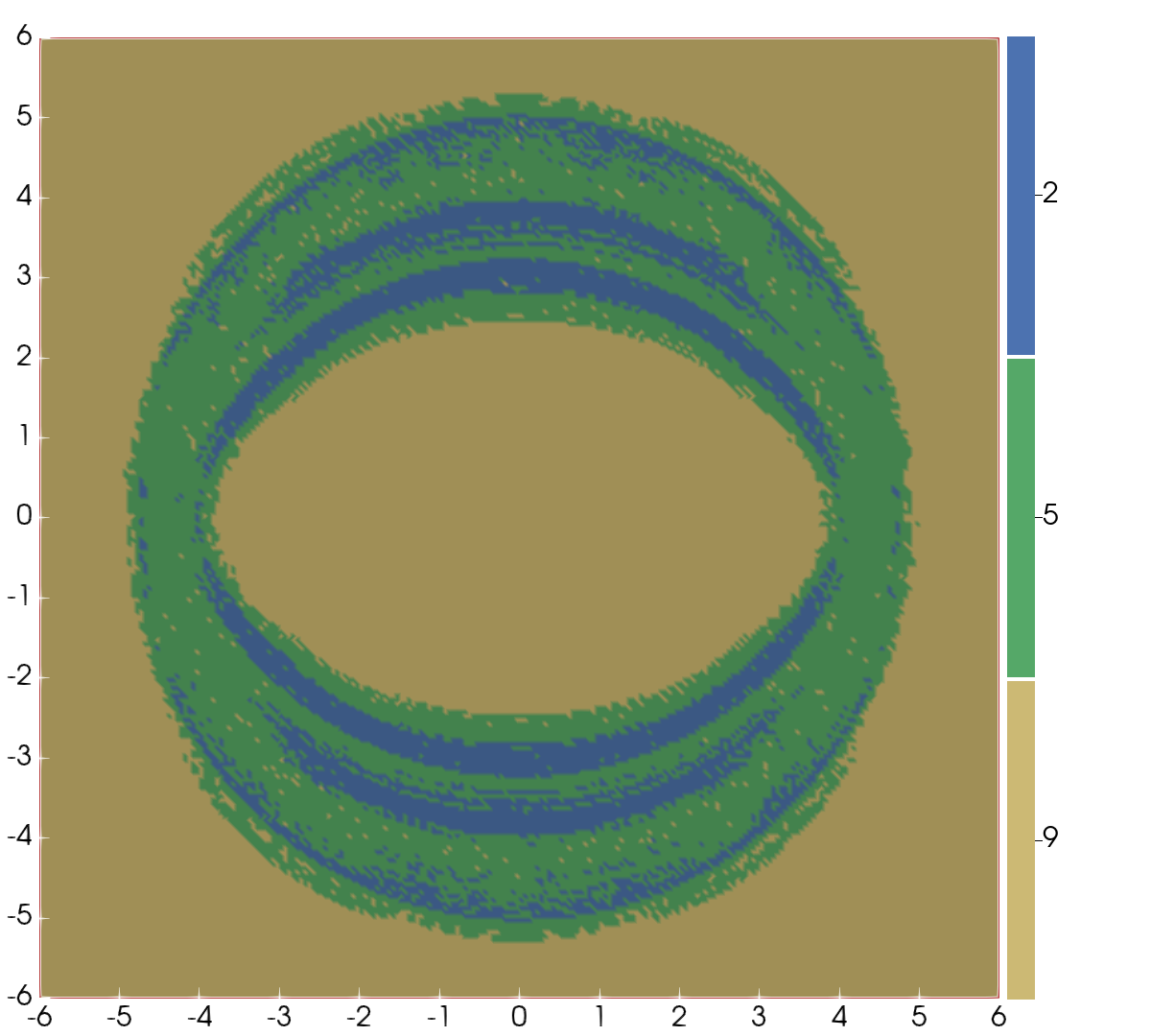}
    \\
    Reconstruction Order in $y$
  \end{minipage}

  \begin{minipage}{0.42\columnwidth}
    \centering
    \includegraphics[width=1.0\textwidth]{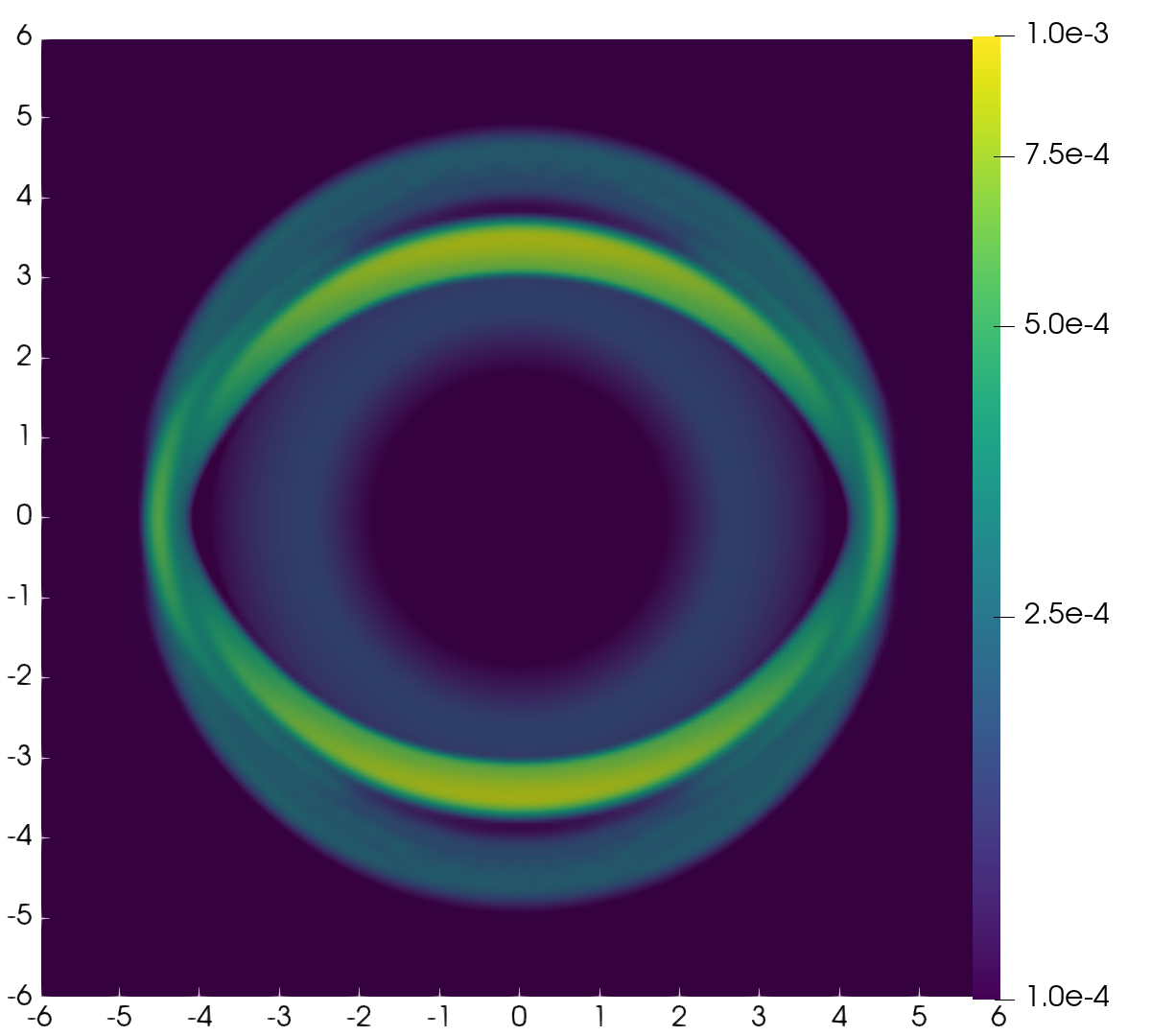}
    \\
    $\rho$
  \end{minipage}
  \begin{minipage}{0.42\columnwidth}
    \centering
    \includegraphics[width=1.0\textwidth]{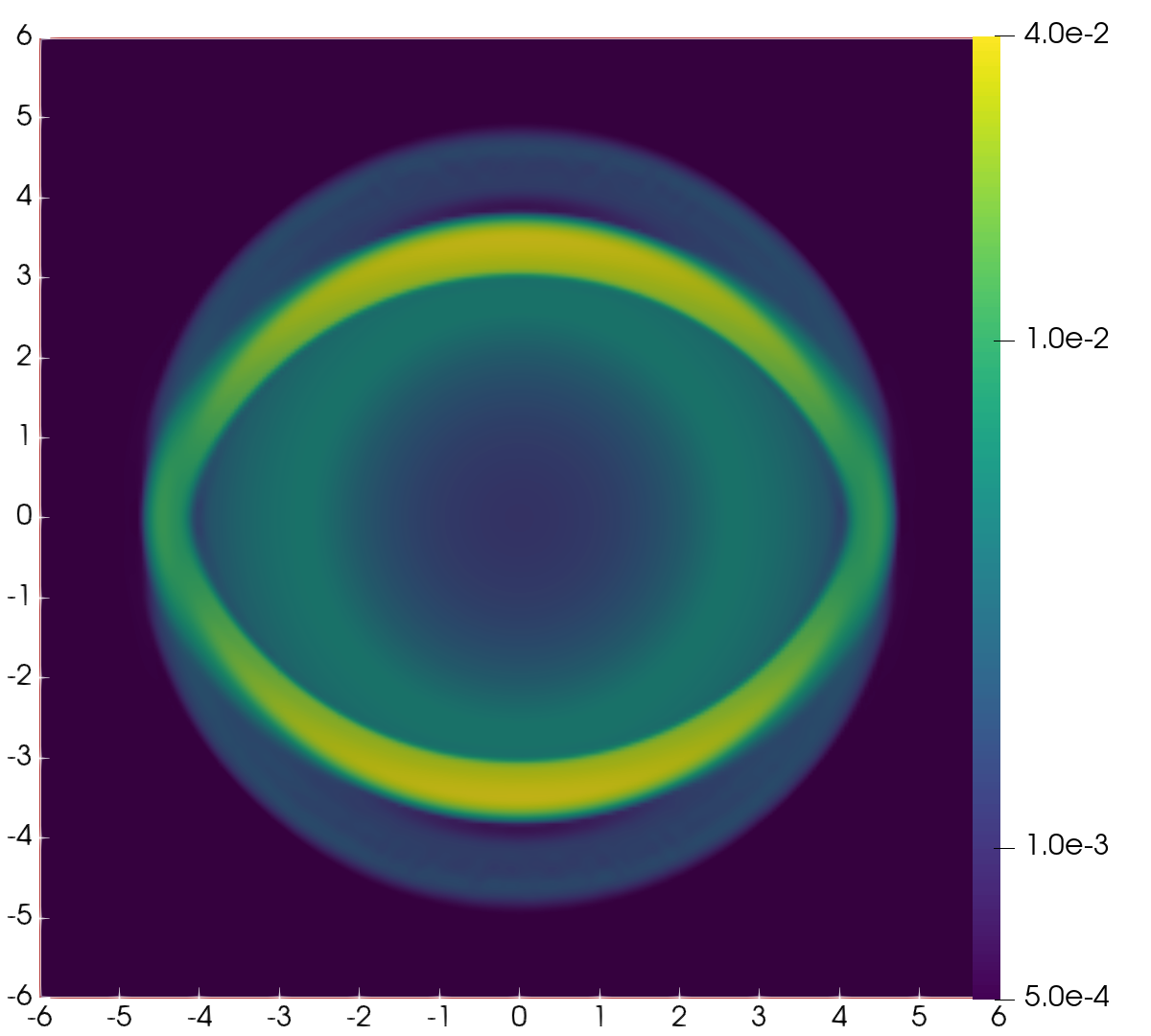}
    \\
    $p$
  \end{minipage}
  \caption{Results from the cylindrical blast wave problem. The panels in the
    top row show the reconstruction and FD derivative order used in the
    $x$-direction (left) and $y$-direction (right) at the final time, while the
    bottom left panel shows the rest mass density and the bottom right the
    pressure at the final time. We see that the adaptive-order FD scheme
    accurately tracks non-smooth features in the solution, specifically the rest
    mass density and pressure, adjusting the order as necessary.
  \label{fig:BlastWaveReconsOrder}}
\end{figure}

We evolve the blast wave to time $t_f=4.0$ using a $240\times240\times15$ FD
grid on a cubical domain of size $[-6,6]^3$ and apply periodic boundary
conditions in all directions. We show the logarithm of the rest mass density
$\rho$ at $t_f$ in figure~\ref{fig:BlastWave} for different FD derivative orders
but always using the PPAO9-5-2-1 reconstruction method. We label the panels
FD-$N$, where $N$ is the FD derivative order. The 10-6-2-2 and 10-4-2-2 use
tenth-order FD derivatives when ninth-order reconstruction is used, sixth
(fourth) order derivatives when fifth-order reconstruction is used, and
second-order derivatives when first- and second-order reconstruction is used.
Despite the high-order nature of the method, it is able to robustly capture the
sharp features that arise, demonstrating that the scheme robustly detects
discontinuities and reduces the reconstruction order in so as to remain stable.
We show the reconstruction order alongside $\rho$ and pressure $p$ in
figure~\ref{fig:BlastWaveReconsOrder}. In this case discontinuous features in
$\rho$ and $p$ are at the spatial locations and the scheme correctly identifies
them. A similar plot for the Kelvin-Helmholtz instability discussed below
(\S~\ref{sec:kh instability}) shows that the code can track both shocks and
contact discontinuities.

\begin{figure}[h]
  \raggedleft
  \begin{minipage}{0.43\columnwidth}
    \centering
    \includegraphics[width=1.0\textwidth]{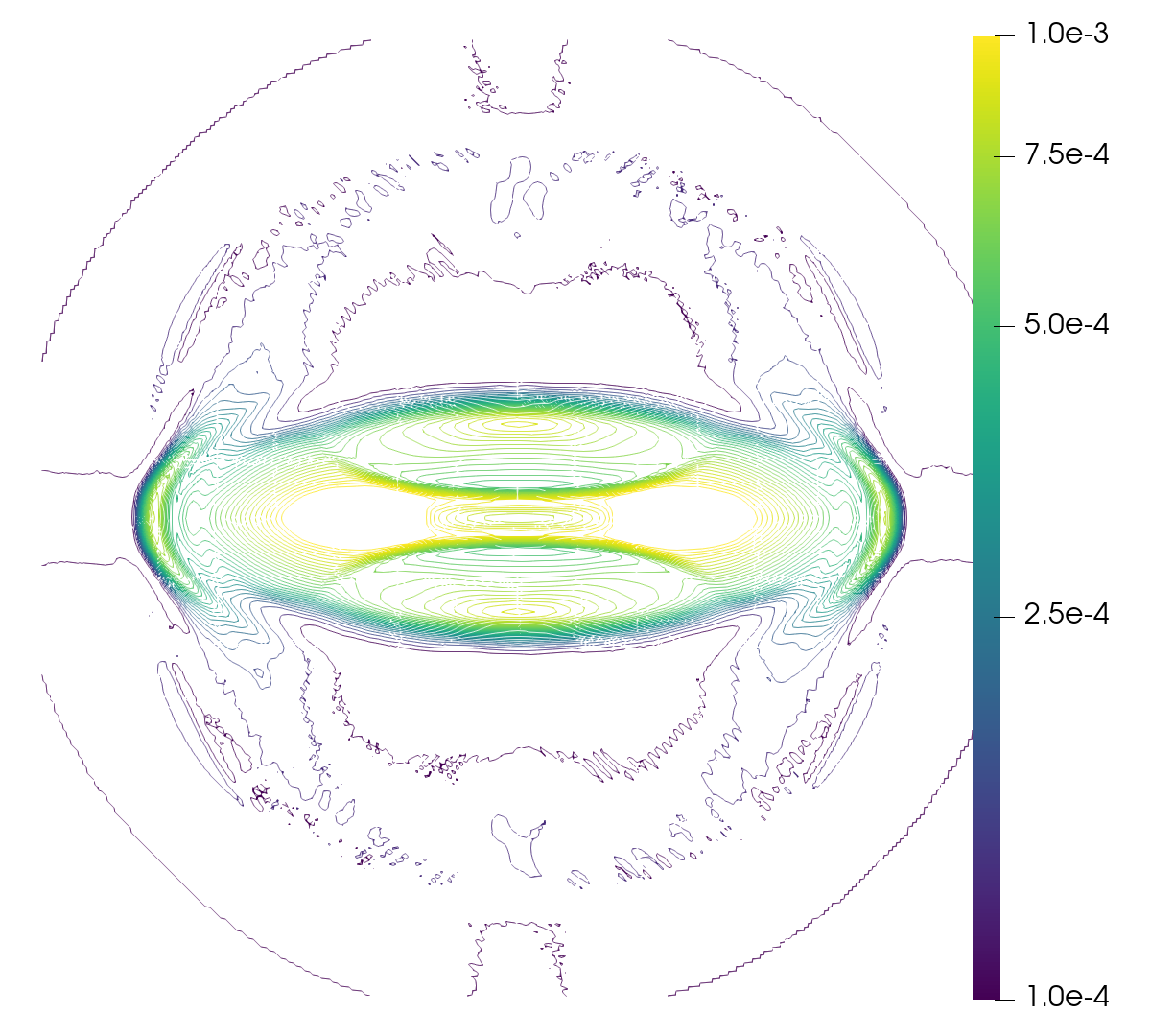}
    \\
    FD-2
  \end{minipage}
  \begin{minipage}{0.43\columnwidth}
    \centering
    \includegraphics[width=1.0\textwidth]{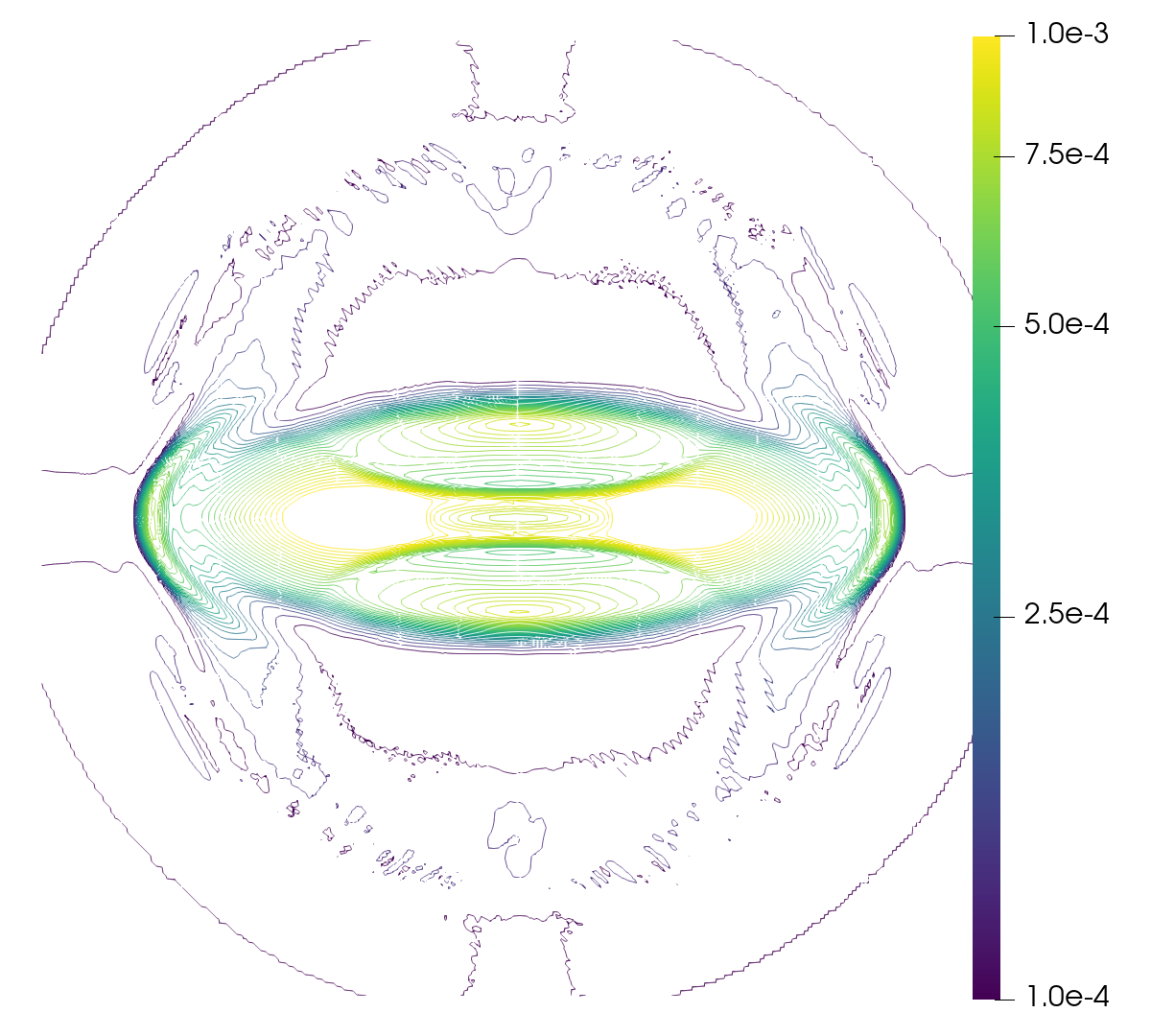}
    \\
    FD-4
  \end{minipage}

  \begin{minipage}{0.43\columnwidth}
    \centering
    \includegraphics[width=1.0\textwidth]{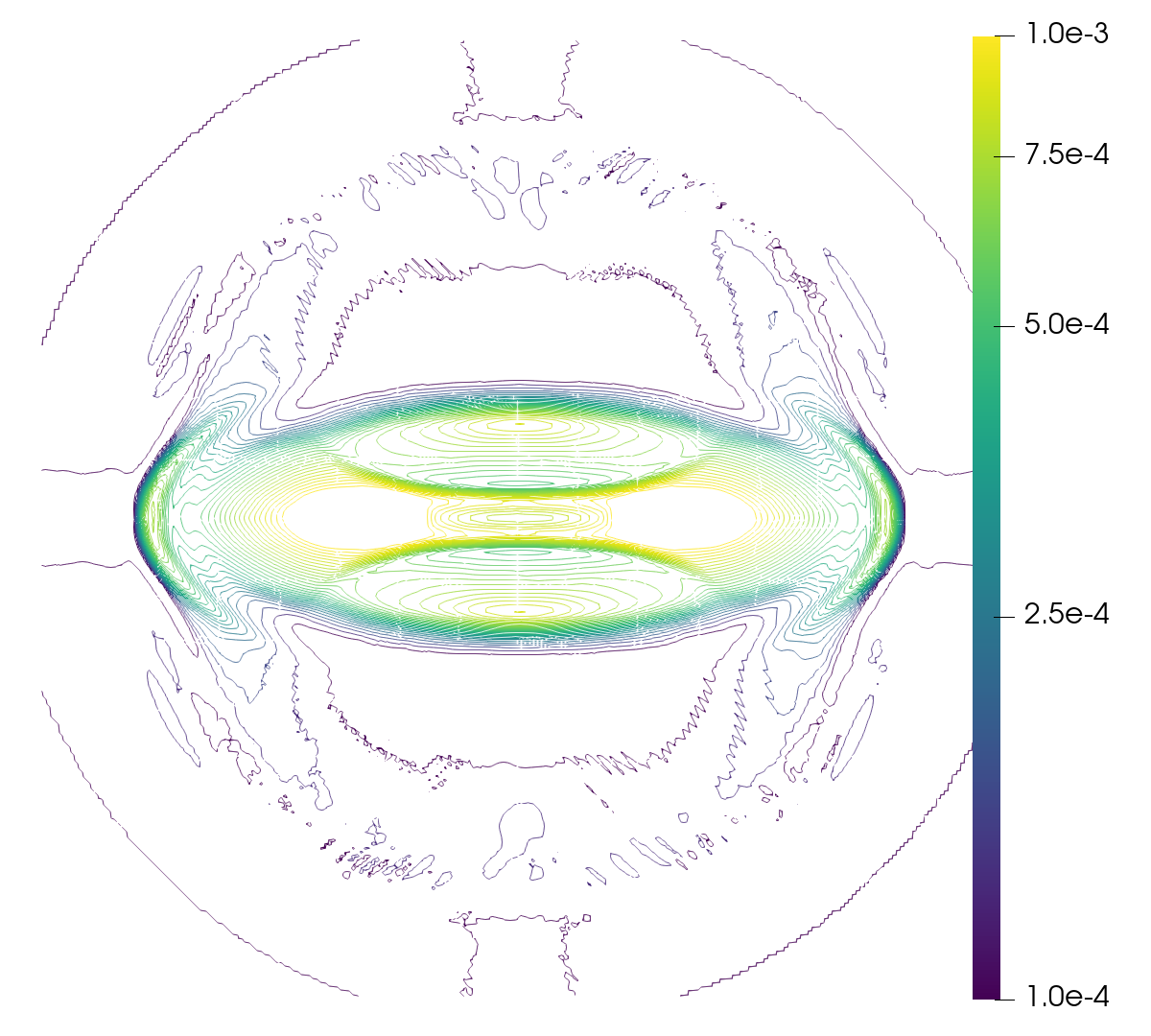}
    \\
    FD-6
  \end{minipage}
  \begin{minipage}{0.43\columnwidth}
    \centering
    \includegraphics[width=1.0\textwidth]{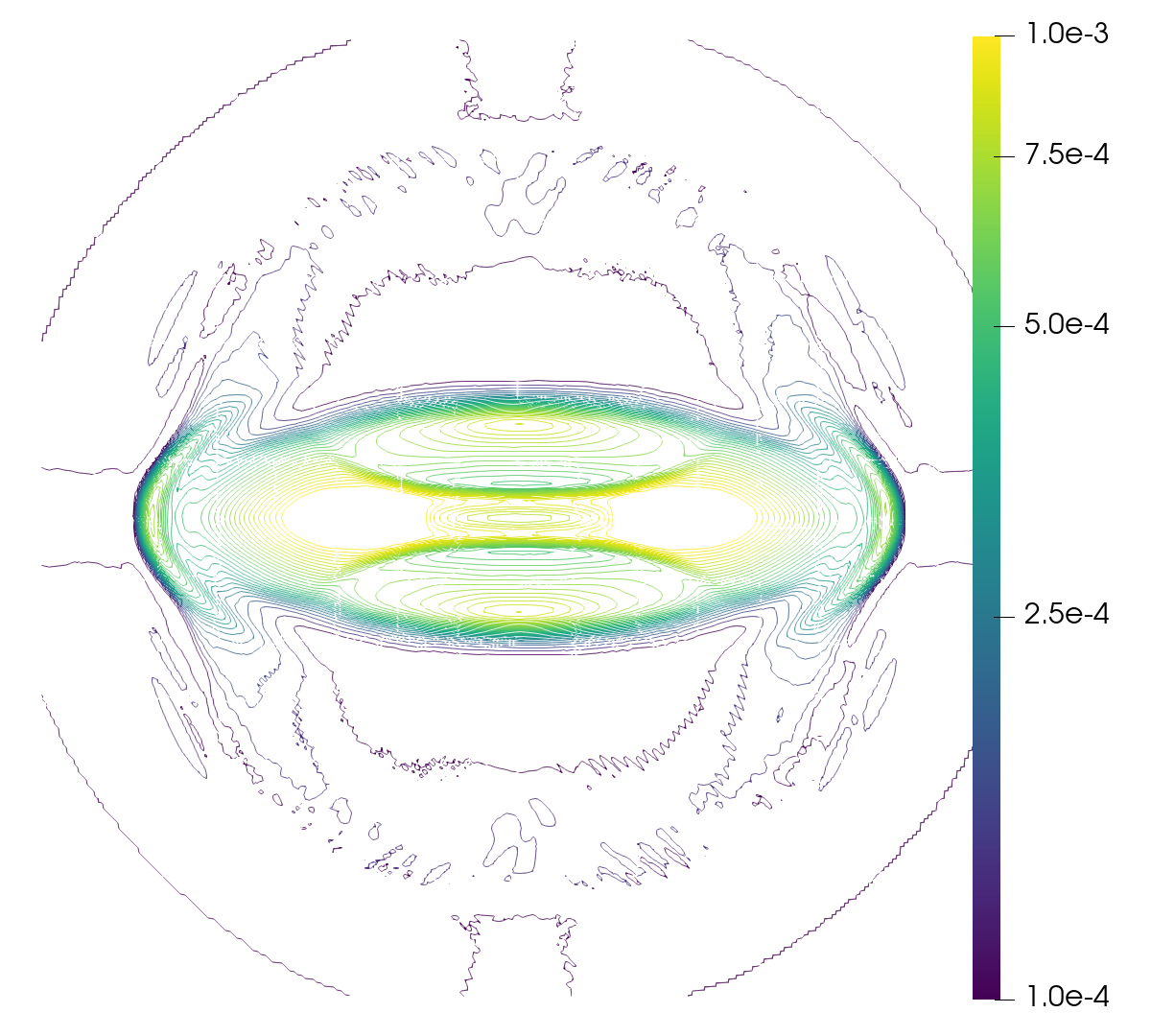}
    \\
    FD-10
  \end{minipage}

  \begin{minipage}{0.43\columnwidth}
    \centering
    \includegraphics[width=1.0\textwidth]{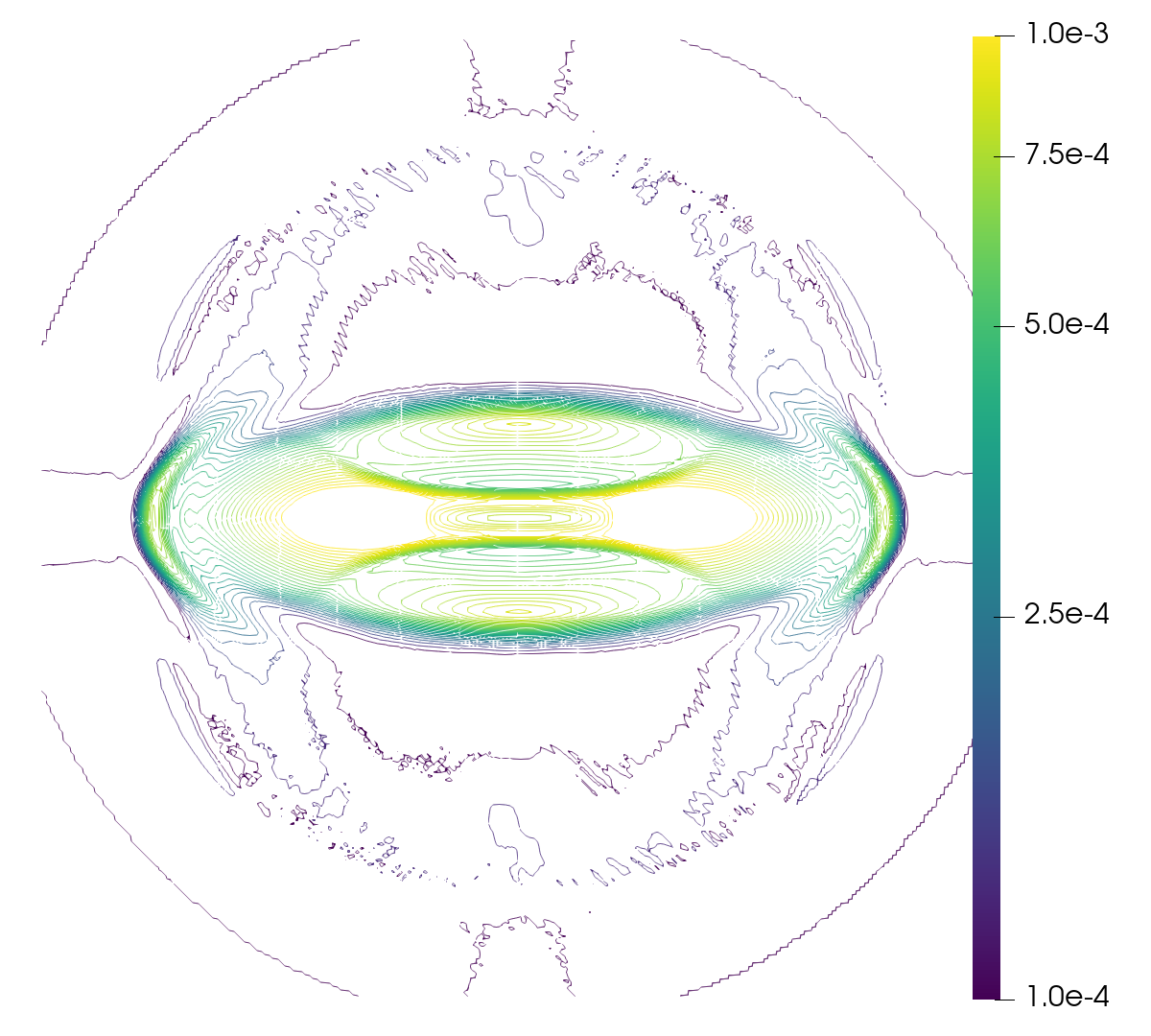}
    \\
    FD-10-6-2-2
  \end{minipage}
  \begin{minipage}{0.43\columnwidth}
    \centering
    \includegraphics[width=1.0\textwidth]{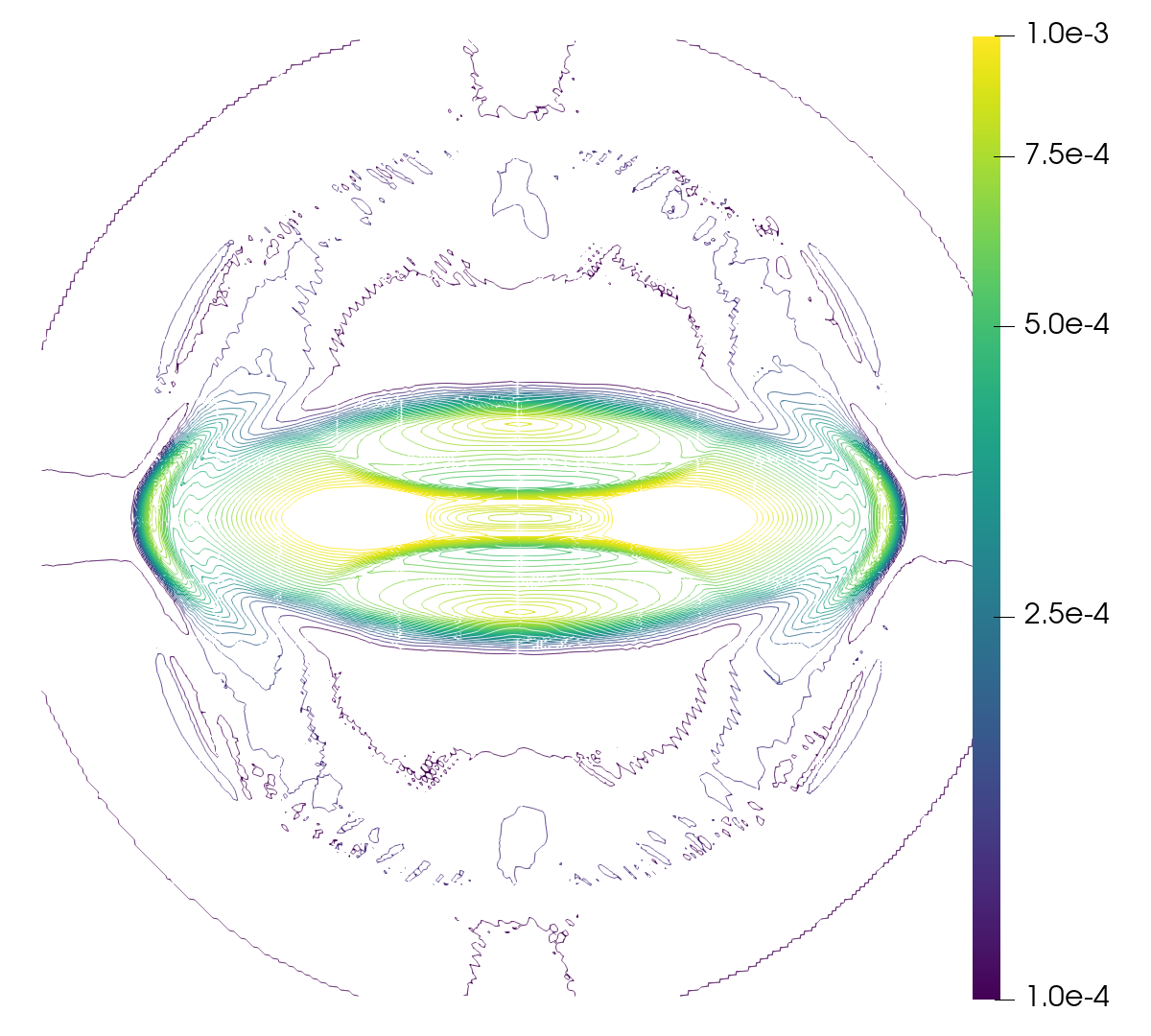}
    \\
    FD-10-4-2-2
  \end{minipage}
  \caption{Strongly magnetized ($B^x=0.5$) cylindrical blast wave $\rho$ at
    $t=4$ showing the results using different FD derivative orders and always
    using the PPAO9-5-2-1 reconstruction method, where at second order minmod
    reconstruction is used. In all cases the scheme is stable. There are 30
    contours linearly spaced between $10^{-3}$ and $10^{-4}$.
  \label{fig:BlastWaveStrong}}
\end{figure}

A particularly challenging case is when the initial magnetic field is increased
to $B^x=0.5$, as was done in~\cite{Fambri:2018udk}. In this case we must use
minmod reconstruction~\cite{Harten:1984} at second order instead of monotonized
central in order for the simulations to remain stable, but the PPAO scheme is
otherwise able to evolve the solution without any challenges. We show results
analogous to the weaker field case in figure~\ref{fig:BlastWaveStrong}.

\subsection{2d Magnetic Rotor}

The 2d magnetic rotor problem was first proposed for non-relativistic
MHD~\cite{1999JCoPh.149..270B, 2000JCoPh.161..605T} and later generalized to the
relativistic case~\cite{2010PhRvD..82h4031E, 2003A&A...400..397D}. A rotating
cylinder of dense fluid is surrounded by a lower density fluid, with uniform
pressure and magnetic field. Magnetic braking ultimately slows the rotor and an
approximately 90 degree rotation is completed by the final time $t_f=0.4$. We
set up the problem on a $272\times272\times17$ grid with domain size
$[-0.5, 0.5]^3$, periodic boundary conditions in all directions, and a time step
size $\Delta t=10^{-3}$. A $\Gamma=5/3$ ideal fluid equation of state is used
with initial conditions:
\begin{eqnarray}
  p&=1 \\
  B^i&=(1,0,0) \\
  v^i&=\left\{
       \begin{array}{ll}
         (-y\Omega, x\Omega, 0),
         & \mathrm{if} \; r \le R_{\mathrm{rotor}}=0.1 \\
         (0,0,0), & \mathrm{otherwise},
       \end{array}\right. \\
  \rho&=\left\{
        \begin{array}{ll}
          10,
          & \mathrm{if} \; r \le R_{\mathrm{rotor}}=0.1 \\
          1, & \mathrm{otherwise},
        \end{array}\right.
\end{eqnarray}
where $\Omega = 9.95$ is the angular velocity, guaranteeing that the maximum
velocity of the fluid (0.995) is below the speed of light.

\begin{figure}[h]
  \raggedleft
  \begin{minipage}{0.43\columnwidth}
    \centering
    \includegraphics[width=1.0\textwidth]{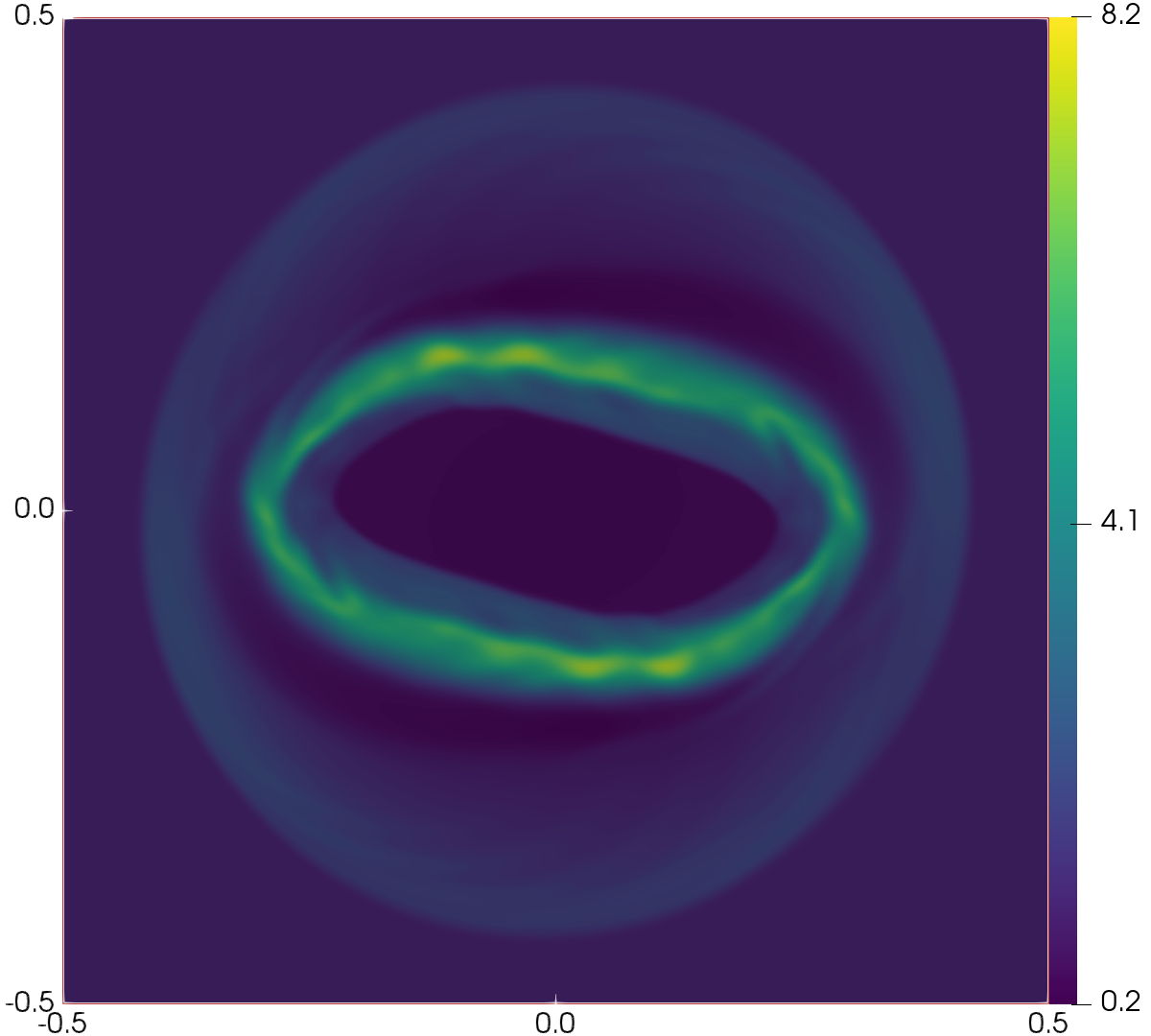}
    \\
    FD-2
  \end{minipage}
  \begin{minipage}{0.43\columnwidth}
    \centering
    \includegraphics[width=1.0\textwidth]{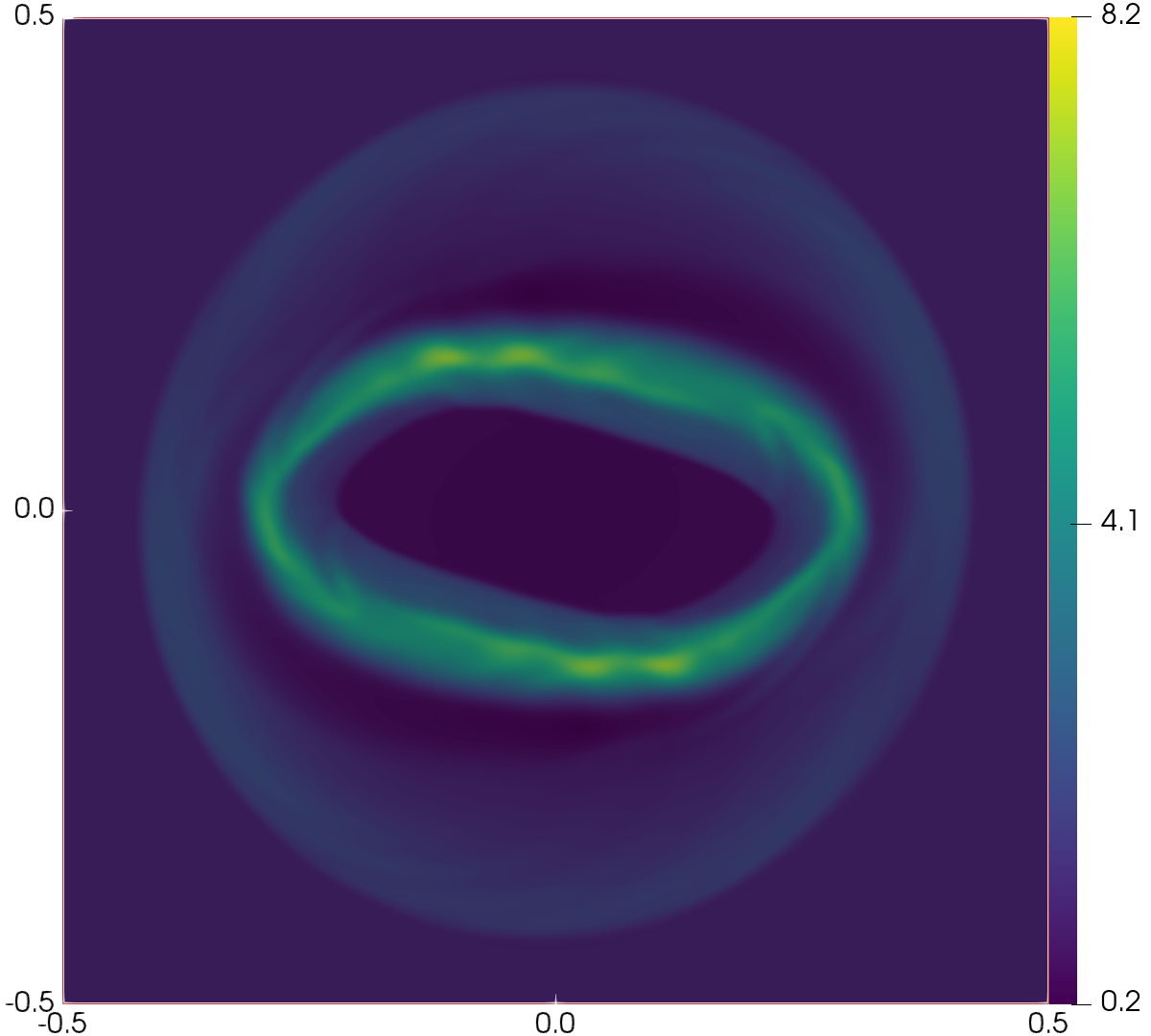}
    \\
    FD-4
  \end{minipage}

  \begin{minipage}{0.43\columnwidth}
    \centering
    \includegraphics[width=1.0\textwidth]{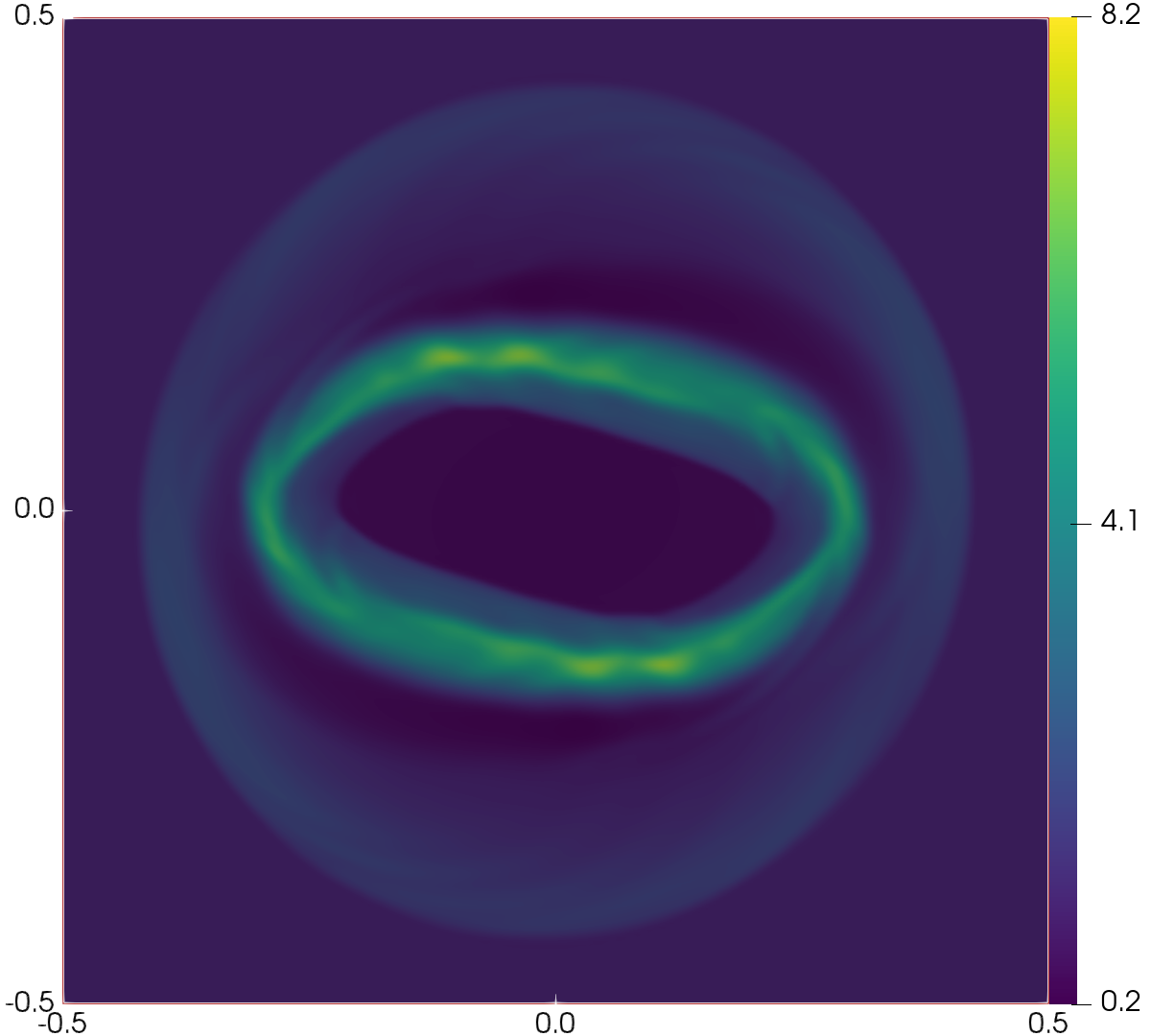}
    \\
    FD-6
  \end{minipage}
  \begin{minipage}{0.43\columnwidth}
    \centering
    \includegraphics[width=1.0\textwidth]{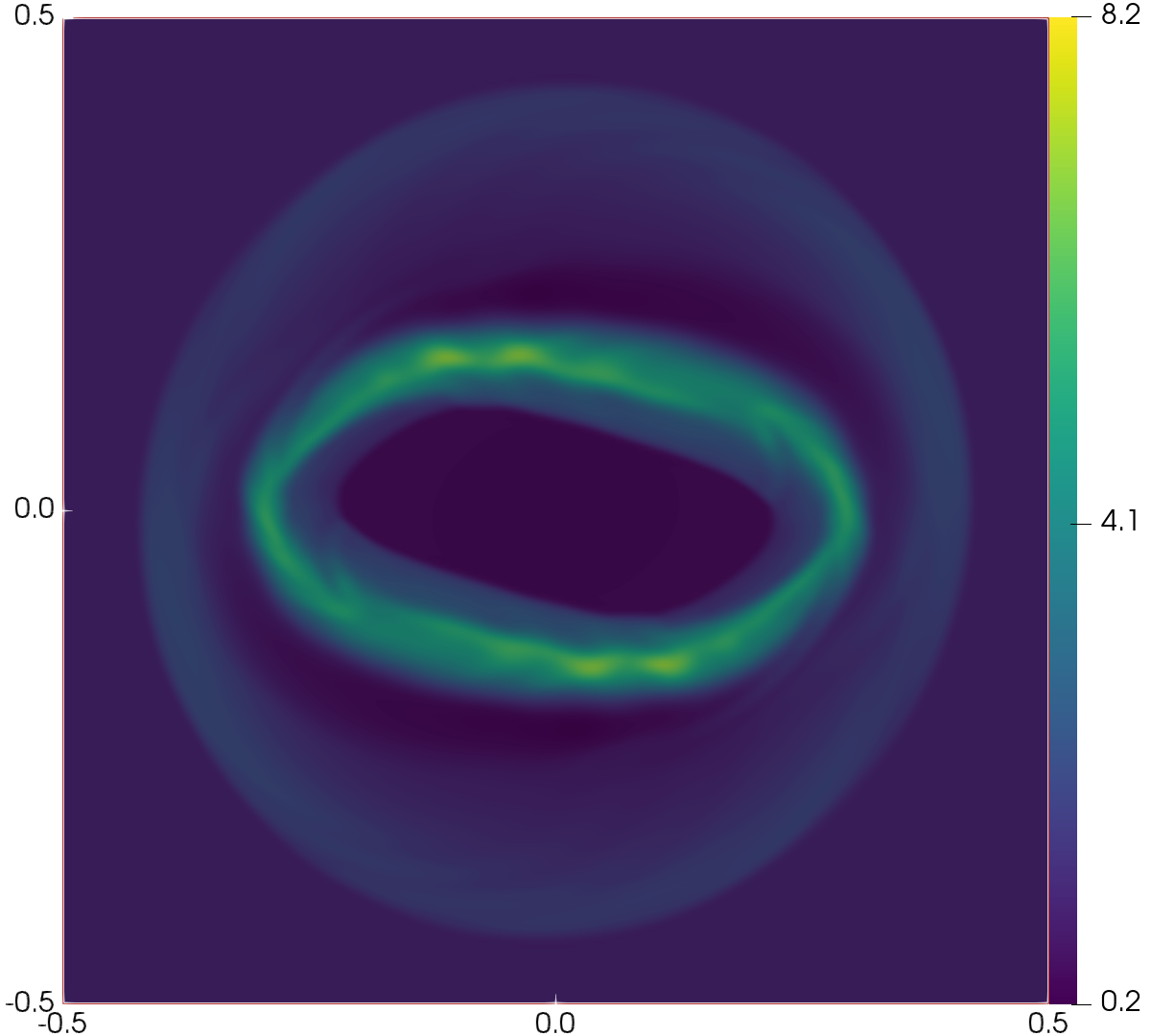}
    \\
    FD-10
  \end{minipage}

  \begin{minipage}{0.43\columnwidth}
    \centering
    \includegraphics[width=1.0\textwidth]{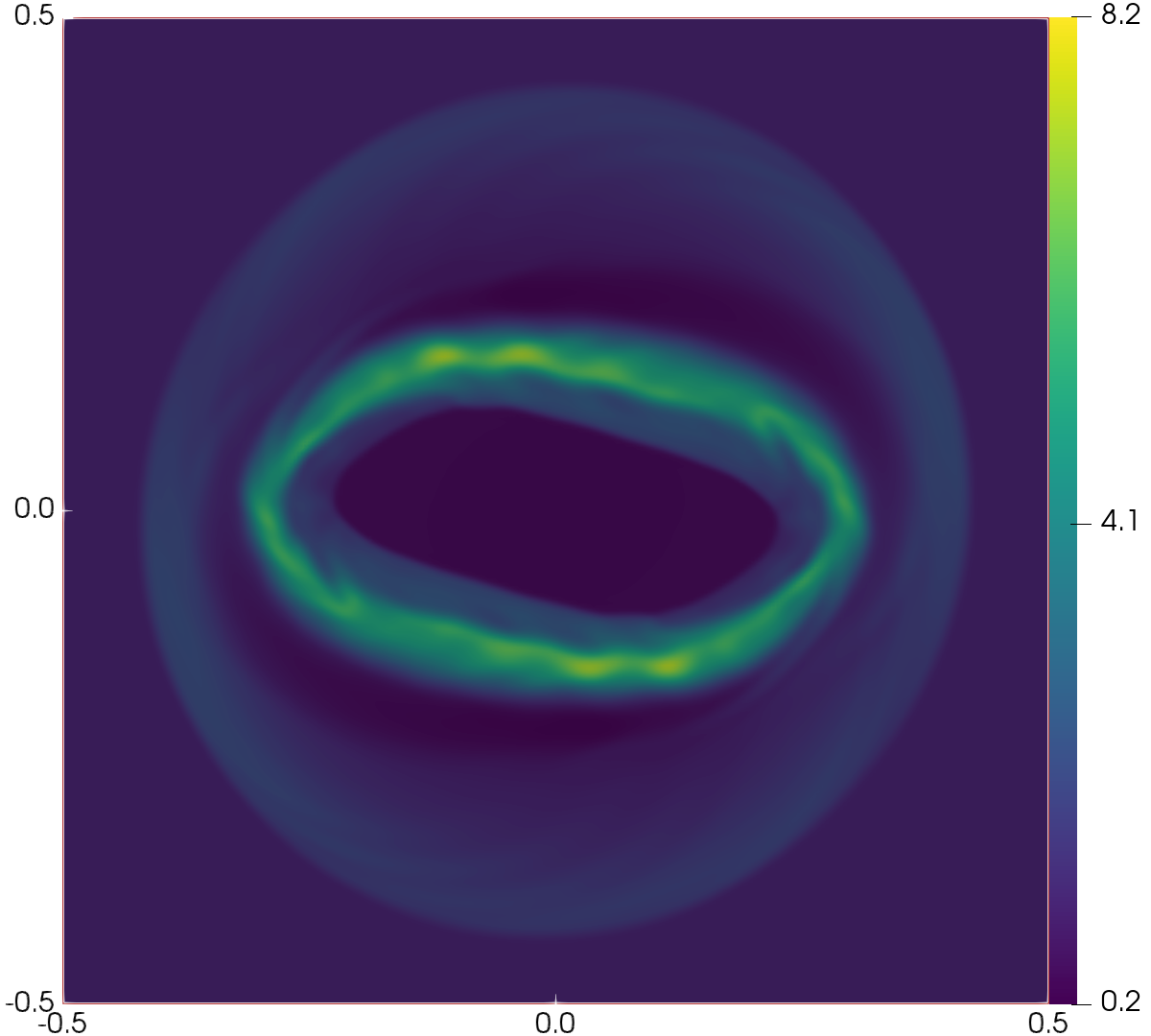}
    \\
    FD-10-6-2-2
  \end{minipage}
  \begin{minipage}{0.43\columnwidth}
    \centering
    \includegraphics[width=1.0\textwidth]{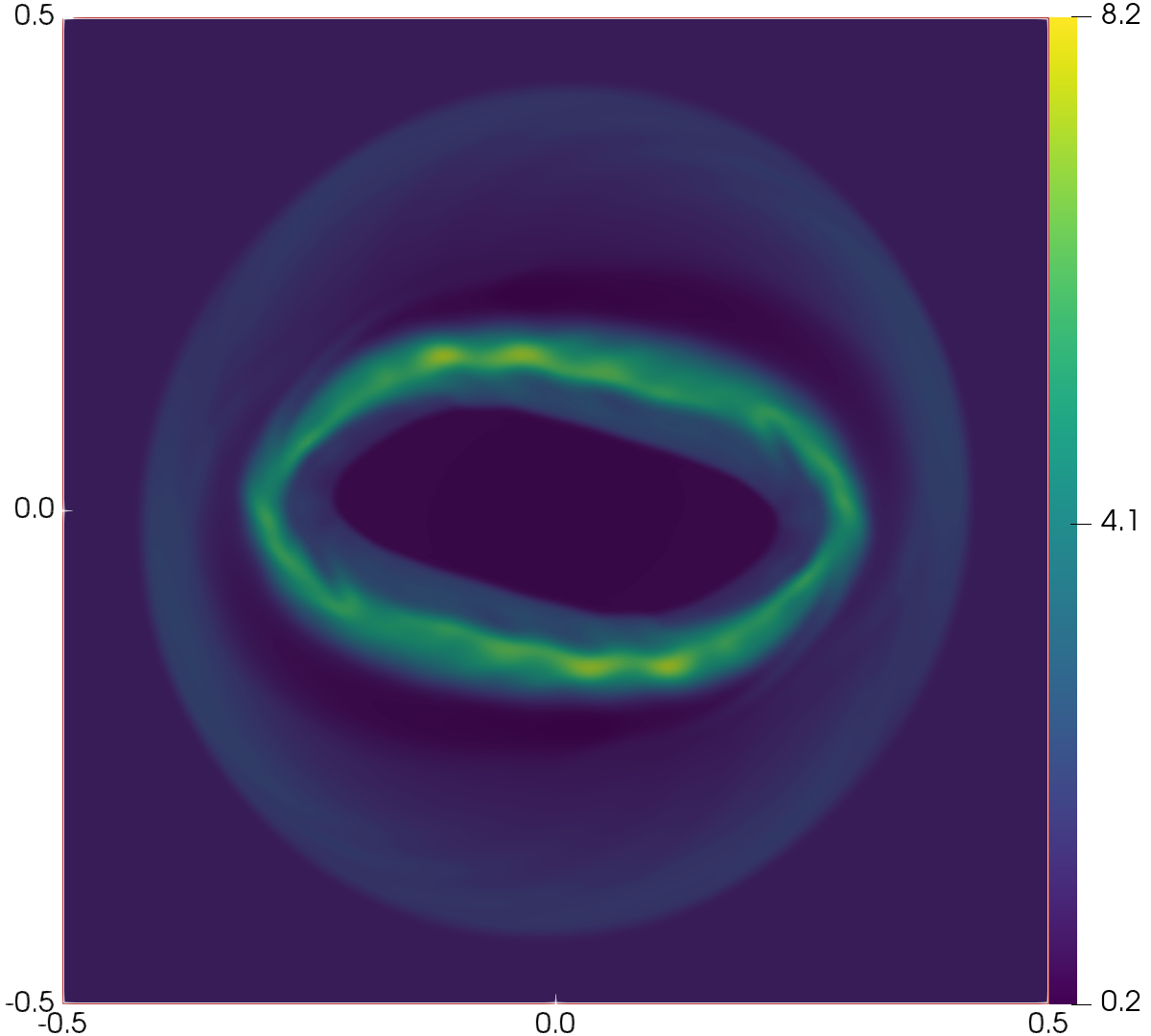}
    \\
    FD-10-4-2-2
  \end{minipage}
  \caption{Magnetic rotor $\rho$ at $t=0.45$ showing the results using
    different FD derivative orders and always using the PPAO9-5-2-1
    reconstruction method. In all cases the scheme is stable, meaning we are
    able to achieve high-order in smooth regions while being robust and stable
    at discontinuities and shocks.
  \label{fig:MagneticRotor}}
\end{figure}

We show the results of our evolutions in figure~\ref{fig:MagneticRotor}.  Again,
we label the panels FD-$N$, where $N$ is the FD derivative order. The 10-6-2-2
and 10-4-2-2 use tenth-order FD derivatives when ninth-order reconstruction is
used, sixth (fourth) order derivatives when fifth-order reconstruction is used,
and second-order derivatives when first- and second-order reconstruction is
used.  Just as in the cylindrical blast wave test case, our adaptive order
scheme is able to robustly evolve the discontinuous parts of the solution while
maintaining high order where the solution is smooth.

\subsection{2d Magnetic Loop Advection}

The 2d magnetic loop advection problem~\cite{1991JCoPh..92..142D} gives a nice
test of how well the method is able to advect magnetic fields through the
domain, and if the divergence cleaning is working properly. In this problem, a
magnetic loop is advected through the domain until it returns to its starting
position. We use an initial configuration very similar to~\cite{Mosta:2013gwu,
  2011ApJS..193....6B, 2005JCoPh.205..509G, 2008ApJS..178..137S}, where
\begin{eqnarray}
  \rho&=1 \\
  p&=3 \\
  v^i &= (1/1.2, 1/2.4, 0) \\
  B^x &= \left\{
        \begin{array}{ll}
          -A_{\mathrm{loop}}y/R_{\mathrm{in}},
          & \mathrm{if} \; r \le R_{\mathrm{in}} \\
          -A_{\mathrm{loop}}y/r,
          & \mathrm{if} \; R_{\mathrm{in}}<r<R_{\mathrm{loop}} \\
          0, & \mathrm{otherwise},
        \end{array}\right. \\
  B^y &= \left\{
        \begin{array}{ll}
          A_{\mathrm{loop}}x / R_{\mathrm{in}},
          & \mathrm{if} \; r \le R_{\mathrm{in}} \\
          A_{\mathrm{loop}}x/r,
          & \mathrm{if} \; R_{\mathrm{in}}<r<R_{\mathrm{loop}} \\
          0, & \mathrm{otherwise},
        \end{array}\right.
\end{eqnarray}
with $R_{\mathrm{loop}}=0.3$, $R_{\mathrm{in}}=0.001$, $A_\mathrm{loop}=10^{-3}$
and an ideal gas equation
of state with $\Gamma=5/3$. The FD grid is $240\times240\times15$ and the domain
size is $[-0.5,0.5]^3$ with periodic boundary conditions. We use a time step
size of $\Delta t=10^{-3}$ and evolve to a final time of $t_f=2.4$, one period.

\begin{figure}[h]
  \raggedleft
  \begin{minipage}{0.42\columnwidth}
    \centering
    \includegraphics[width=1.0\textwidth]{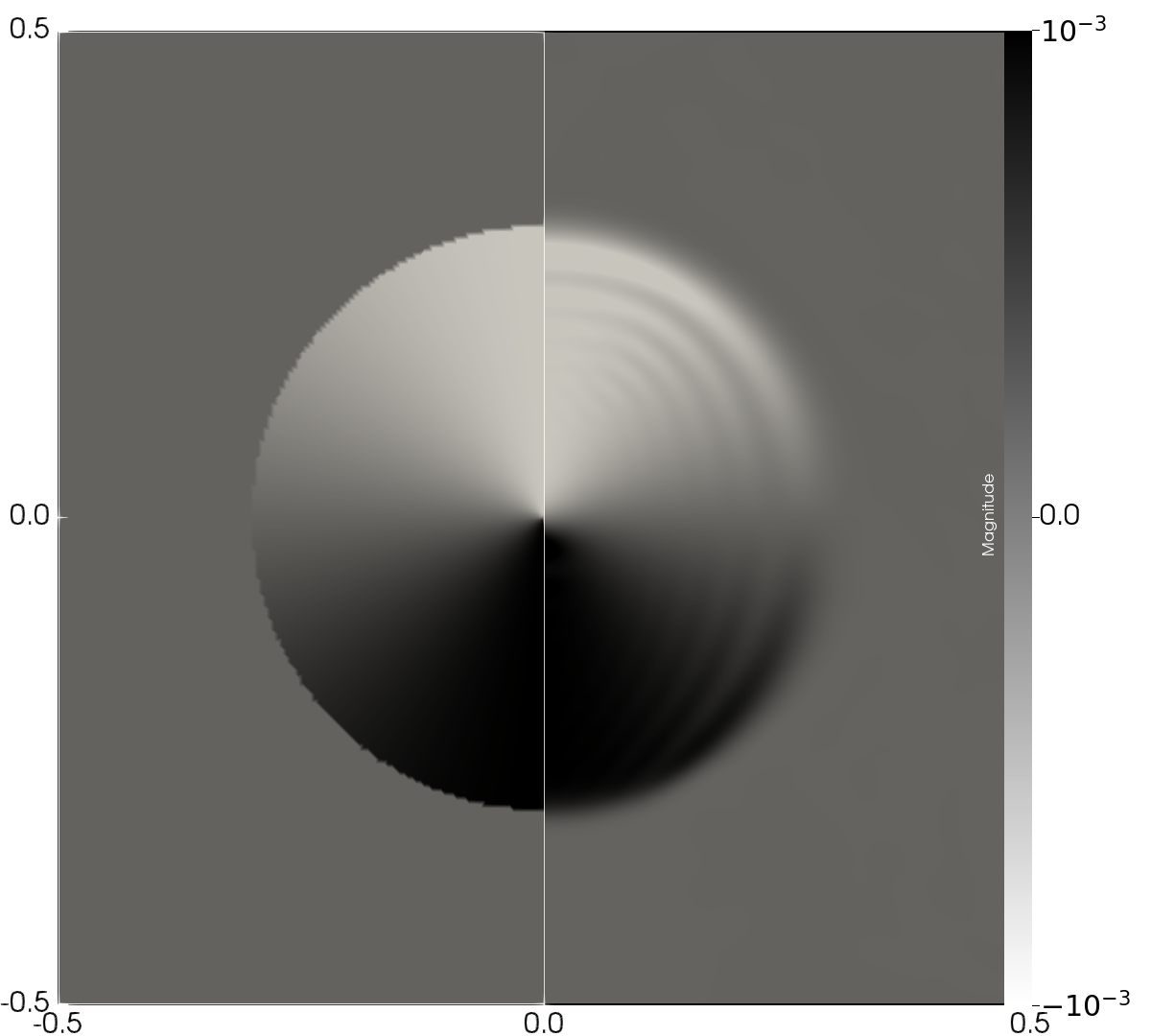}
    \\
    FD-2
  \end{minipage}
  \begin{minipage}{0.42\columnwidth}
    \centering
    \includegraphics[width=1.0\textwidth]{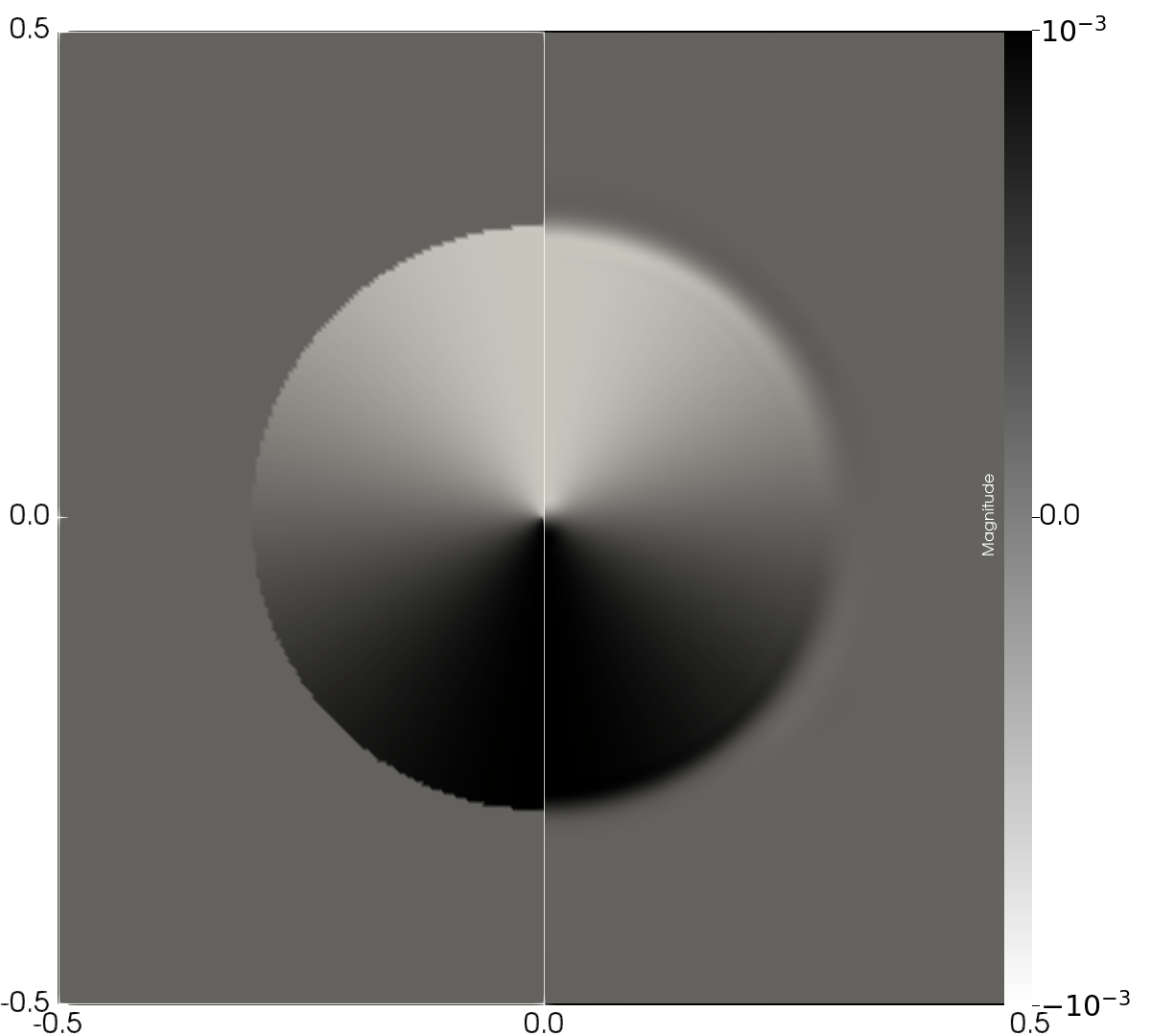}
    \\
    FD-4
  \end{minipage}

  \begin{minipage}{0.42\columnwidth}
    \centering
    \includegraphics[width=1.0\textwidth]{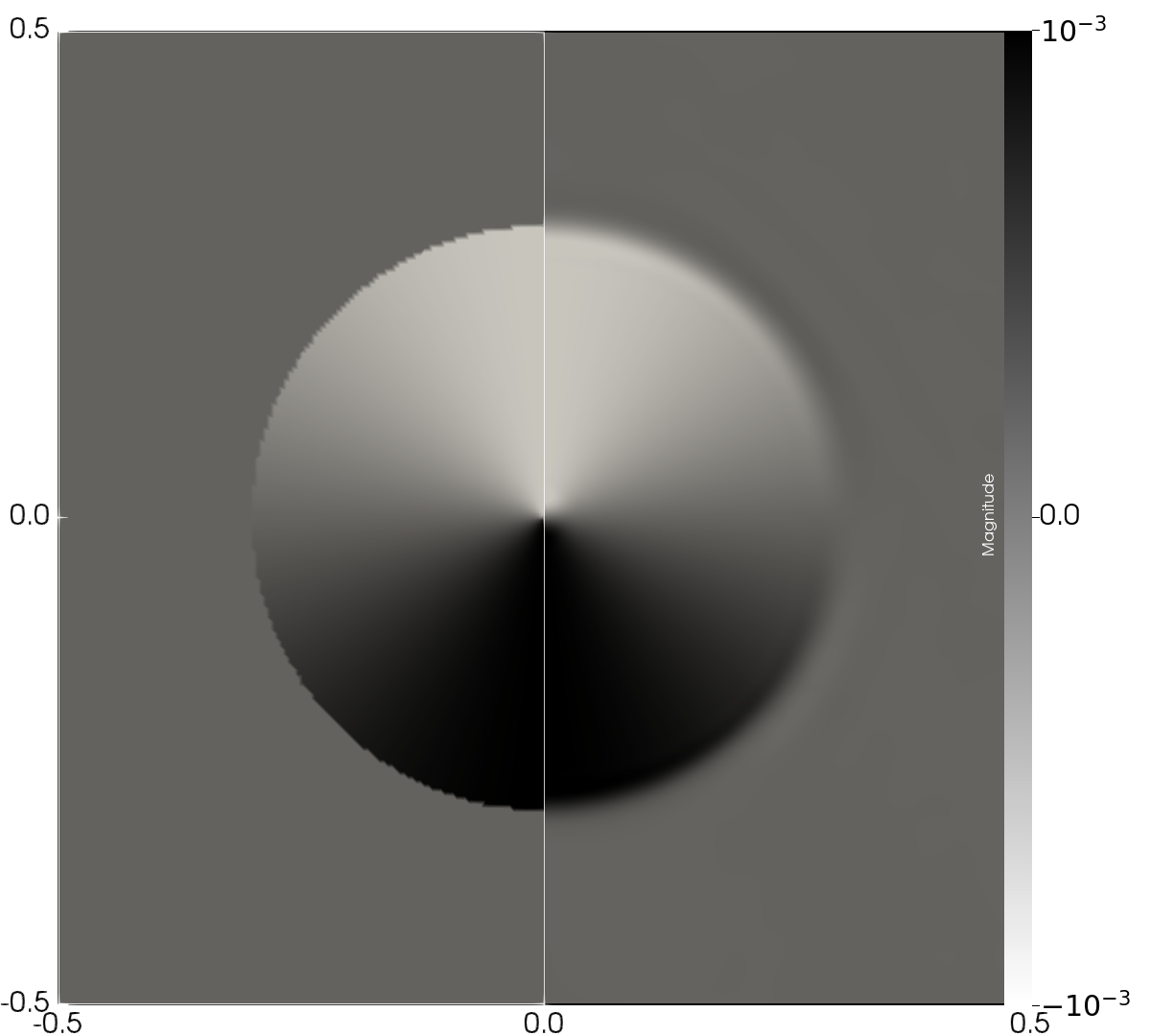}
    \\
    FD-6
  \end{minipage}
  \begin{minipage}{0.42\columnwidth}
    \centering
    \includegraphics[width=1.0\textwidth]{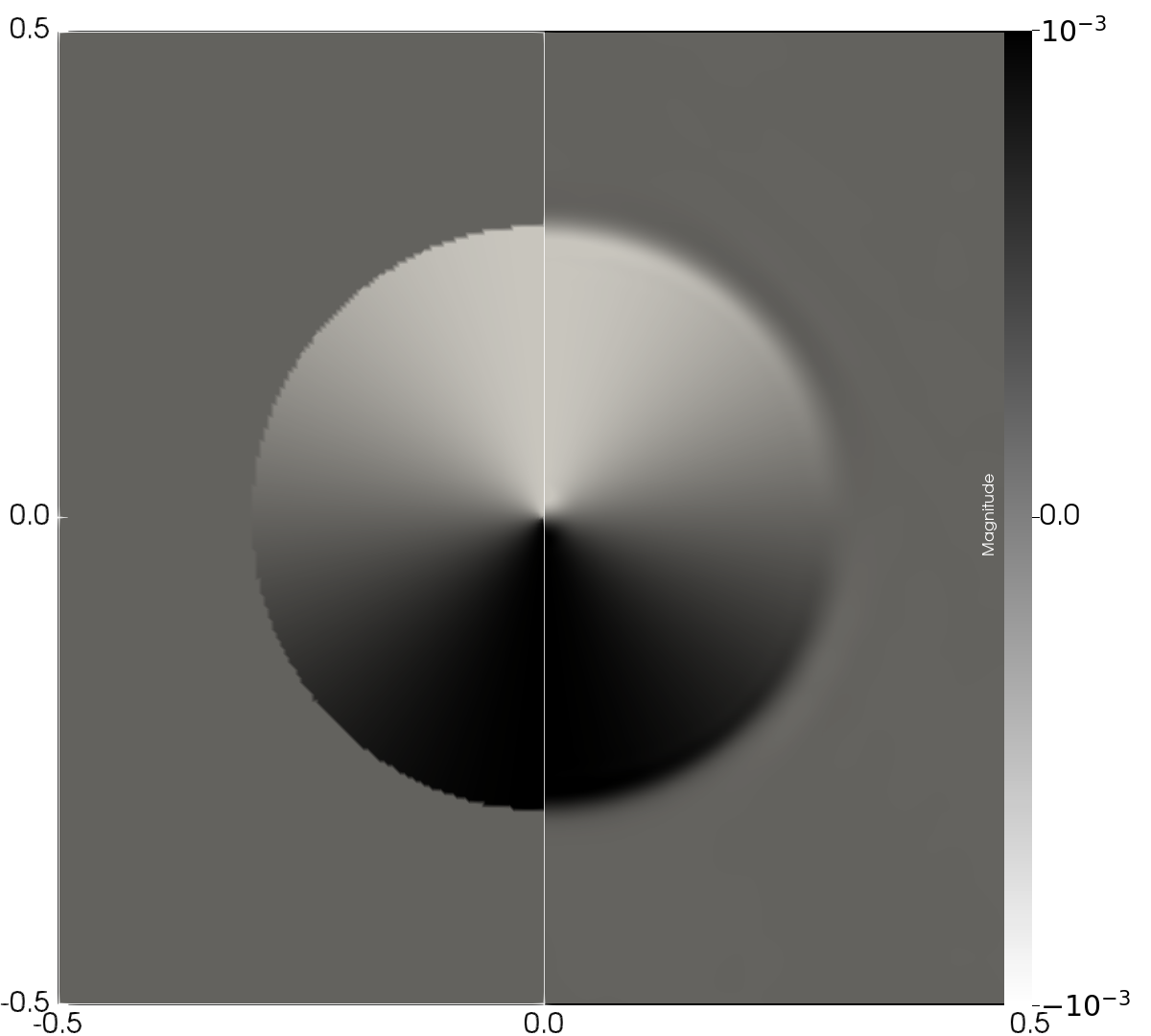}
    \\
    FD-10
  \end{minipage}

  \begin{minipage}{0.42\columnwidth}
    \centering
    \includegraphics[width=1.0\textwidth]{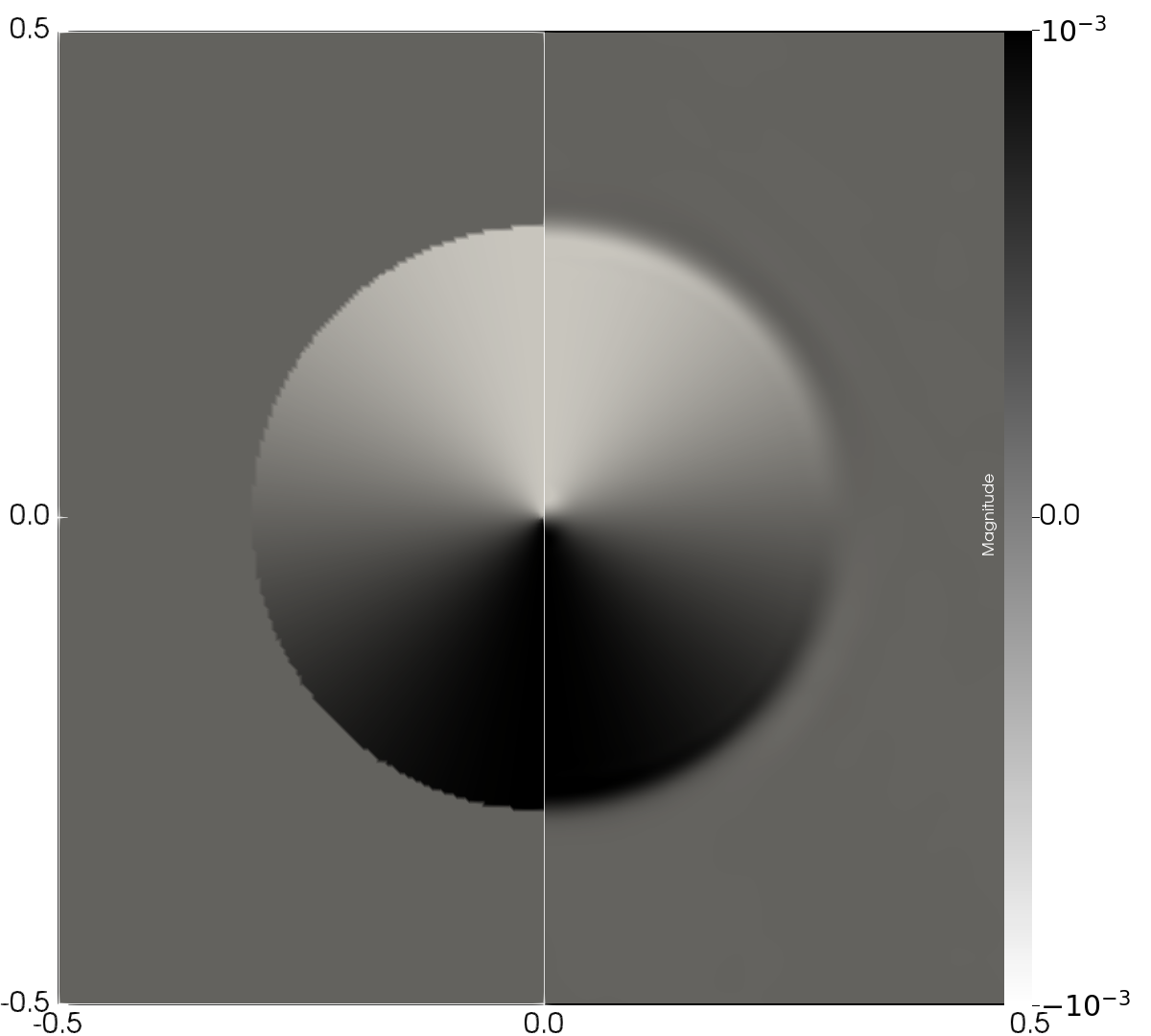}
    \\
    FD-10-6-2-2
  \end{minipage}
  \begin{minipage}{0.42\columnwidth}
    \centering
    \includegraphics[width=1.0\textwidth]{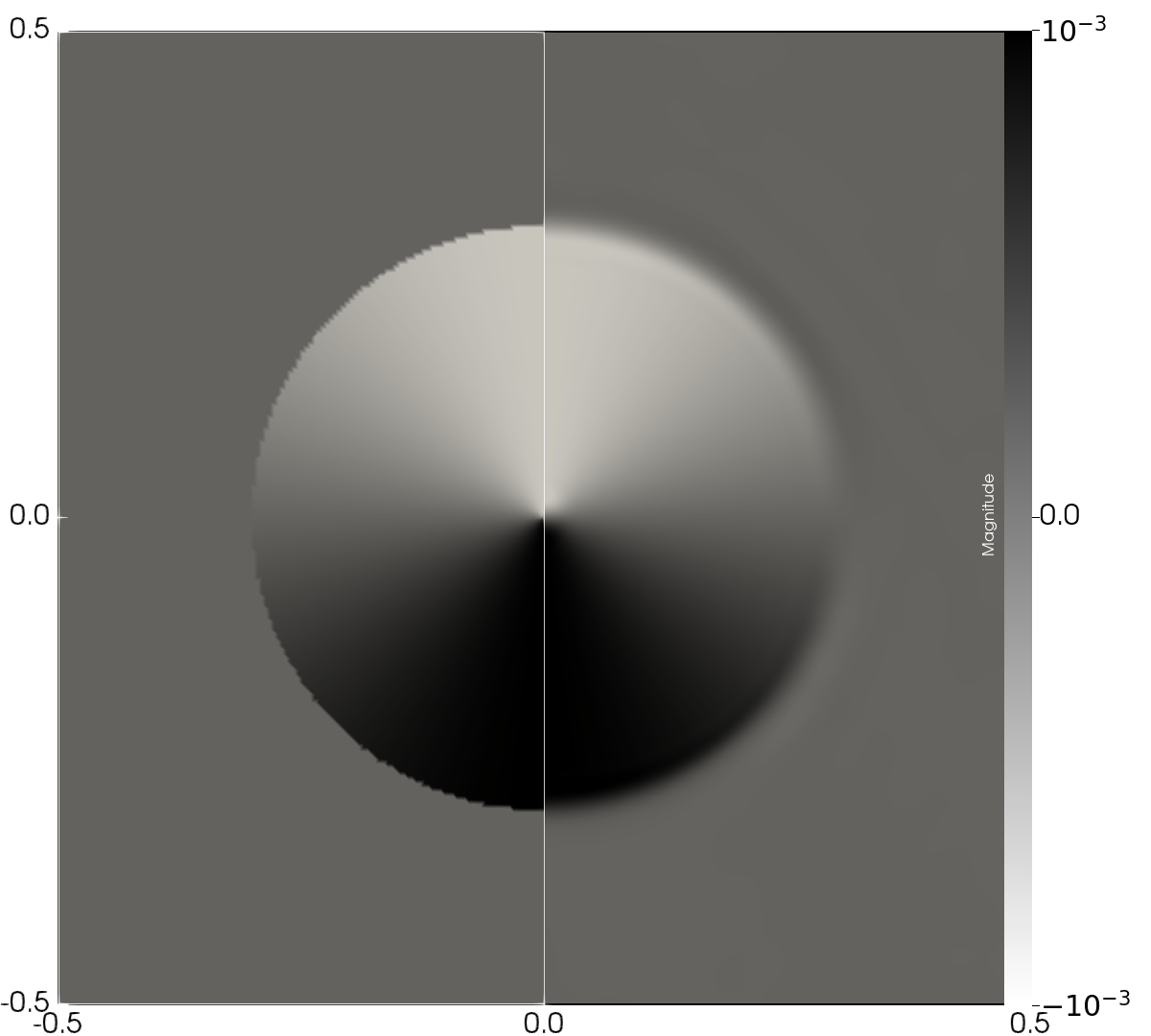}
    \\
    FD-10-4-2-2
  \end{minipage}
  \caption{$B^x$ for the magnetic loop advection problem. The left half of each
    plot is at the initial time, while the right half is after one period
    ($t_f=2.4$). The different panels all use the same
    PPAO9-5-2-1 reconstruction method but use different derivative orders. We
    see that the second-order FD derivative generates spurious oscillations
    throughout the loop, while the high-order and adaptive-order FD derivatives
    produce a cleaner solution.
  \label{fig:MagneticLoopBx}}
\end{figure}

\begin{figure}[h]
  \raggedleft
  \begin{minipage}{0.42\columnwidth}
    \centering
    \includegraphics[width=1.0\textwidth]{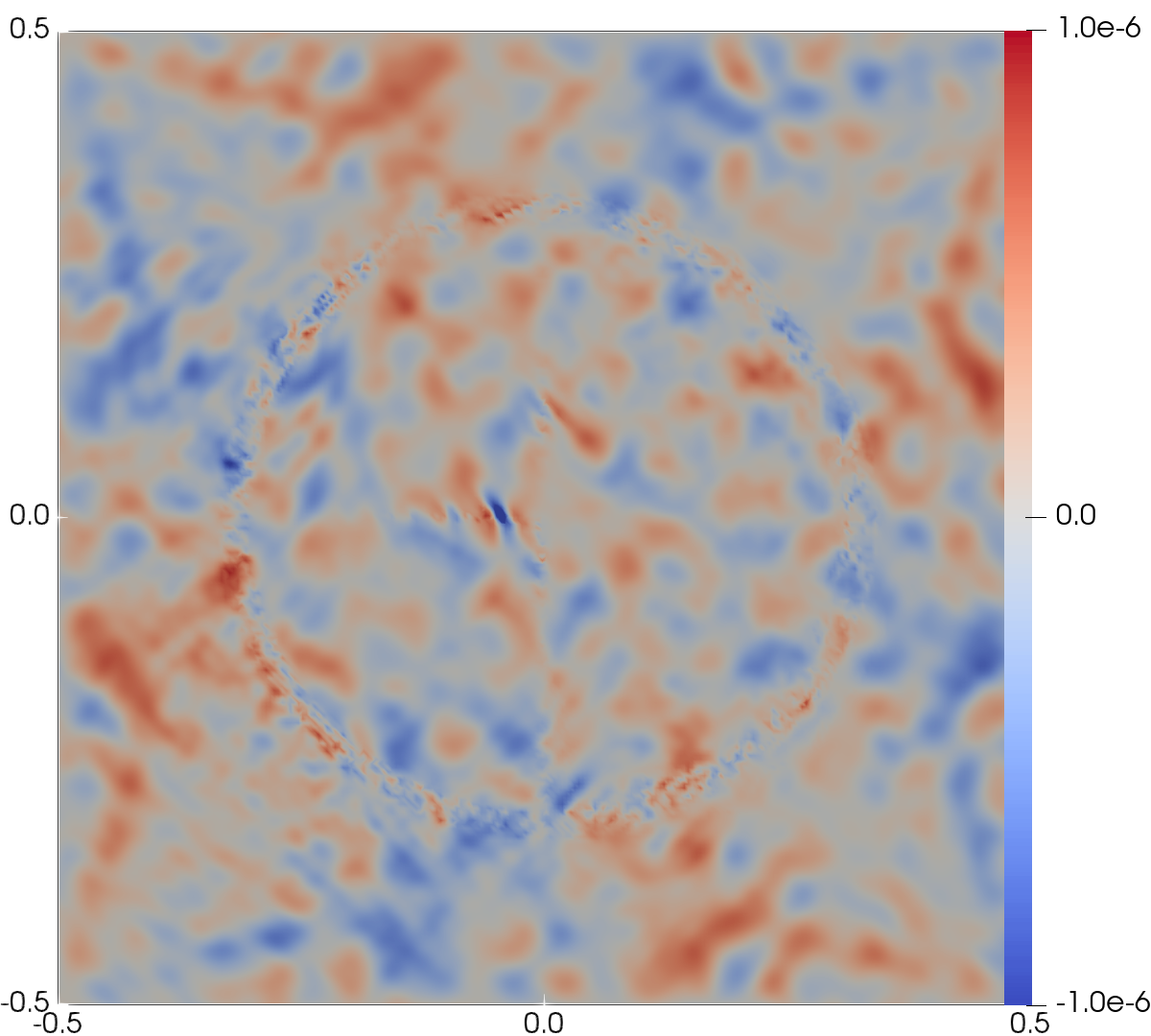}
    \\
    FD-2
  \end{minipage}
  \begin{minipage}{0.42\columnwidth}
    \centering
    \includegraphics[width=1.0\textwidth]{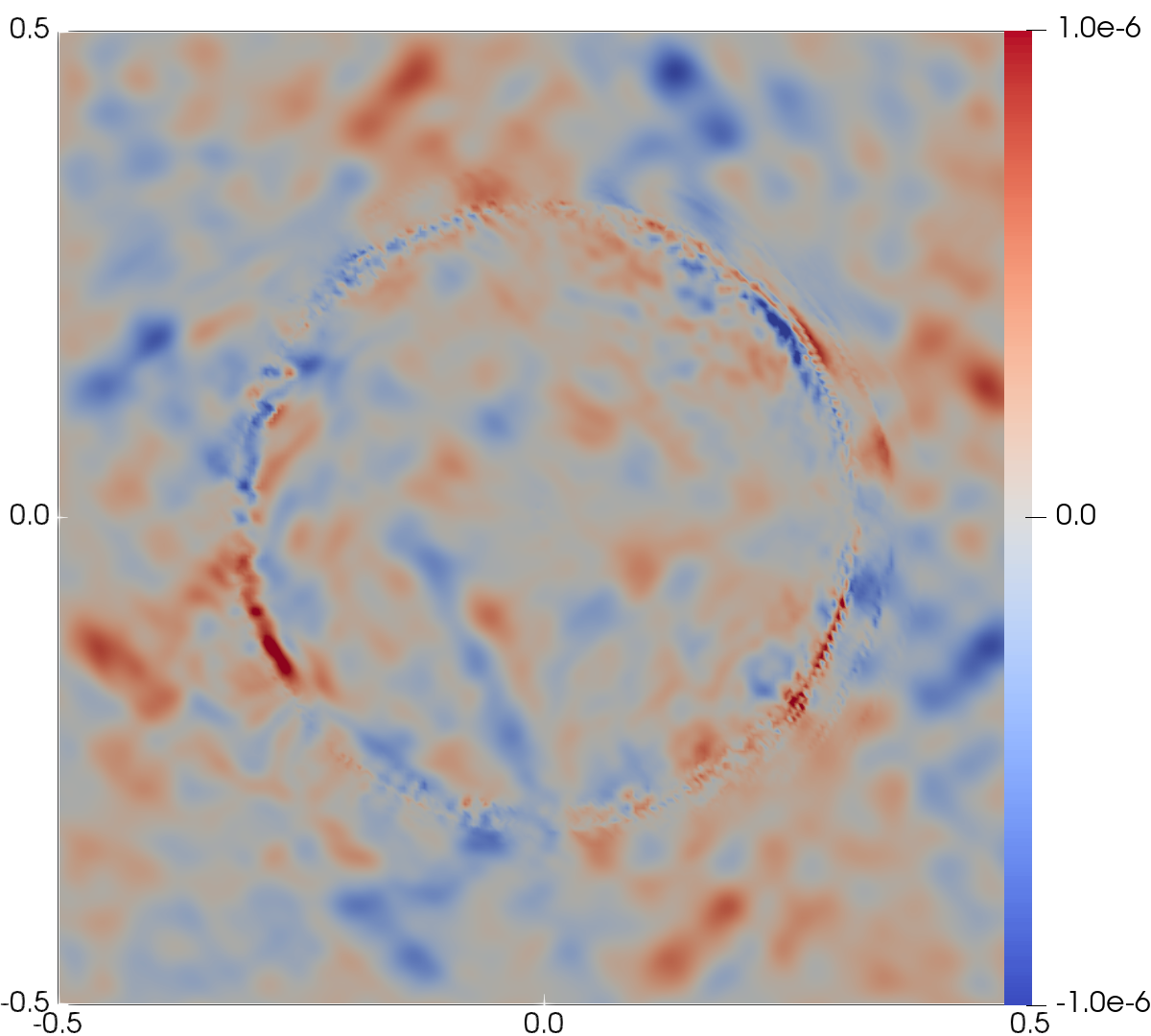}
    \\
    FD-4
  \end{minipage}

  \begin{minipage}{0.42\columnwidth}
    \centering
    \includegraphics[width=1.0\textwidth]{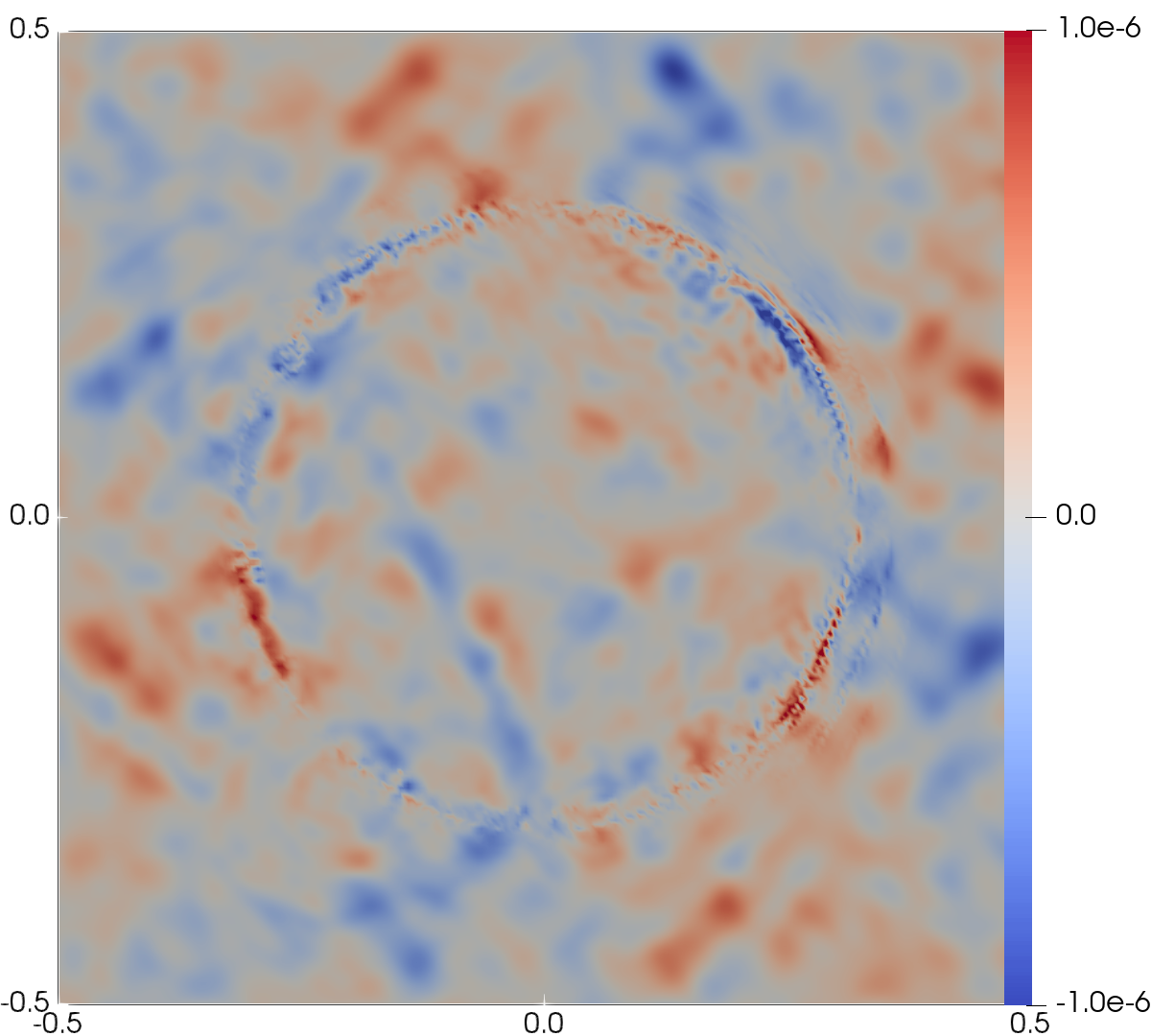}
    \\
    FD-6
  \end{minipage}
  \begin{minipage}{0.42\columnwidth}
    \centering
    \includegraphics[width=1.0\textwidth]{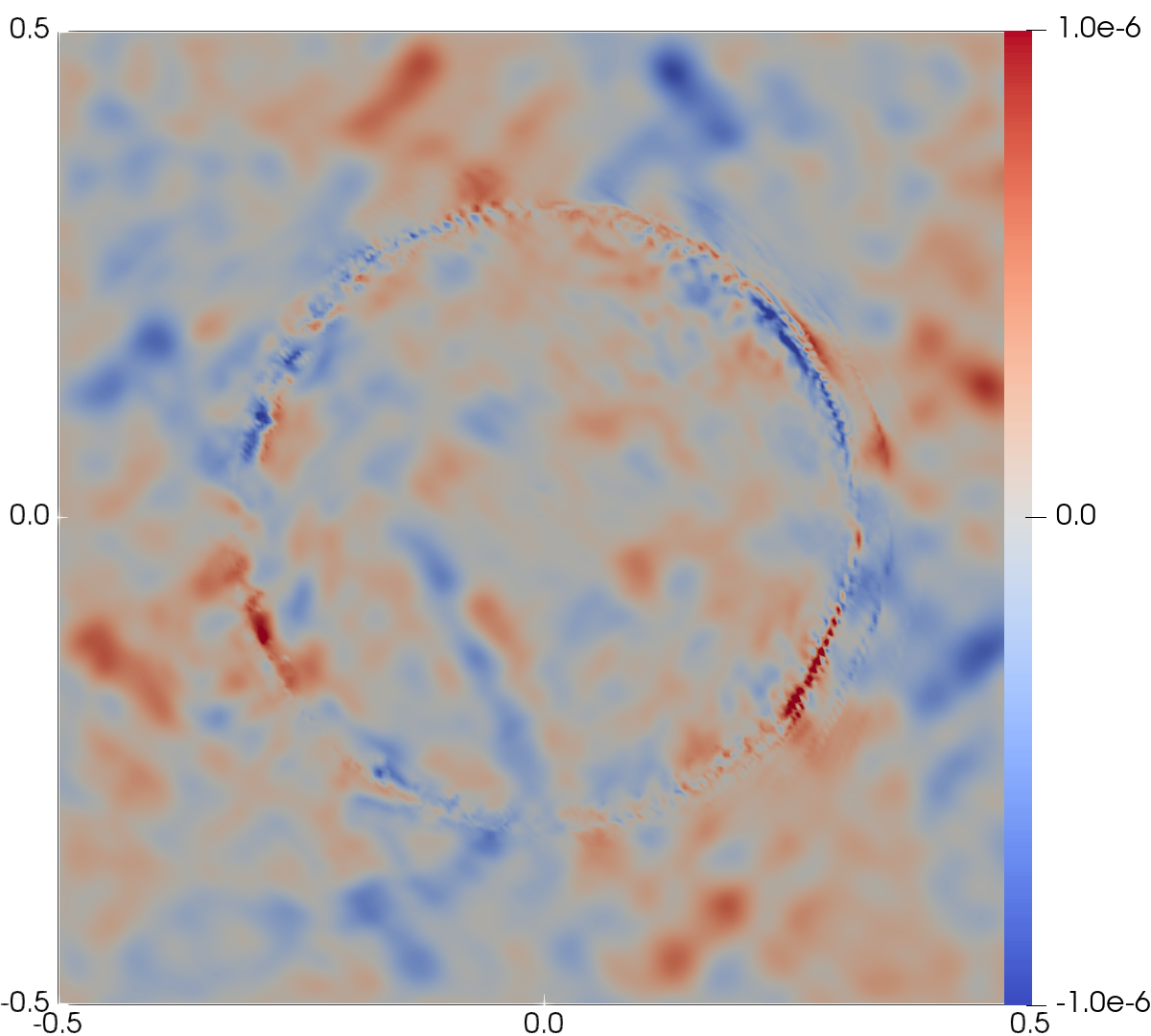}
    \\
    FD-10
  \end{minipage}

  \begin{minipage}{0.42\columnwidth}
    \centering
    \includegraphics[width=1.0\textwidth]{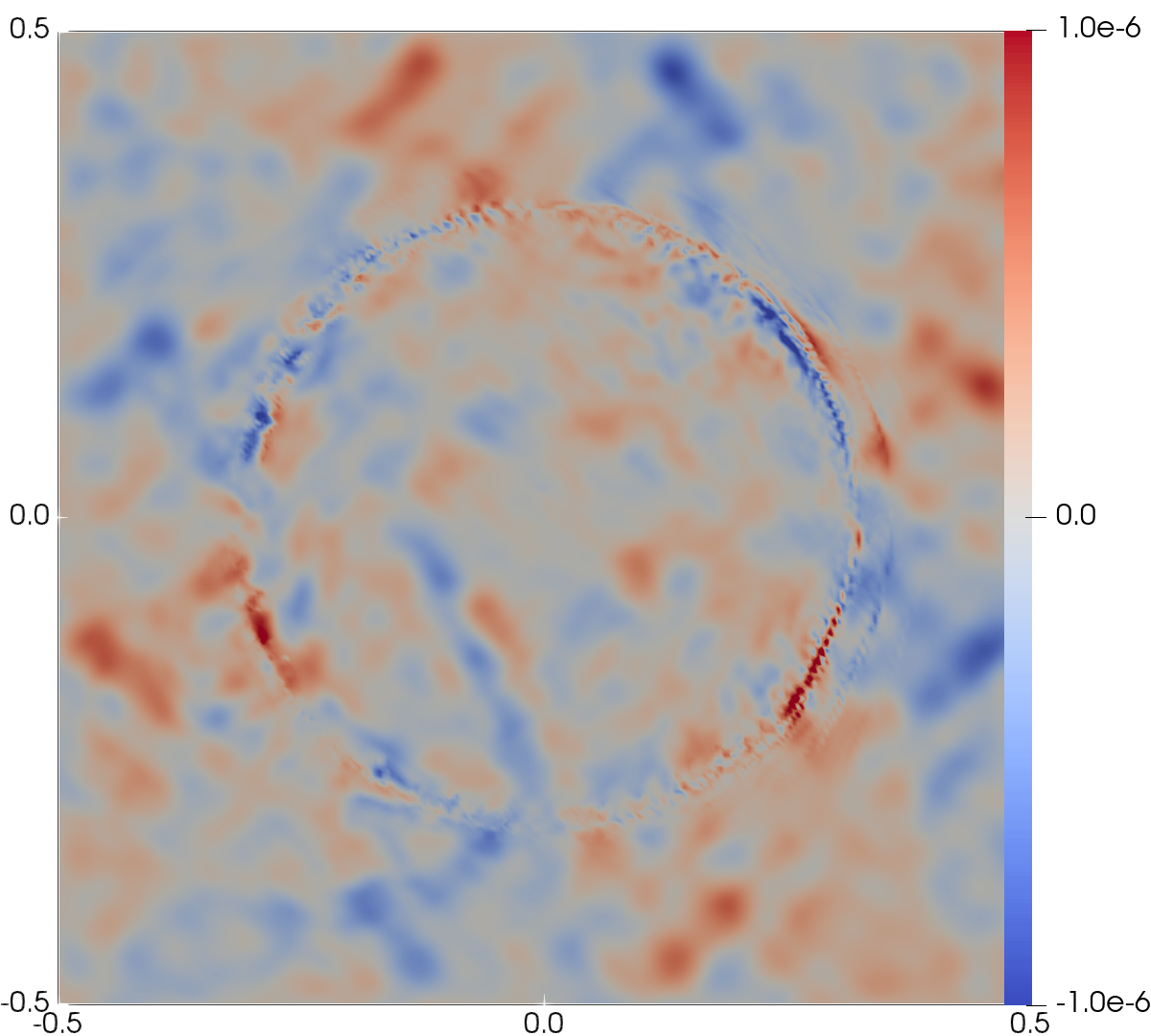}
    \\
    FD-10-6-2-2
  \end{minipage}
  \begin{minipage}{0.42\columnwidth}
    \centering
    \includegraphics[width=1.0\textwidth]{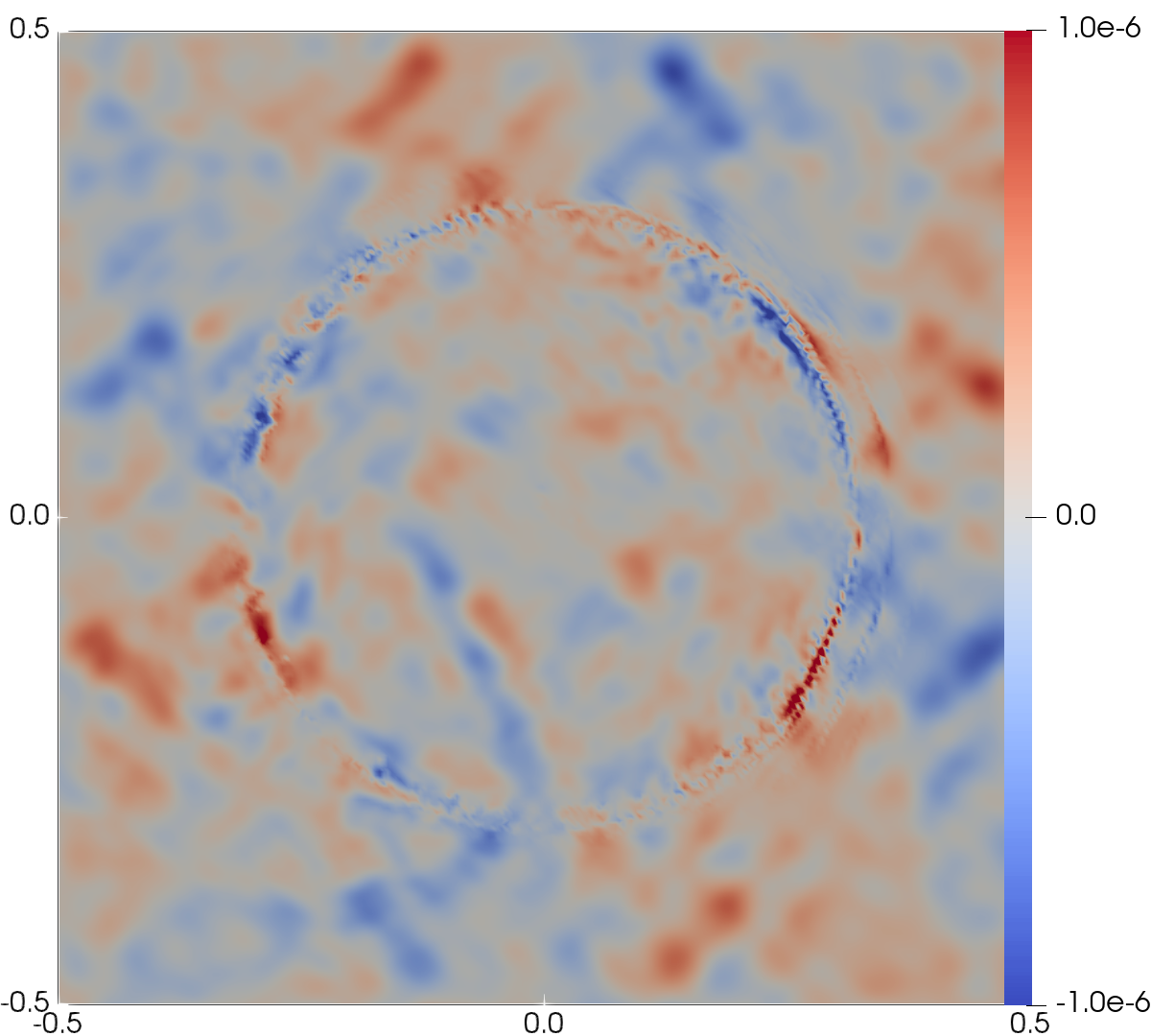}
    \\
    FD-10-4-2-2
  \end{minipage}
  \caption{The divergence cleaning field $\Phi$ for the magnetic loop advection
    problem after one period ($t_f=2.4$). The different panels all use the same
    PPAO9-5-2-1 reconstruction method but use different derivative orders. We
    see that high-order FD derivatives generate additional $\partial_i B^i=0$
    constraint violations at the edge of the loop, but at an amplitude
    comparable to that of the background noise.
  \label{fig:MagneticLoopPhi}}
\end{figure}

In the left half of each plot in figure~\ref{fig:MagneticLoopBx} we plot $B^x$
at $t=0$, while in the right half we plot $B^x$ at $t_f=2.4$. If the numerical
method perfectly preserved the structure the two halves would look
identical. However, we see that using second-order derivatives everywhere (top
left panel) creates additional oscillations in the cone that are not present
when high-order (labeled FD-4, FD-6, and FD-10) or adaptive-order (labeled
FD-10-6-2-2 and FD-10-4-2-2) derivatives are used. We show the divergence
cleaning field $\Phi$ in figure~\ref{fig:MagneticLoopPhi}, which is a direct
measure of the $\partial_i B^i=0$ constraint violation. We see that in addition
to a random background, the high-order derivatives have larger constraint
violations at the outer edge of the loop ($r=R_{\mathrm{loop}}=0.3$). However,
the violations at $R_{\mathrm{loop}}$ are still of the same scale as the
background violations and so do not adversely affect the overall solution. This
demonstrates that our high-order and adaptive-order methods do not adversely
affect the divergence cleaning properties of the GRMHD system.

\subsection{Kelvin-Helmhotz instability}\label{sec:kh instability}

Next we study a magnetized Kelvin-Helmholtz (KH) instability, similar
to~\cite{Beckwith:2011iy}. The domain is $[-0.5,0.5]^3$, periodic boundary
conditions are applied in all directions, and we use initial conditions similar
to~\cite{Schaal:2015ila}:
\begin{eqnarray}
  \rho&=\left\{\begin{array}{ll}
    1, & \left|y - 0.5\right| < 0.25\\
    10^{-2}, & \mathrm{otherwise},
  \end{array}\right. \\
  p &= 1.0, \\
  v^x &= \left\{\begin{array}{ll}
    0.5, & \left|y - 0.5\right| < 0.25\\
    -0.5, & \mathrm{otherwise},
  \end{array}\right. \\
  v^y &= 0.1\sin(4\pi x)
          \left[\exp\left(-\frac{(y - 0.75)^2}{0.0707^2}\right)
        +\exp\left(-\frac{(y - 0.25)^2}{0.0707^2}\right)\right], \\
  v^z &= 0.0, \\
  B^i &= \left(10^{-3}, 0, 0\right).
\end{eqnarray}
An ideal gas equation of state is used with $\Gamma=4/3$ and simulation to a
final time $t_f=1.6$ with a CFL factor of 0.7. We find that using second-order
FD derivatives and the adaptive-order derivatives the simulations are stable
with a CFL of 0.9, while derivatives orders four through ten are unstable with
such a large CFL.

\begin{figure}[h]
  \centering \includegraphics[width=0.5\textwidth]{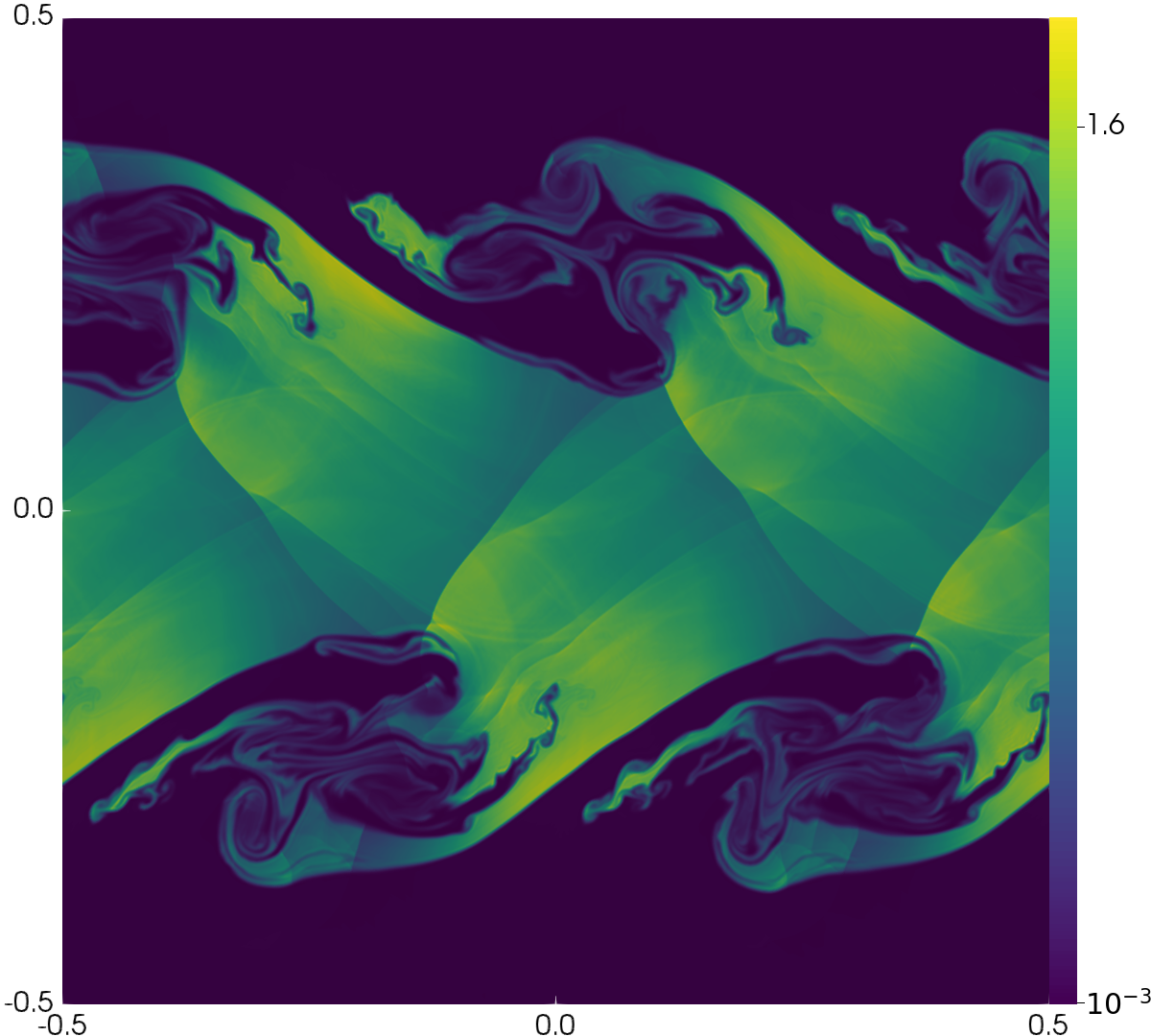}
  \caption{\blue{Rest mass density $\rho$ from a high-resolution, $1408^2$,
    Kelvin-Helmholtz instability simulation at $t=1.6$. The results in
    figure~\ref{fig:KhInstability} below should be compared to this.}
  \label{fig:KhInstabilityHighRes}}
\end{figure}

\begin{figure}[h]
  \raggedleft
  \begin{minipage}{0.42\columnwidth}
    \centering \includegraphics[width=1.0\textwidth]{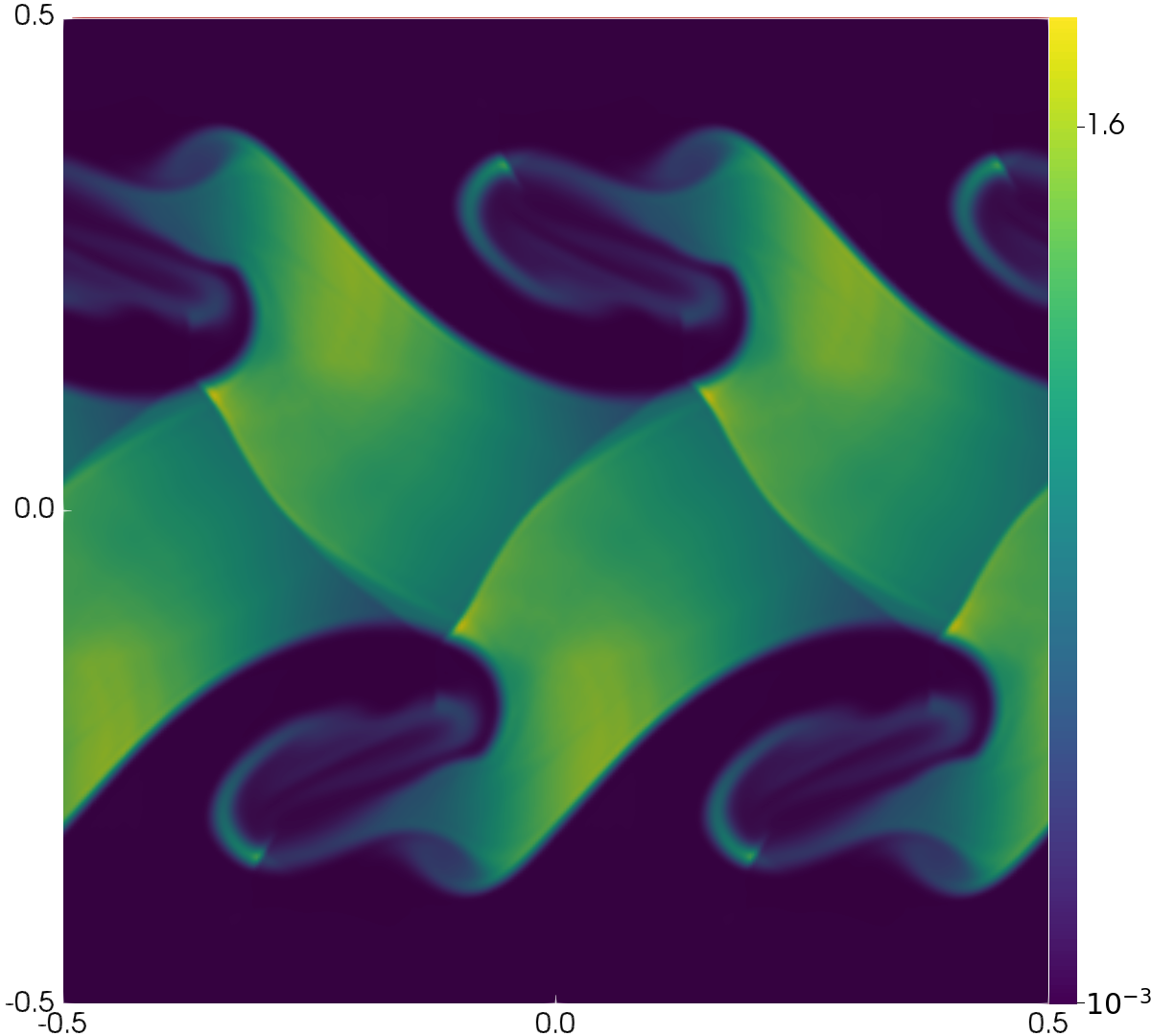}
    \\
    FD-2
  \end{minipage}
  \begin{minipage}{0.42\columnwidth}
    \centering \includegraphics[width=1.0\textwidth]{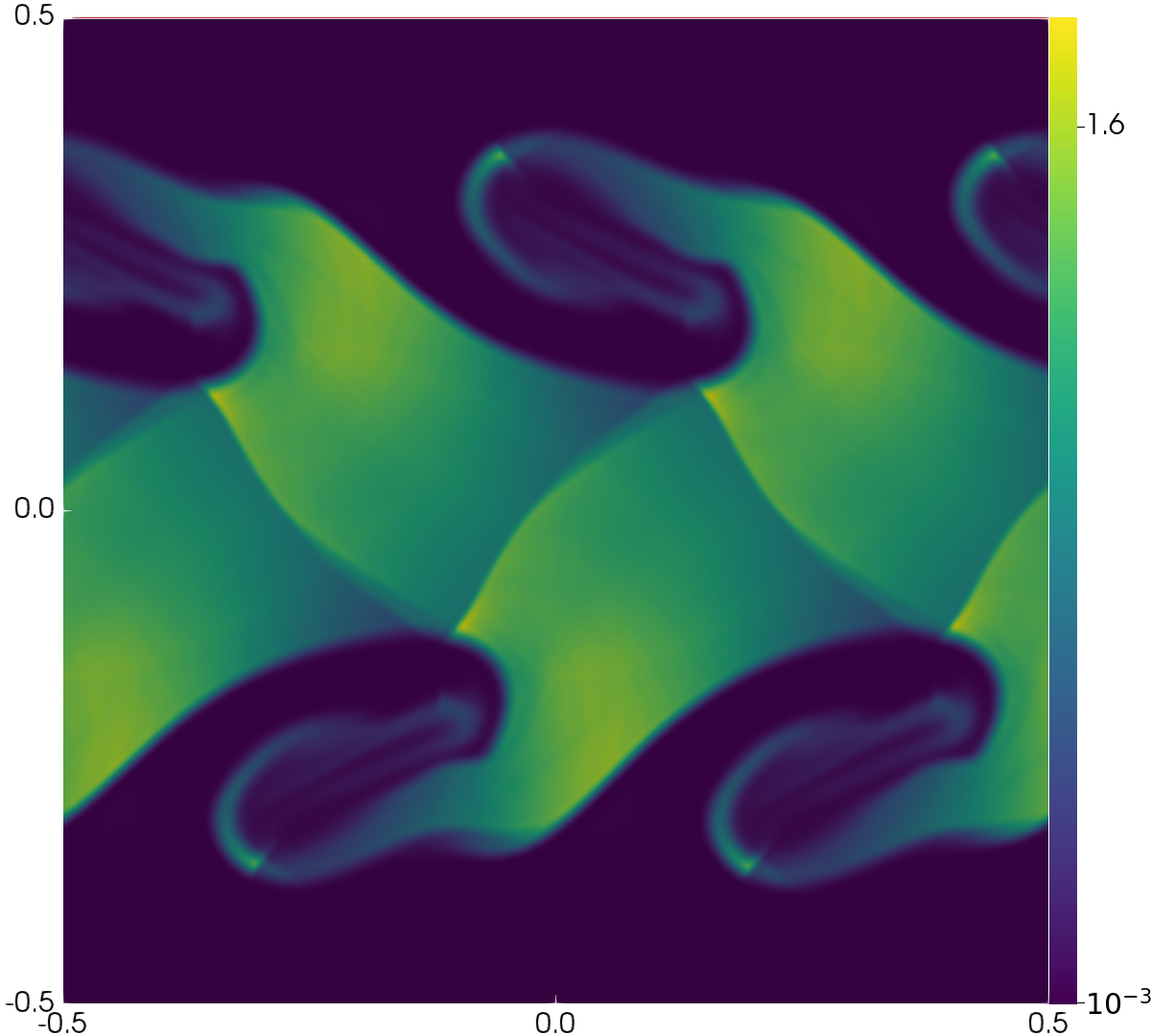}
    \\
    FD-4
  \end{minipage}

  \begin{minipage}{0.42\columnwidth}
    \centering \includegraphics[width=1.0\textwidth]{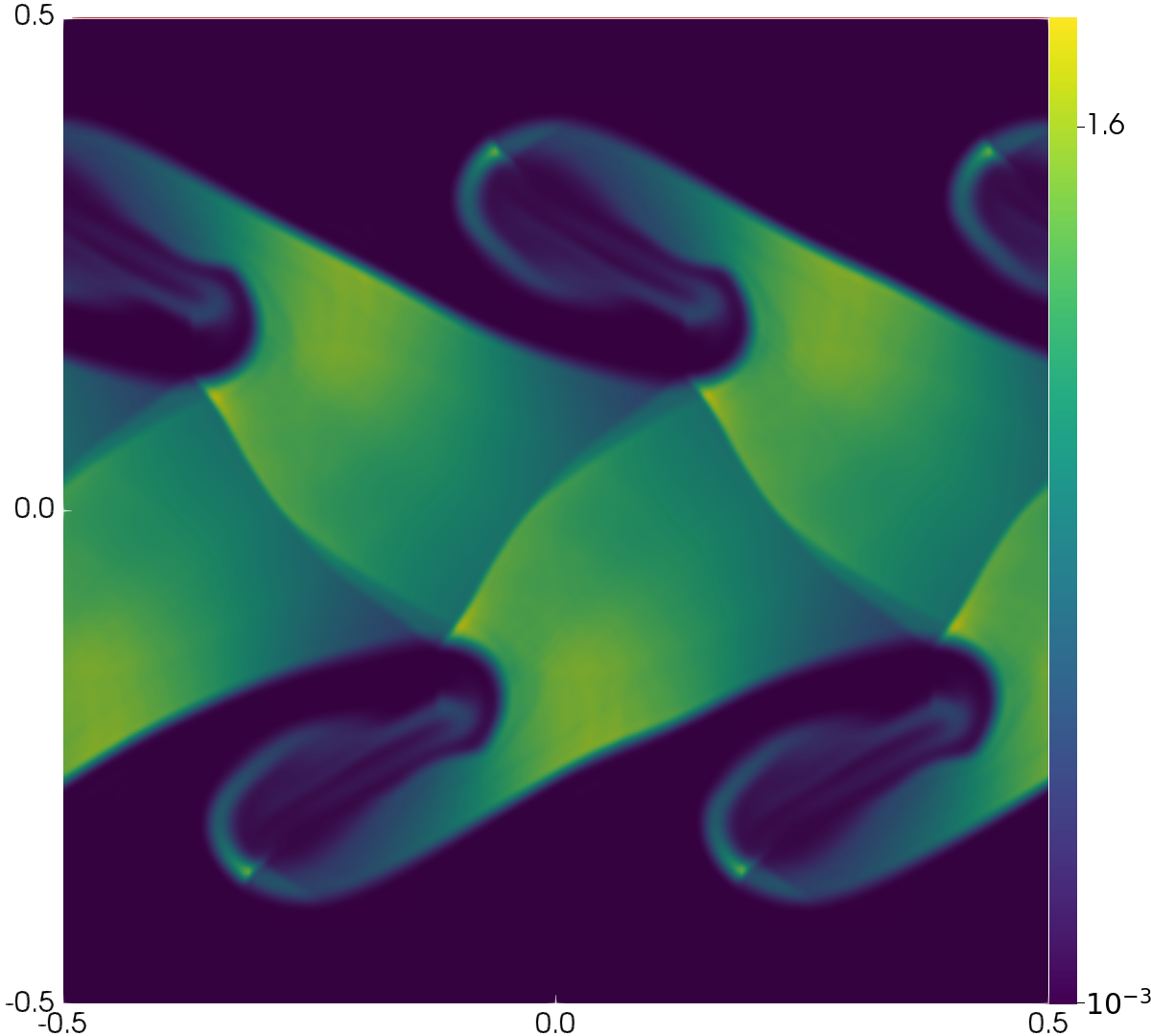}
    \\
    FD-6
  \end{minipage}
  \begin{minipage}{0.42\columnwidth}
    \centering \includegraphics[width=1.0\textwidth]{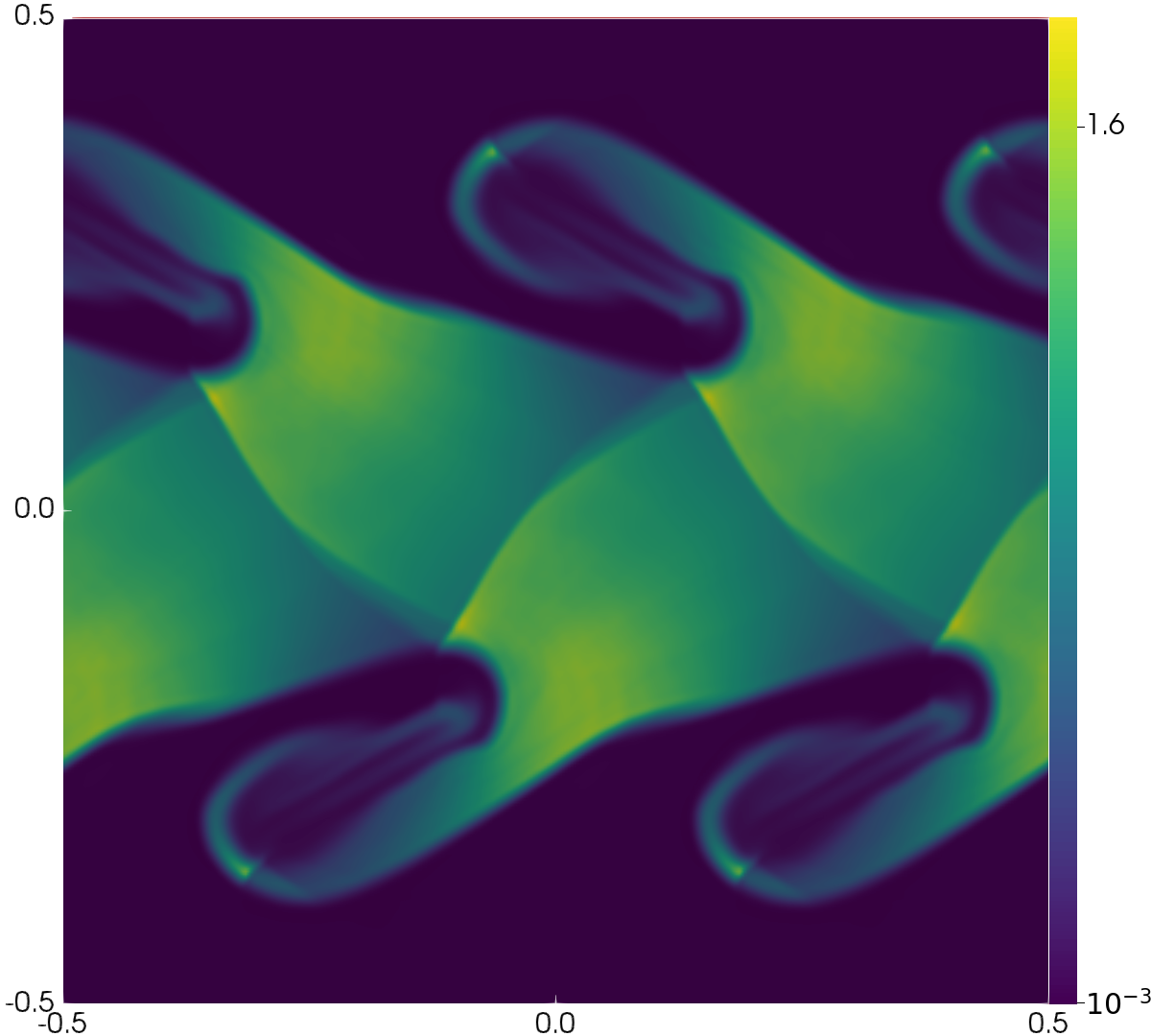}
    \\
    FD-8
  \end{minipage}

  \begin{minipage}{0.42\columnwidth}
    \centering \includegraphics[width=1.0\textwidth]{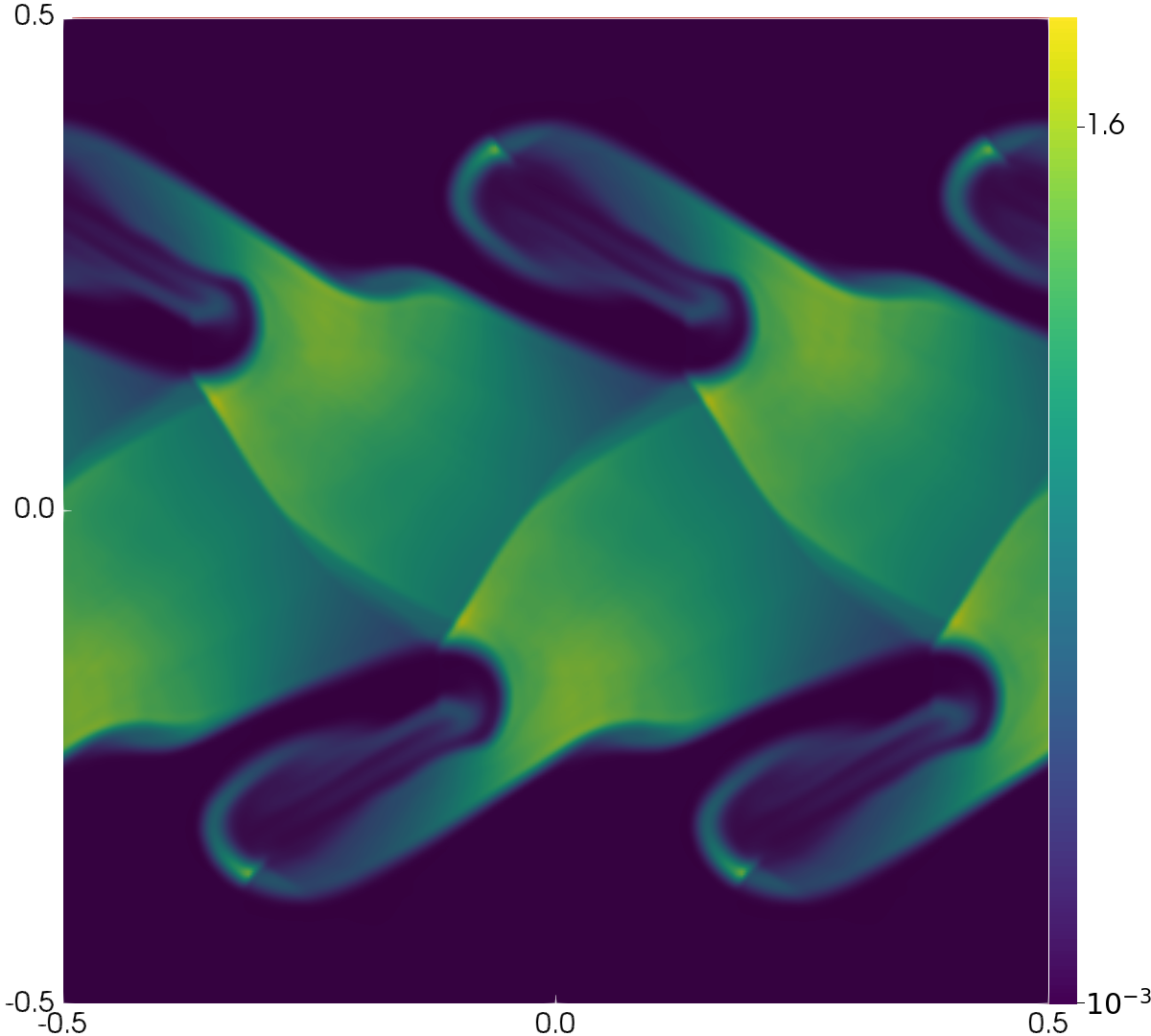}
    \\
    FD-10
  \end{minipage}
  \begin{minipage}{0.42\columnwidth}
    \centering
    \includegraphics[width=1.0\textwidth]{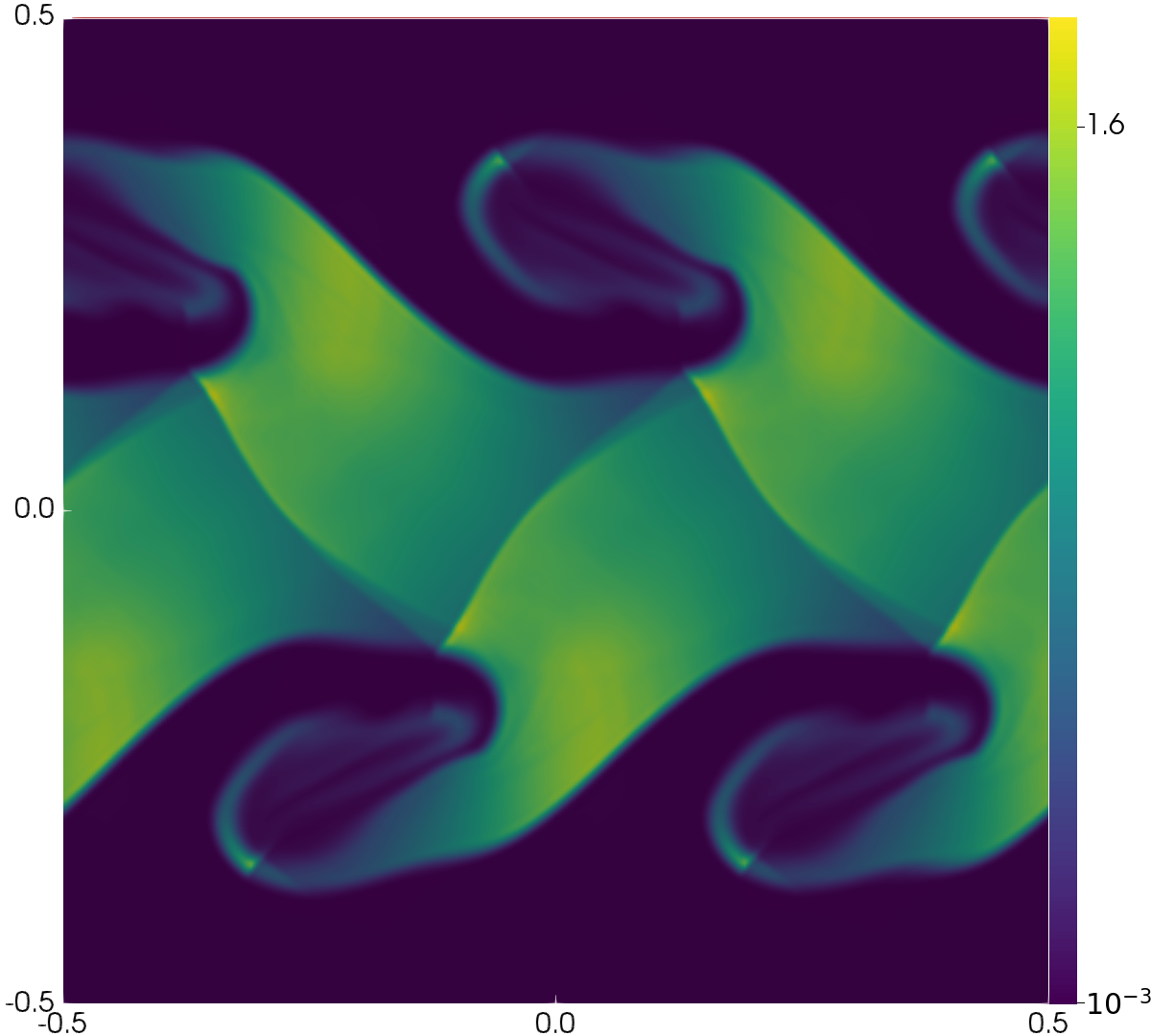}
    \\
    FD-10-6-2-2
  \end{minipage}
  \caption{Kelvin-Helmholtz instability $\rho$ at $t=1.6$ showing the results
    using different FD derivative orders and always using the PPAO9-5-2-1
    reconstruction method. In all cases the scheme is stable, though the FD-2
    case is able to resolve additional small-scale vortices and the FD-10-6-2-2
    case is almost able to resolve them. Reducing the CFL factor to 0.5 allows
    the FD-10-6-2-2 scheme to resolve the additional vortices.
    \label{fig:KhInstability}}
\end{figure}

\begin{figure}[h]
  \raggedleft
  \begin{minipage}{0.42\columnwidth}
    \centering
    \includegraphics[width=1.0\textwidth]{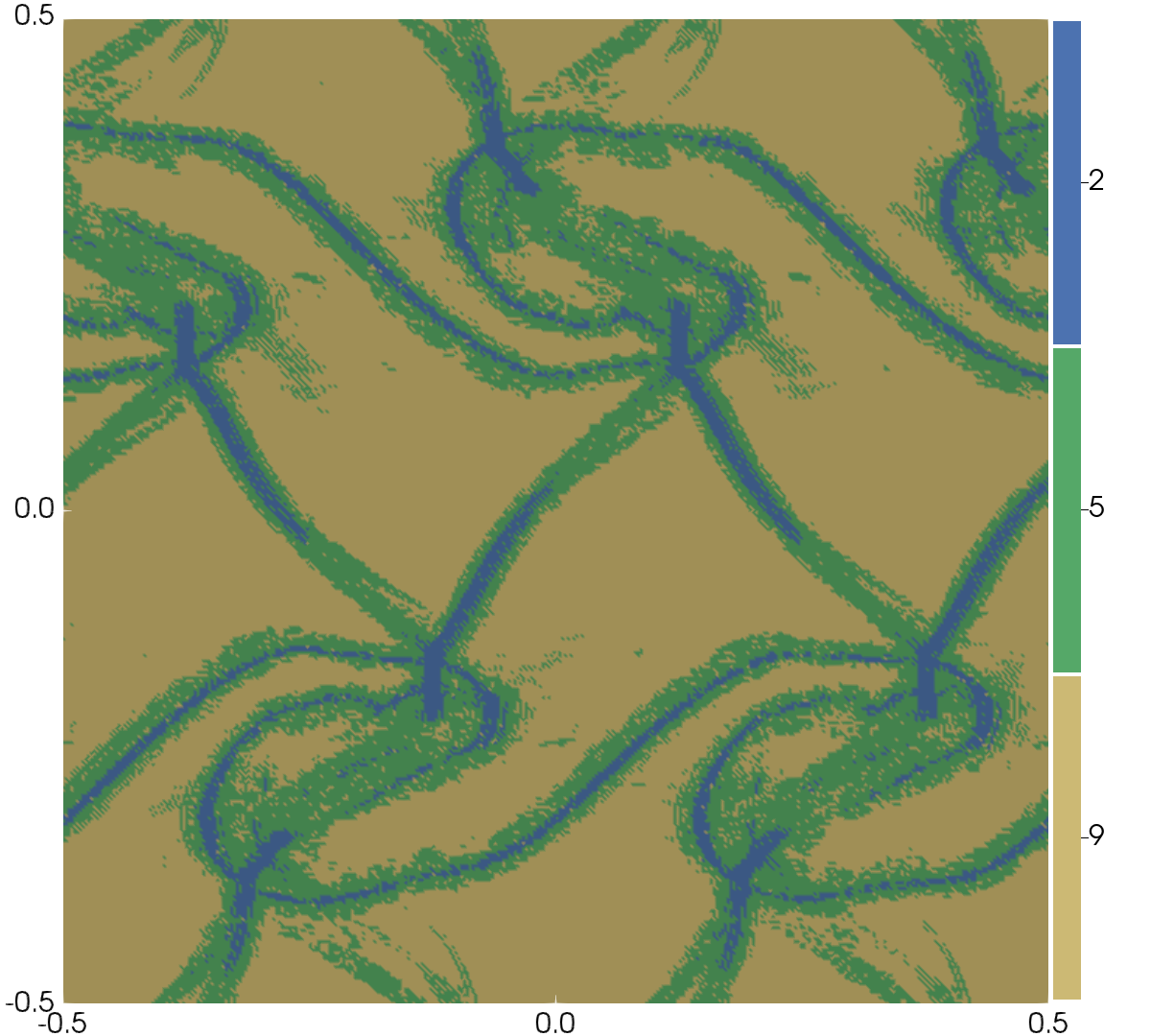}
    \\
    Reconstruction Order in $x$
  \end{minipage}
  \begin{minipage}{0.42\columnwidth}
    \centering
    \includegraphics[width=1.0\textwidth]{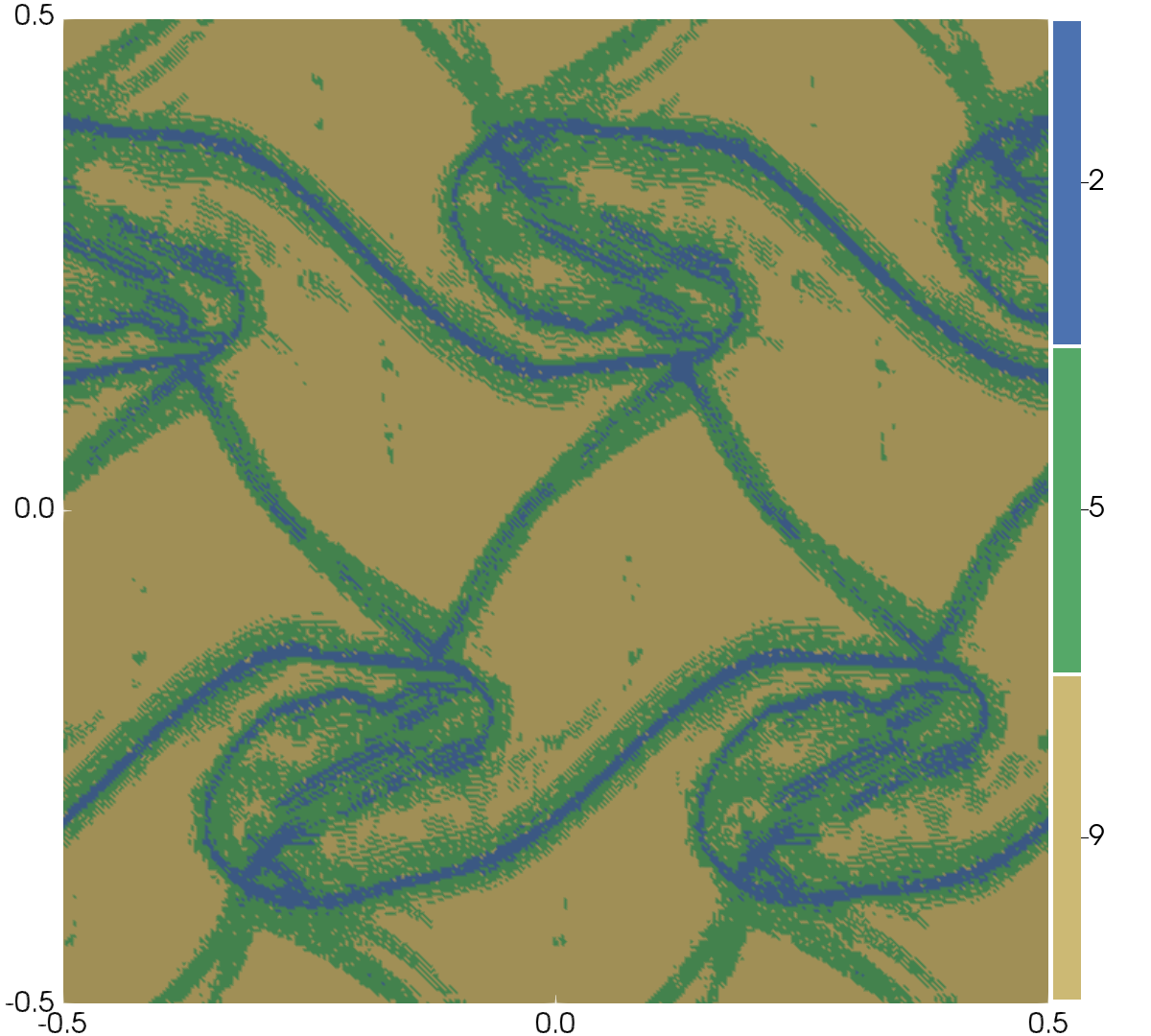}
    \\
    Reconstruction Order in $y$
  \end{minipage}

    \begin{minipage}{0.42\columnwidth}
    \centering
    \includegraphics[width=1.0\textwidth]{./KhInstabilityDOneHigherThanRecons.png}
    \\
    $\rho$
  \end{minipage}
  \begin{minipage}{0.42\columnwidth}
    \centering
    \includegraphics[width=1.0\textwidth]{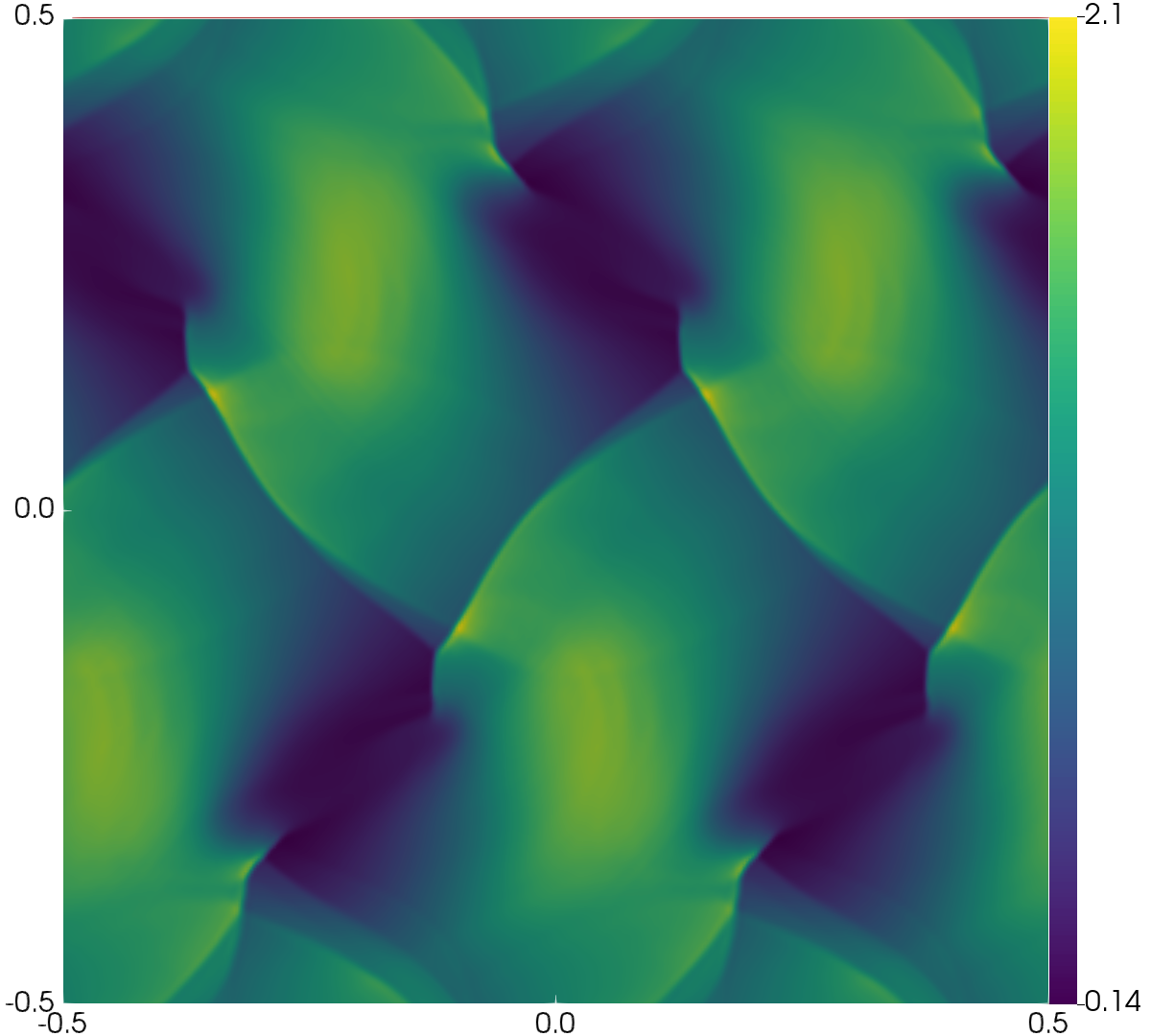}
    \\
    $p$
  \end{minipage}
  \caption{Results from the Kelvin-Helmholtz instability problem. The panels in
    the top row show the reconstruction and FD derivative order used in the
    $x$-direction (left) and $y$-direction (right) at the final time, while the
    bottom left panel shows the rest mass density and the bottom right the
    pressure at the final time. We see that the adaptive-order FD scheme
    accurately tracks non-smooth features in the solution, specifically the rest
    mass density and pressure, adjusting the order as necessary.
  \label{fig:KhInstabilityReconsOrder}}
\end{figure}

\blue{We plot the rest mass density $\rho$ from a high-resolution, $1408^2$,
  reference simulation using PPAO9-5-2-1 and FD-2 in
  figure~\ref{fig:KhInstabilityHighRes} to compare our results to. The
  high-resolution simulation has four times as many points per dimension as our
  standard resolution.}  We plot the rest mass density $\rho$ at the final time
$t_f=1.6$ in figure~\ref{fig:KhInstability} for simulations using the
PPAO9-5-2-1 reconstruction scheme but using different order FD derivatives. All
schemes are stable, though the FD-2 is able to resolve additional small-scale
vortices at $(x,y)\sim(0.5, 0.25)$ and FD-10-6-2-2 is almost able to resolve
these. We find that reducing the CFL factor to 0.5 allows the FD-10-6-2-2 scheme
to resolve the additional vortices, \blue{similar to those seen in the
  high-resolution simulation shown in
  figure~\ref{fig:KhInstabilityHighRes}}. We show the reconstruction order
alongside $\rho$ and pressure $p$ in figure~\ref{fig:KhInstabilityReconsOrder},
nicely demonstrating that the PPAO scheme accurately tracks non-smooth features
in both $\rho$ and $p$. This ultimately means that both shocks and contact
discontinuities are tracked and resolved. \blue{While there is not a clear best
  scheme, the important aspect for us is that the FD-10-6-2-2 method
  maintains reasonable shock capturing ability and is not significantly more
  diffuse than FD-2.}

\subsection{Orszag-Tang vortex}\label{sec:ot vortex}

The relativistic version of the Orszag-Tang vortex is a 2-dimensional test case
for GRMHD systems~\cite{Beckwith:2011iy}. The initial conditions (and hence the
states at later times) are periodic in both $x$ and $y$ with period 1.  The
initial conditions are:
\begin{eqnarray}
  \rho &= \frac{25}{36 \pi}, \\
  p &= \frac{5}{12 \pi}, \\
  v^i &= \left[-\frac{1}{2} \sin(2 \pi y), \frac{1}{2} \sin(2 \pi x), 0\right] \\
  B^i &=\left[-\frac{1}{\sqrt{4 \pi}} \sin(2 \pi y),
        \frac{1}{\sqrt{4 \pi}} \sin(4 \pi x), 0\right]
\end{eqnarray}
closed by an ideal equation of state with $\Gamma=5/3$. We use a domain
$[0,1]^3$ with periodic boundary conditions and evolve until a final time
$t_f=1$ using a CFL factor of 0.7.

\begin{figure}[h]
  \raggedleft
  \begin{minipage}{0.42\columnwidth}
    \centering
    \includegraphics[width=1.0\textwidth]{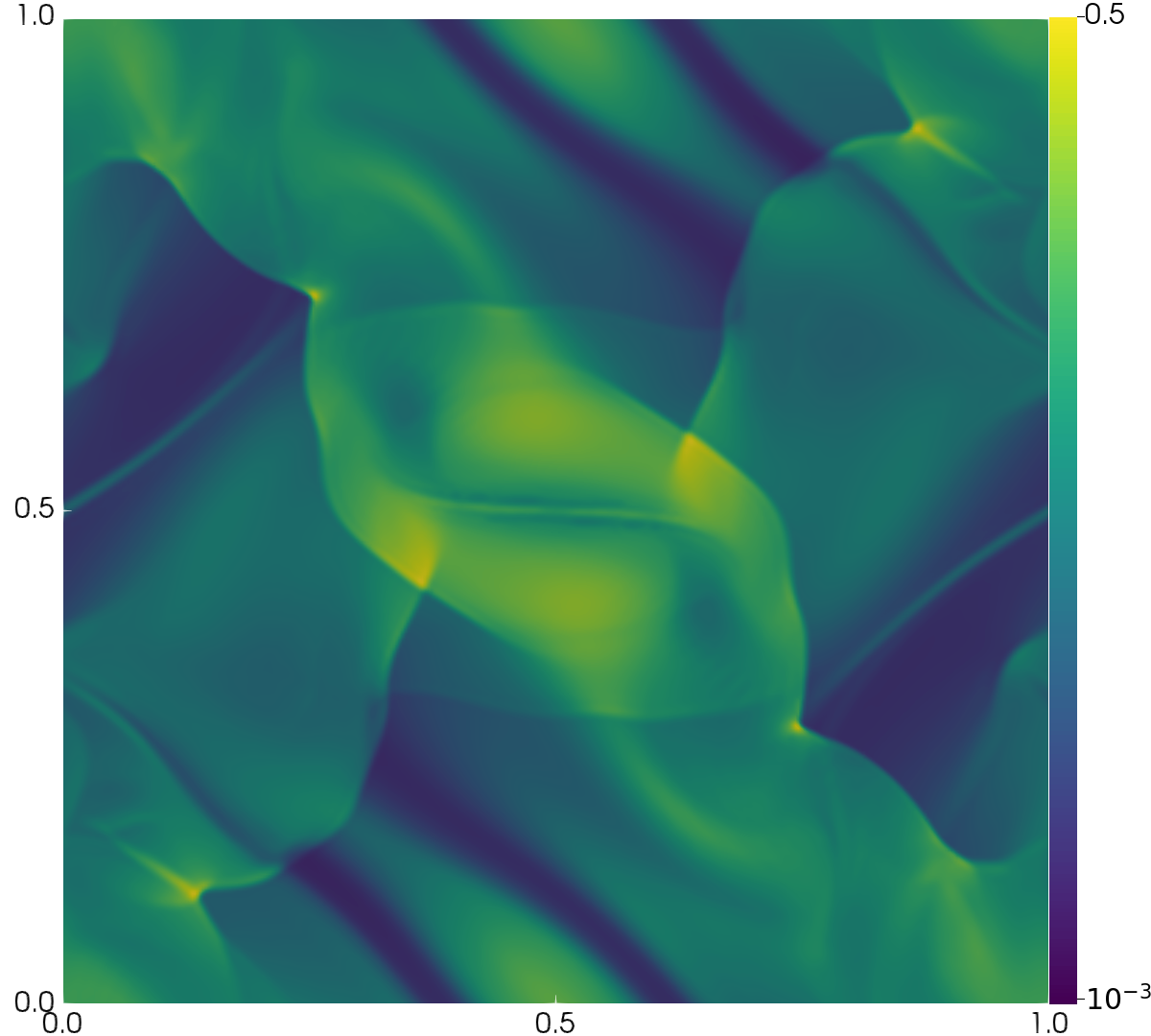}
    \\
    FD-2
  \end{minipage}
  \begin{minipage}{0.42\columnwidth}
    \centering
    \includegraphics[width=1.0\textwidth]{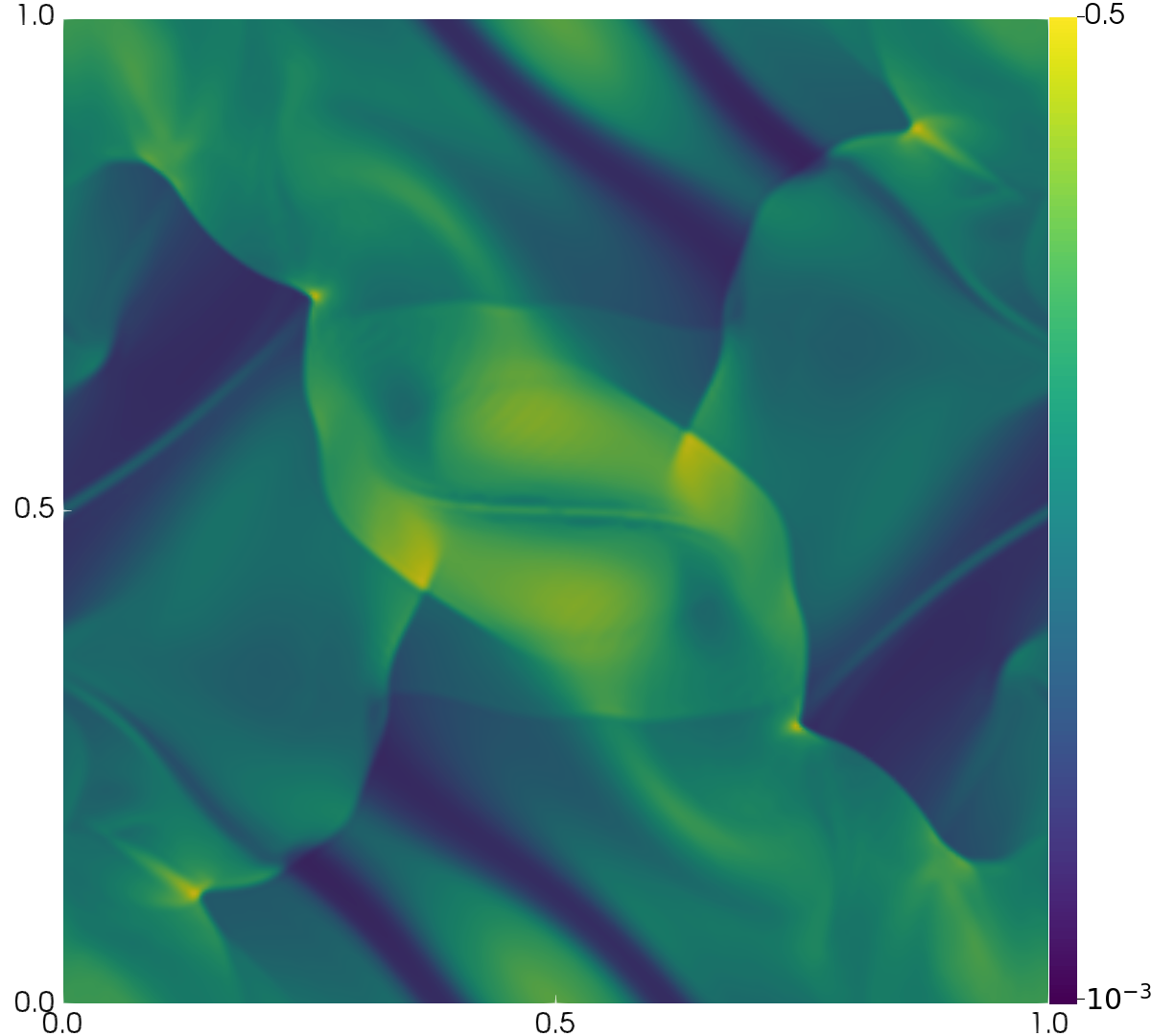}
    \\
    FD-4
  \end{minipage}

  \begin{minipage}{0.42\columnwidth}
    \centering
    \includegraphics[width=1.0\textwidth]{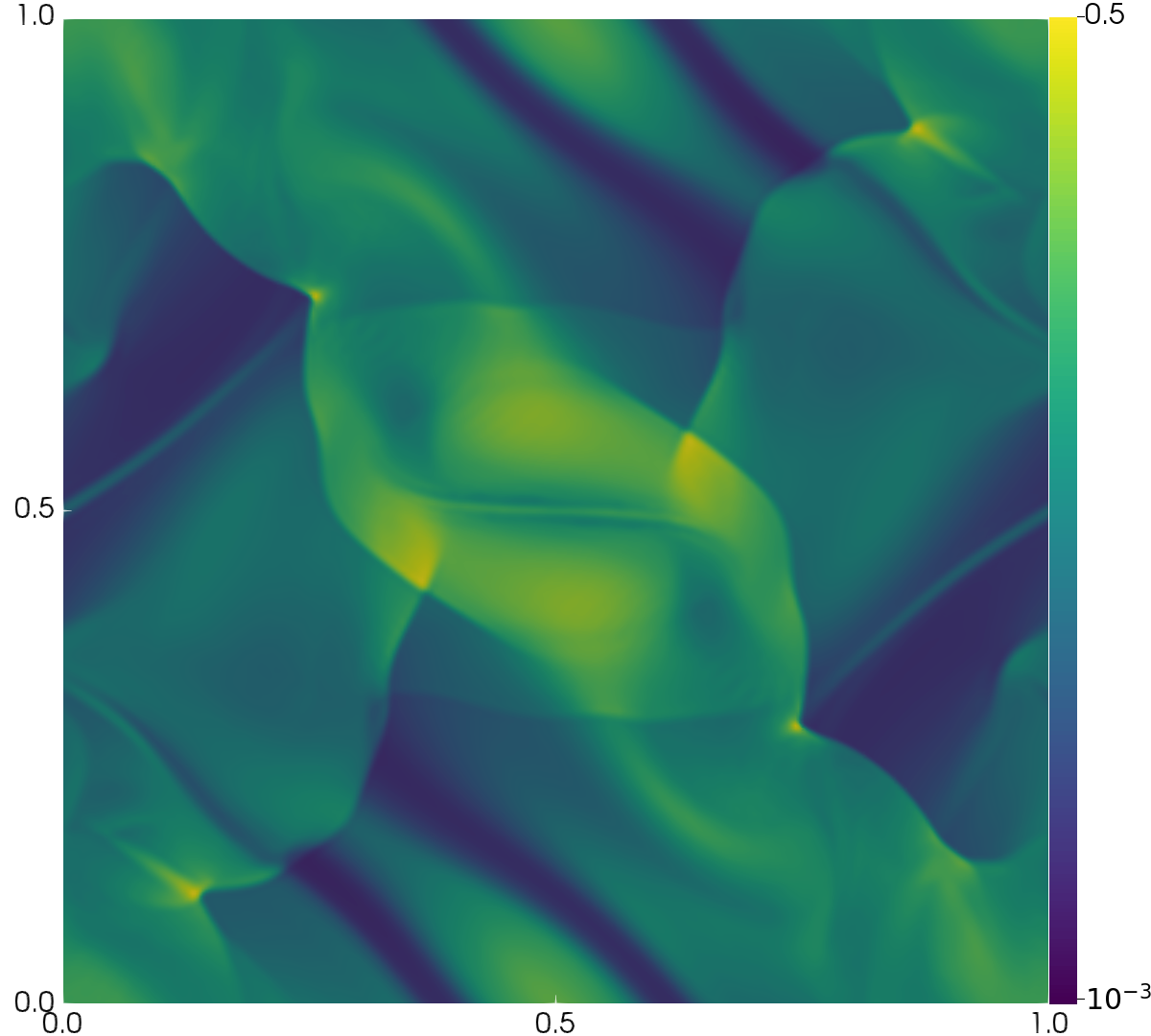}
    \\
    FD-6
  \end{minipage}
  \begin{minipage}{0.42\columnwidth}
    \centering
    \includegraphics[width=1.0\textwidth]{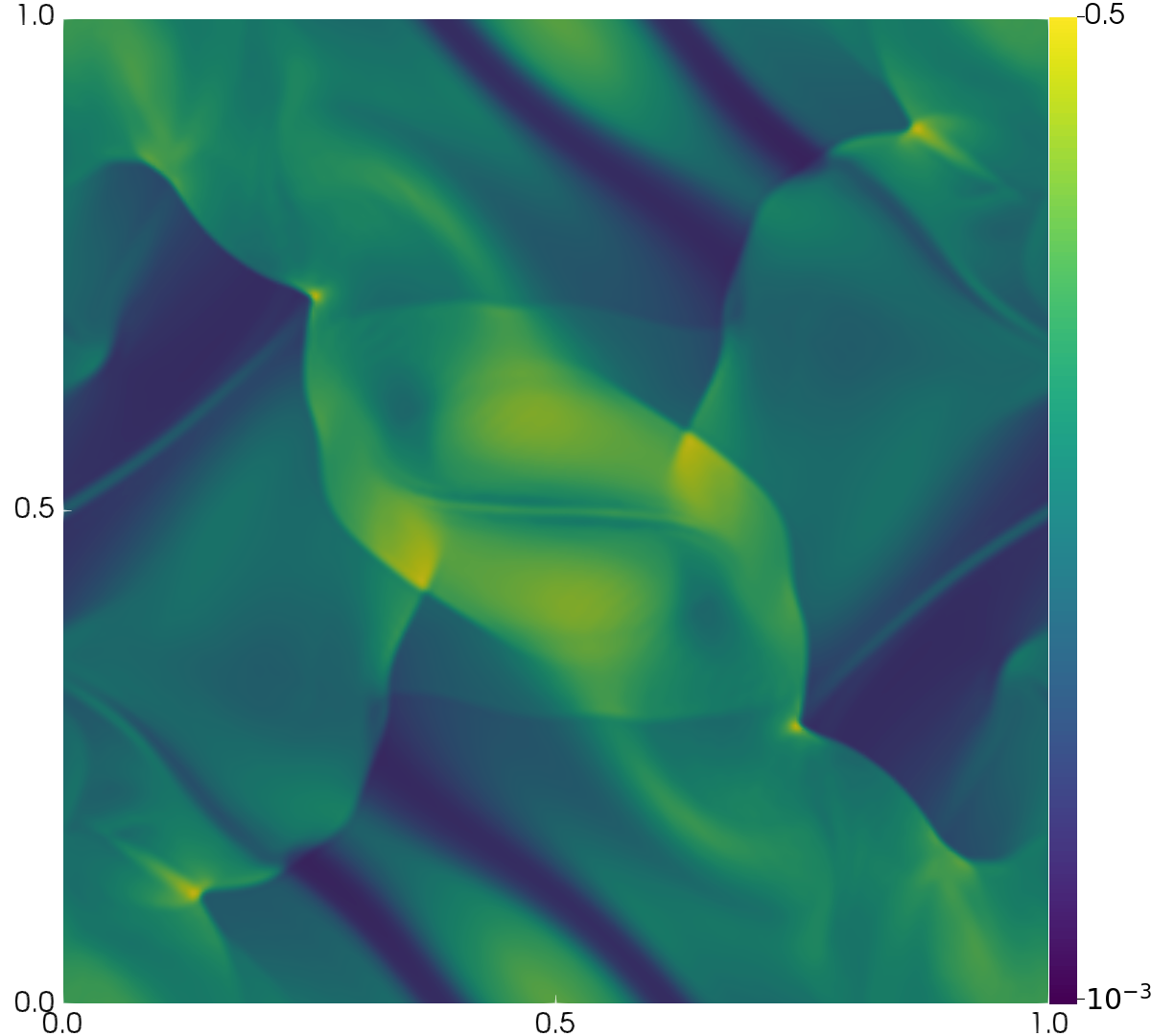}
    \\
    FD-8
  \end{minipage}

  \begin{minipage}{0.42\columnwidth}
    \centering
    \includegraphics[width=1.0\textwidth]{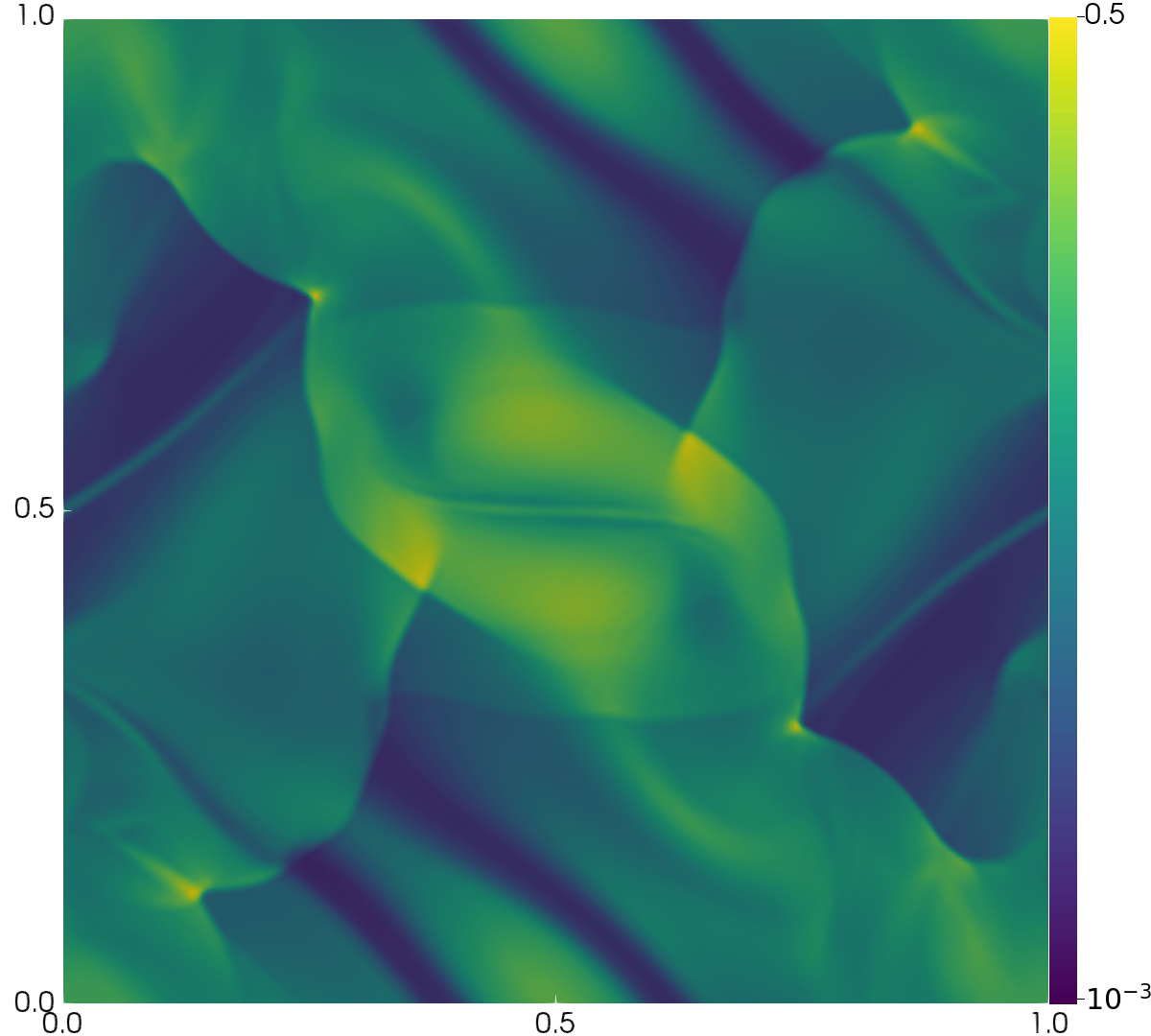}
    \\
    FD-10
  \end{minipage}
  \begin{minipage}{0.42\columnwidth}
    \centering
    \includegraphics[width=1.0\textwidth]{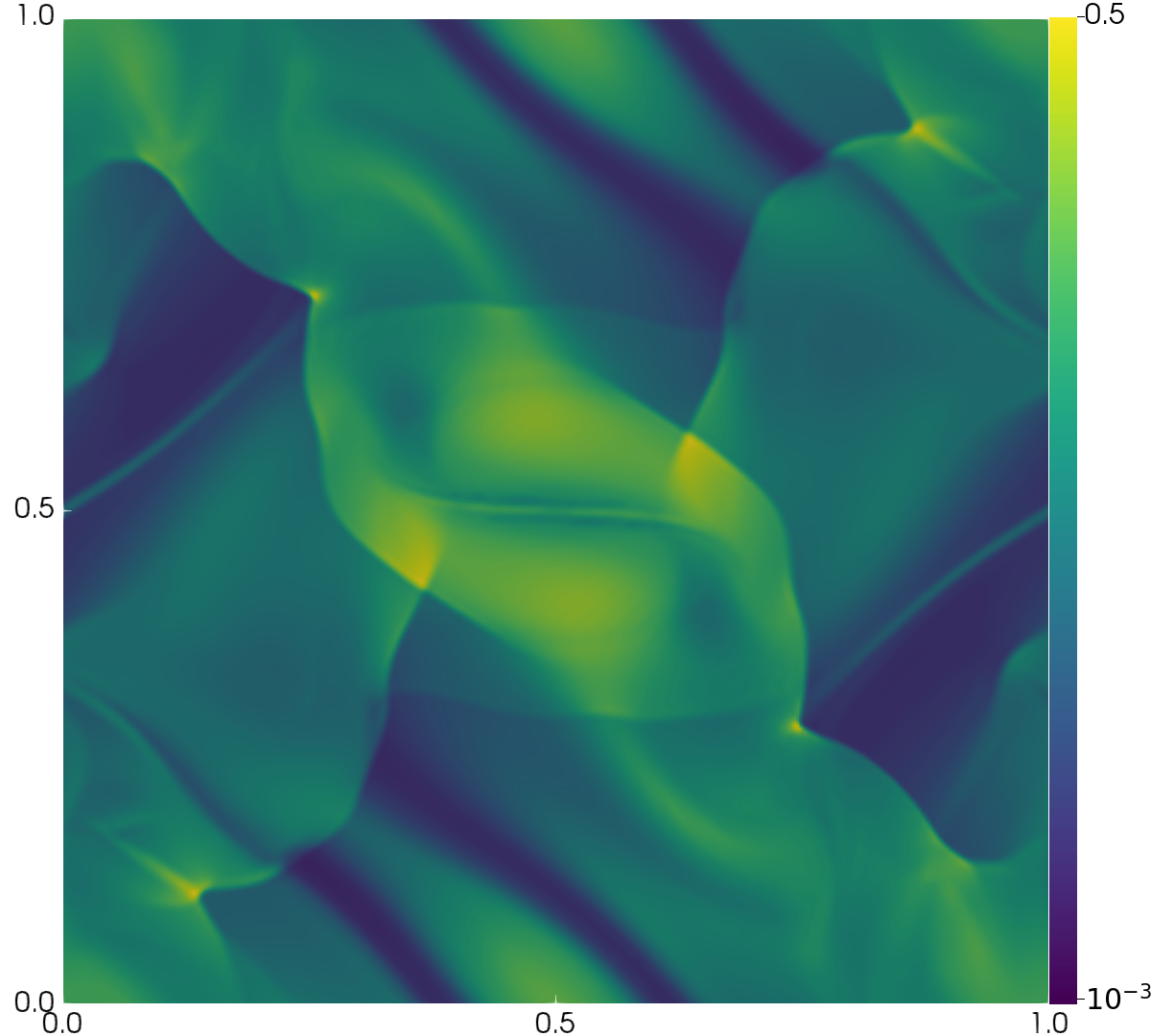}
    \\
    FD-10-6-2-2
  \end{minipage}
  \caption{Orszag-Tang vortex $\rho$ at $t=1$ showing the results using
    different FD derivative orders and always using the PPAO9-5-2-1
    reconstruction method. In all cases the scheme perform equally well for
    resolving the rest mass density.
    \label{fig:OrszagTangVortex}}
\end{figure}

\begin{figure}[h]
  \raggedleft
  \begin{minipage}{0.42\columnwidth}
    \centering
    \includegraphics[width=1.0\textwidth]{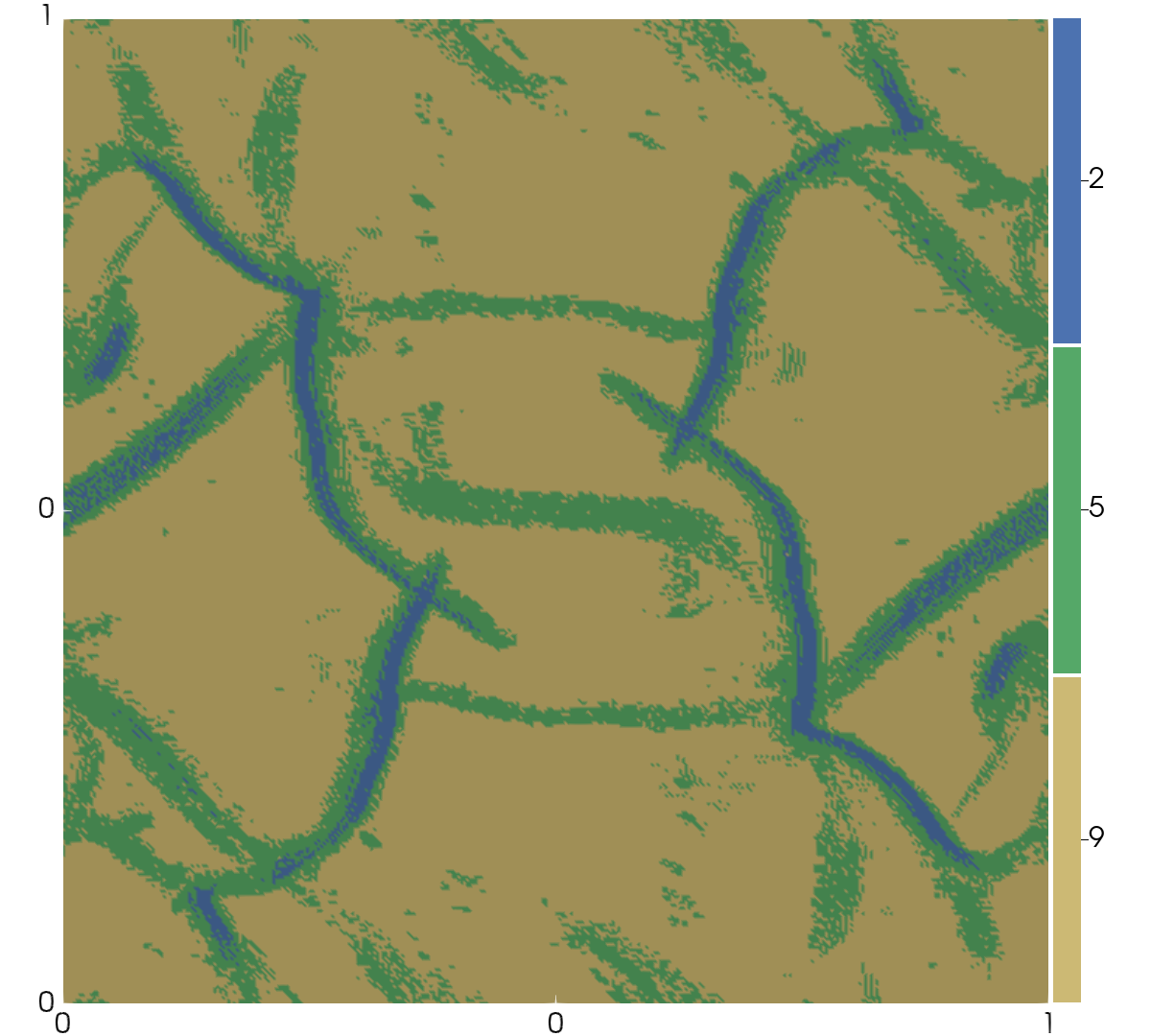}
    \\
    Reconstruction Order in $x$
  \end{minipage}
  \begin{minipage}{0.42\columnwidth}
    \centering
    \includegraphics[width=1.0\textwidth]{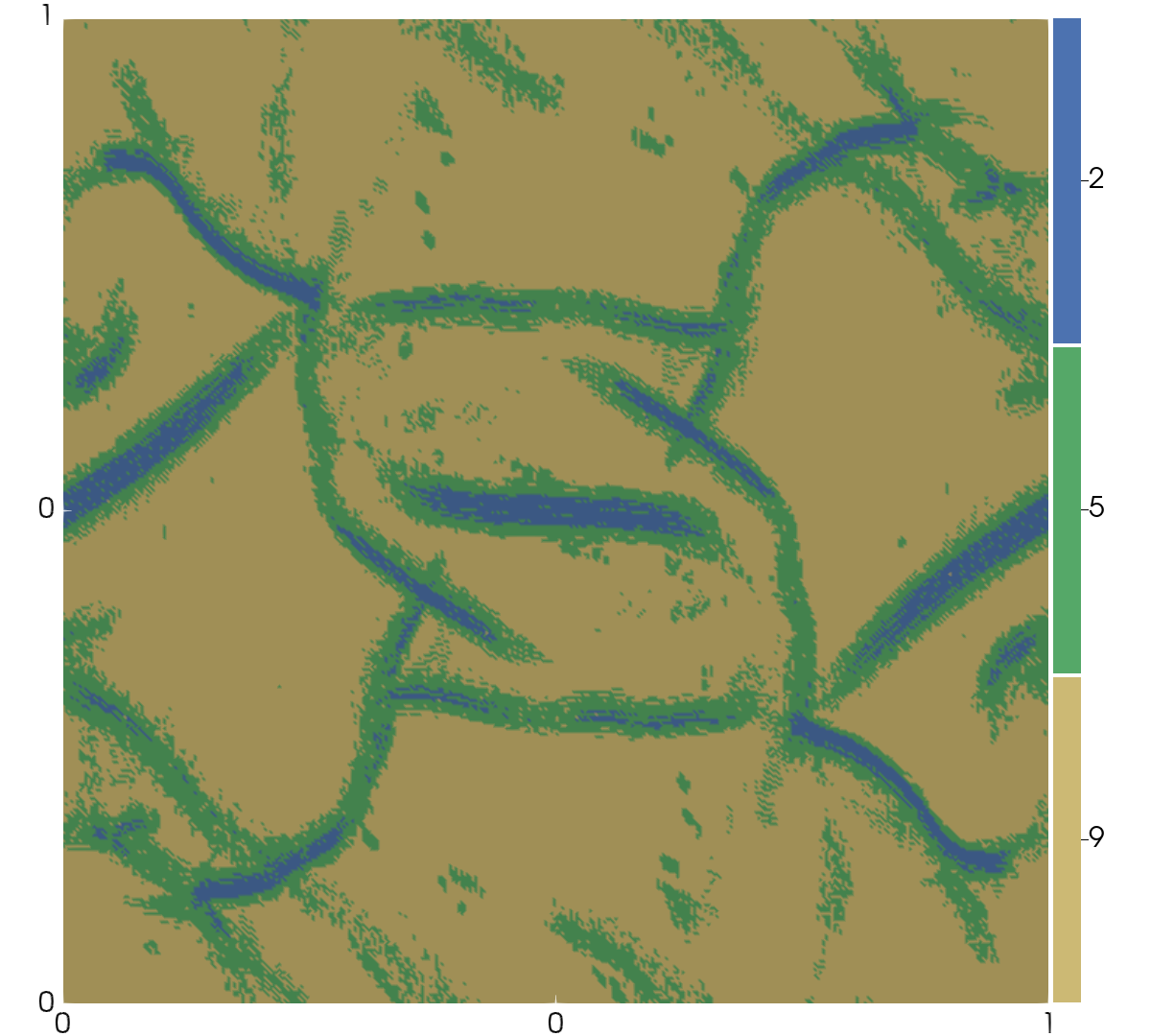}
    \\
    Reconstruction Order in $y$
  \end{minipage}

    \begin{minipage}{0.42\columnwidth}
    \centering
    \includegraphics[width=1.0\textwidth]{./OrszagTangVortexDOneHigherThanRecons.png}
    \\
    $\rho$
  \end{minipage}
  \begin{minipage}{0.42\columnwidth}
    \centering
    \includegraphics[width=1.0\textwidth]{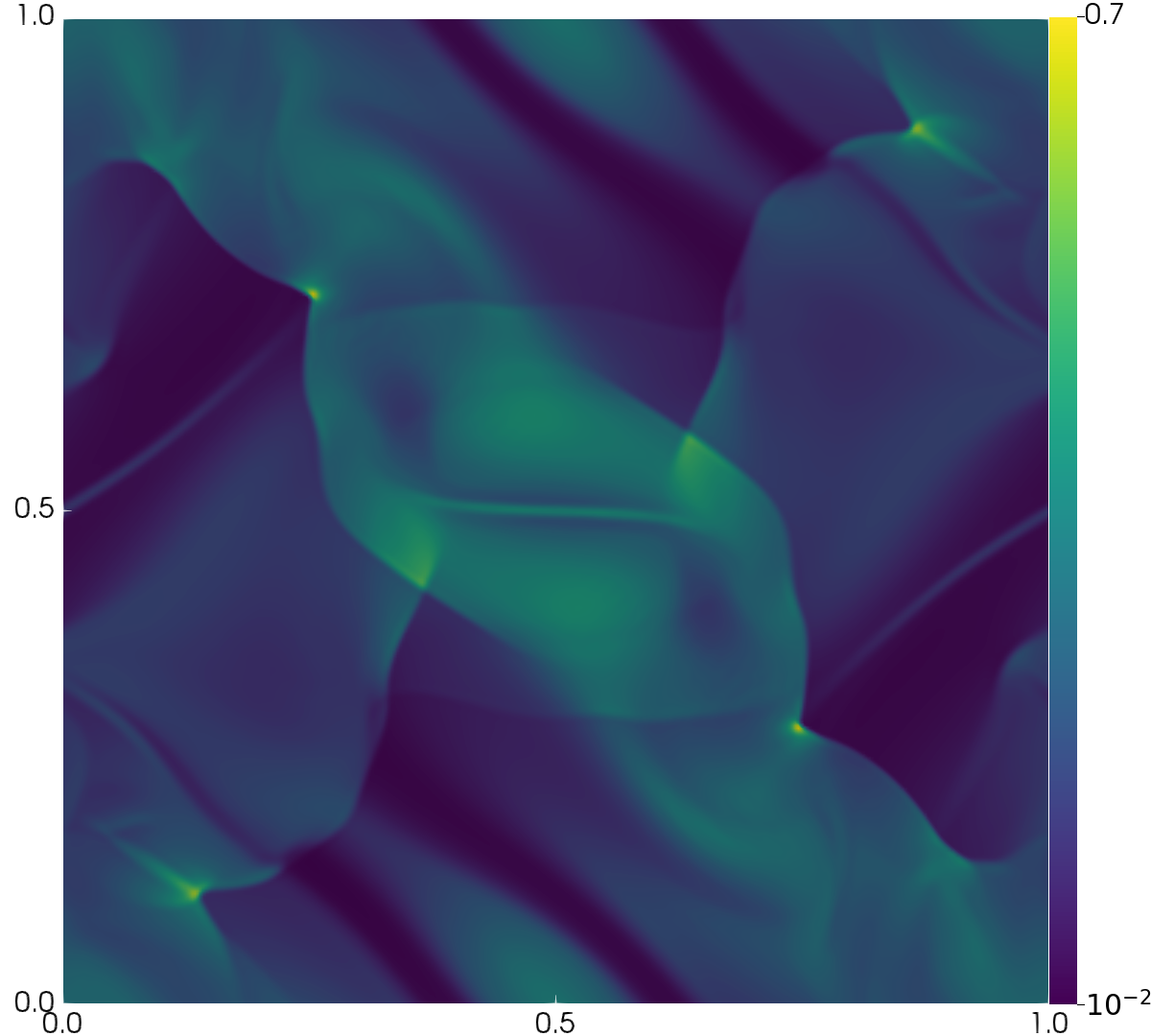}
    \\
    $p$
  \end{minipage}
  \caption{Results from the Orszag-Tang vortex. The panels in
    the top row show the reconstruction and FD derivative order used in the
    $x$-direction (left) and $y$-direction (right) at the final time, while the
    bottom left panel shows the rest mass density and the bottom right the
    pressure at the final time. We see that the adaptive-order FD scheme
    accurately tracks non-smooth features in the solution, specifically the rest
    mass density and pressure, adjusting the order as necessary.
  \label{fig:OrszagTangReconsOrder}}
\end{figure}

We plot the rest mass density $\rho$ at the final time $t_f=1$ in
figure~\ref{fig:OrszagTangVortex} for simulations using the PPAO9-5-2-1
reconstruction scheme but using different order FD derivatives. All schemes
perform equally well with no discernible differences between them.
We show the reconstruction order alongside $\rho$ and pressure $p$ in
figure~\ref{fig:OrszagTangReconsOrder}, nicely demonstrating that the PPAO
scheme accurately tracks non-smooth features in both $\rho$ and $p$ for this
problem as well.

\subsection{Slab jet}\label{sec:slab jet}

To study our scheme's ability to handle explosions akin to what one would
encounter in core-collapse supernova simulations, we run a slab jet simulation
similar to that of~\cite{Komissarov1999}. We use the domain
$[-0.5,20]\times[-14,14]\times[-0.5,0.5]$ with resolution
$176\times352\times11$. We impose periodic boundary conditions in the
$z$-direction and use Dirichlet boundary conditions fixed to the initial
conditions in $x$ and $y$. The initial conditions are given by
\begin{eqnarray}
  \rho&=\left\{\begin{array}{ll}
    10, & \left|y\right| < 0.5\;\mathrm{and}\;x\le0\\
    0.1, & \mathrm{otherwise},
  \end{array}\right. \\
  p &= 0.01, \\
  u^i &= \left\{\begin{array}{ll}
    (20, 0, 0), & \left|y\right| < 0.5\;\mathrm{and}\;x\le0\\
    (0,0,0, & \mathrm{otherwise},
  \end{array}\right. \\
  B^i &= \left(1, 0, 0\right),
\end{eqnarray}
and an ideal fluid equation of state is used with $\Gamma=4/3$. We evolve to a
final time of $t_f=27$ with a CFL factor of 0.5. We find that for larger CFLs
the FD-4, FD-6, FD-8, and FD-10 algorithms become unstable.

\begin{figure}[h]
  \centering
  \begin{minipage}{0.33\columnwidth}
    \centering
    \includegraphics[width=1.0\textwidth]{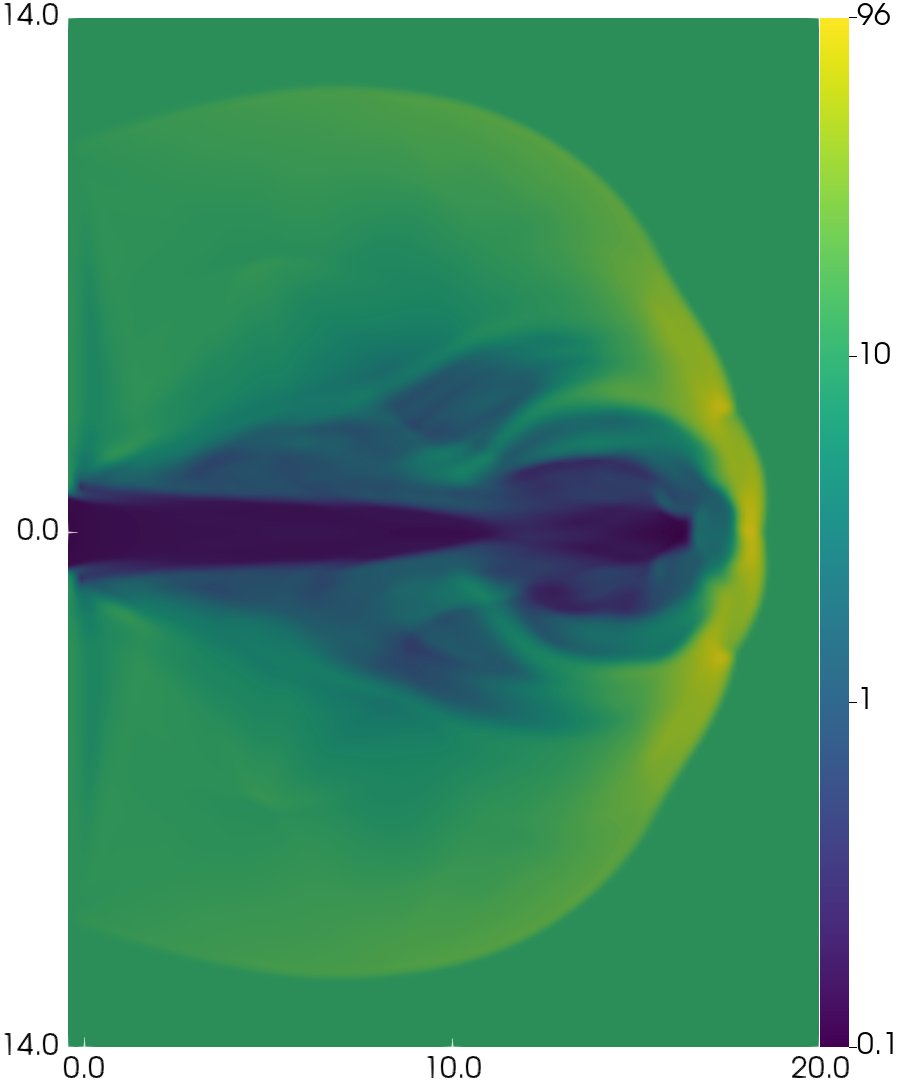}
    \\
    FD-2
  \end{minipage}
  \begin{minipage}{0.33\columnwidth}
    \centering
    \includegraphics[width=1.0\textwidth]{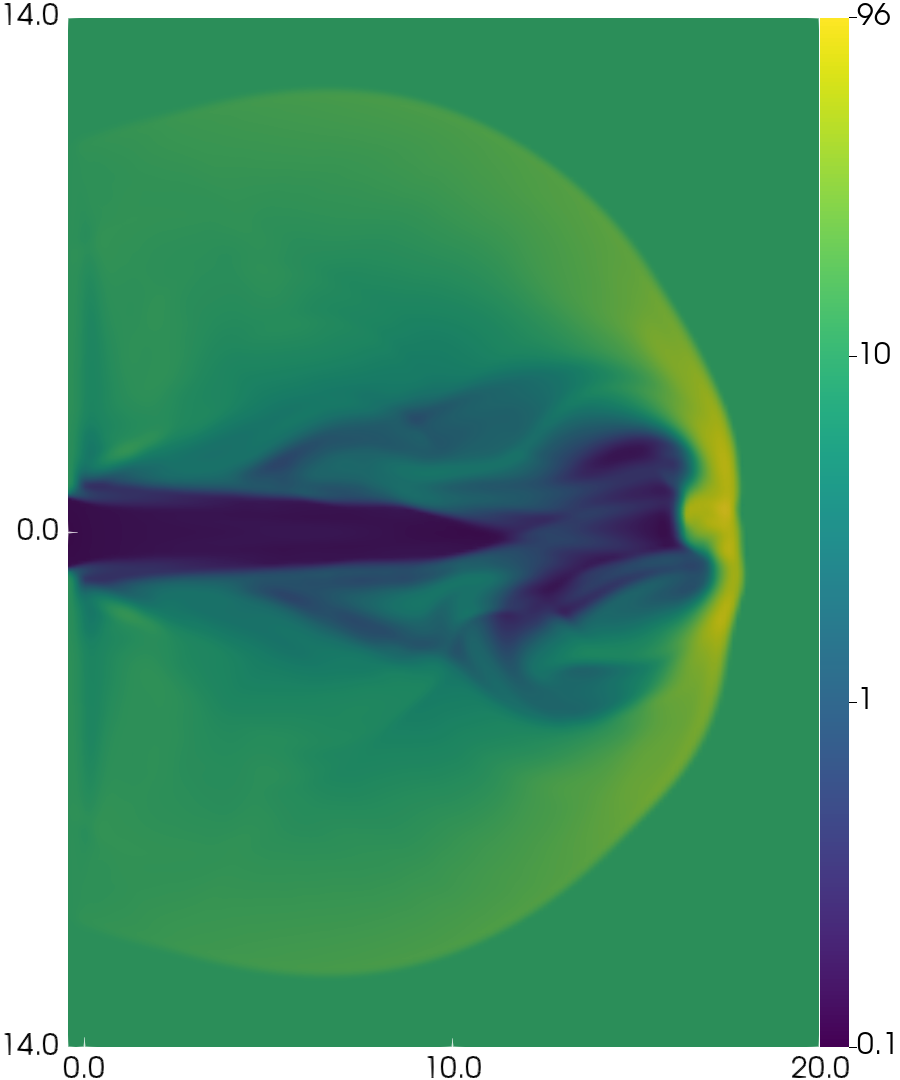}
    \\
    FD-4
  \end{minipage}

  \begin{minipage}{0.33\columnwidth}
    \centering
    \includegraphics[width=1.0\textwidth]{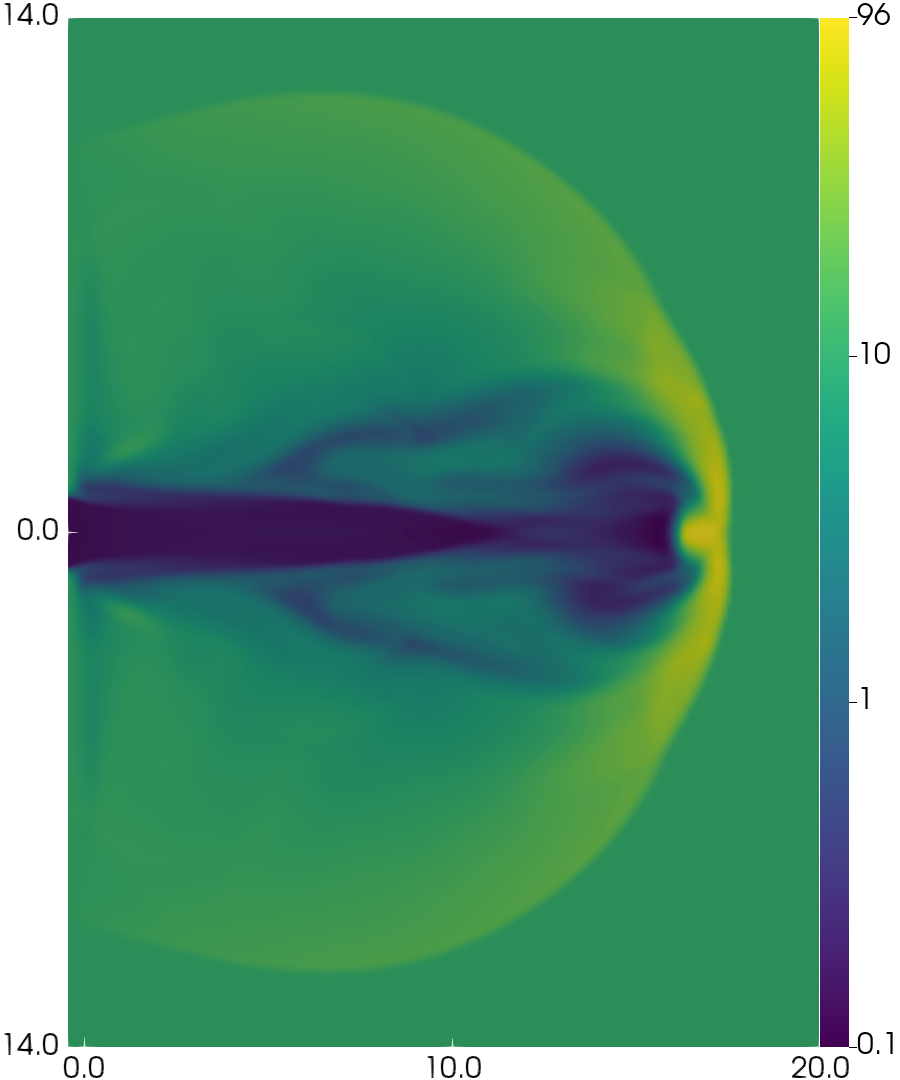}
    \\
    FD-6
  \end{minipage}
  \begin{minipage}{0.33\columnwidth}
    \centering
    \includegraphics[width=1.0\textwidth]{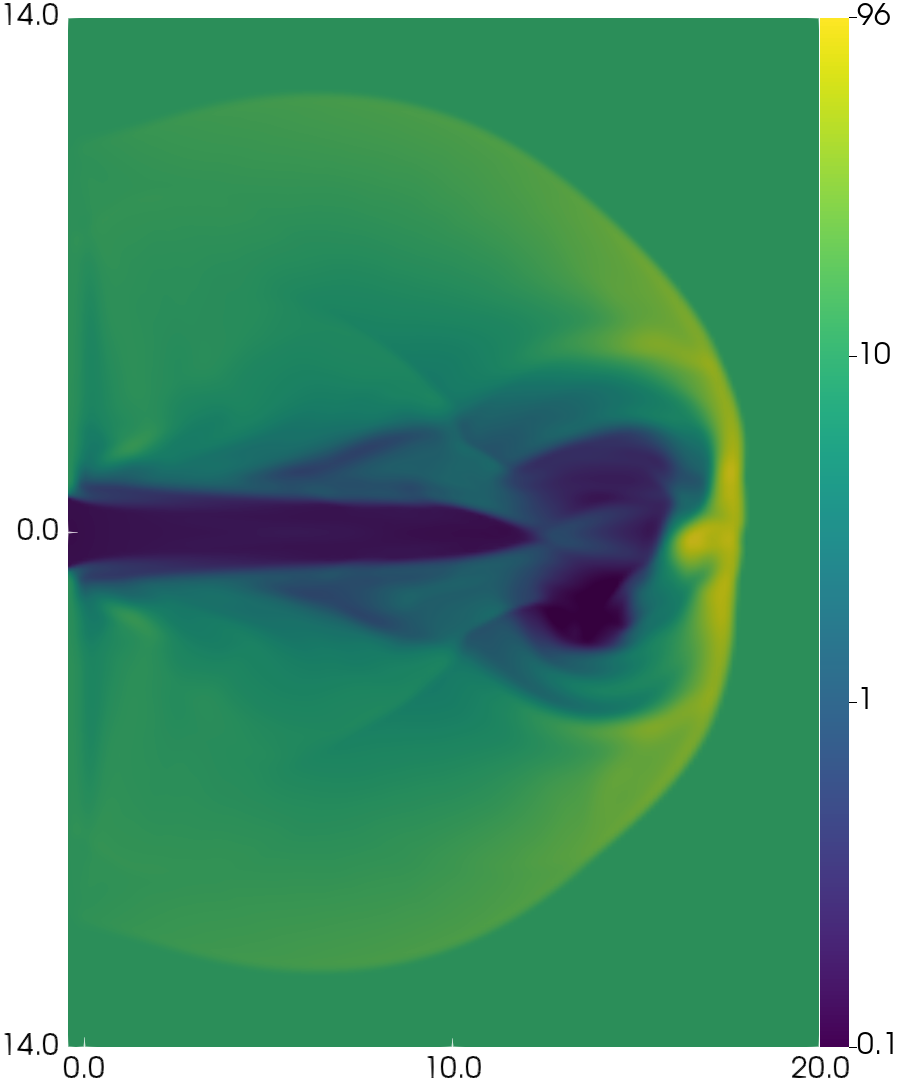}
    \\
    FD-8
  \end{minipage}

  \begin{minipage}{0.33\columnwidth}
    \centering
    \includegraphics[width=1.0\textwidth]{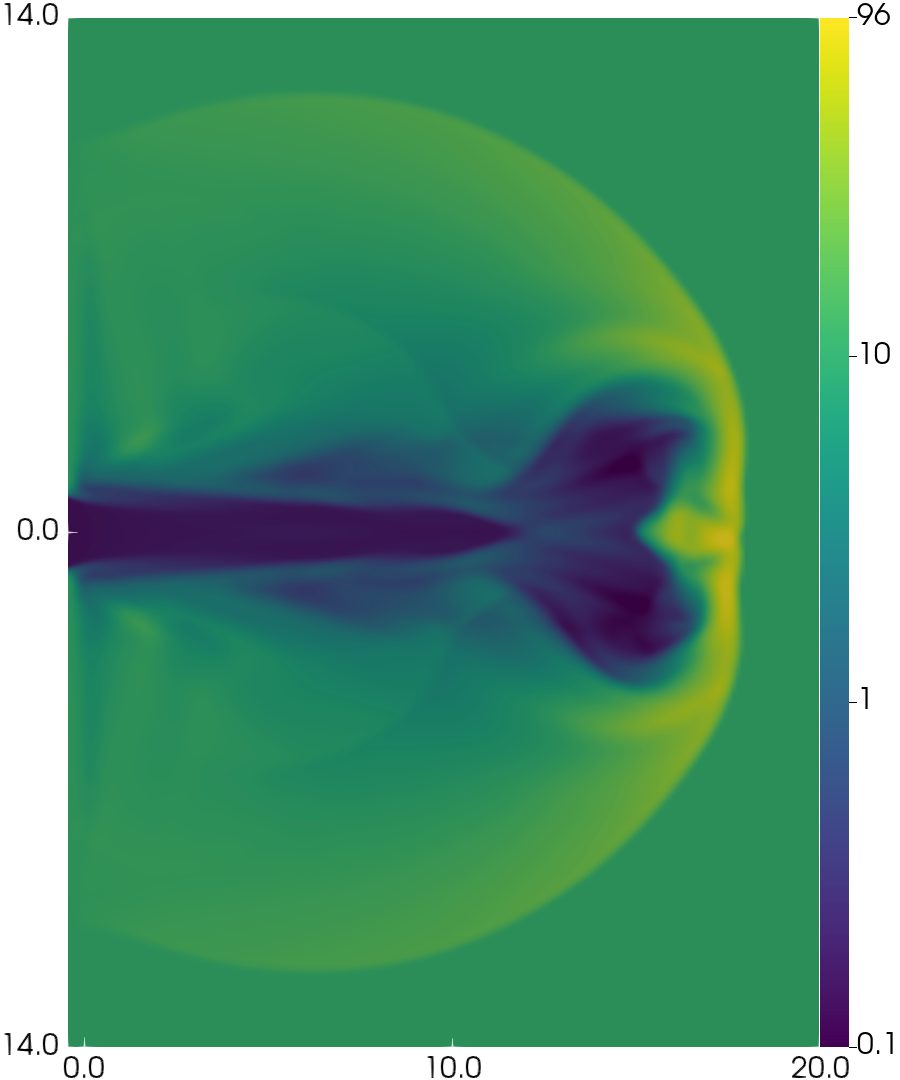}
    \\
    FD-10
  \end{minipage}
  \begin{minipage}{0.33\columnwidth}
    \centering
    \includegraphics[width=1.0\textwidth]{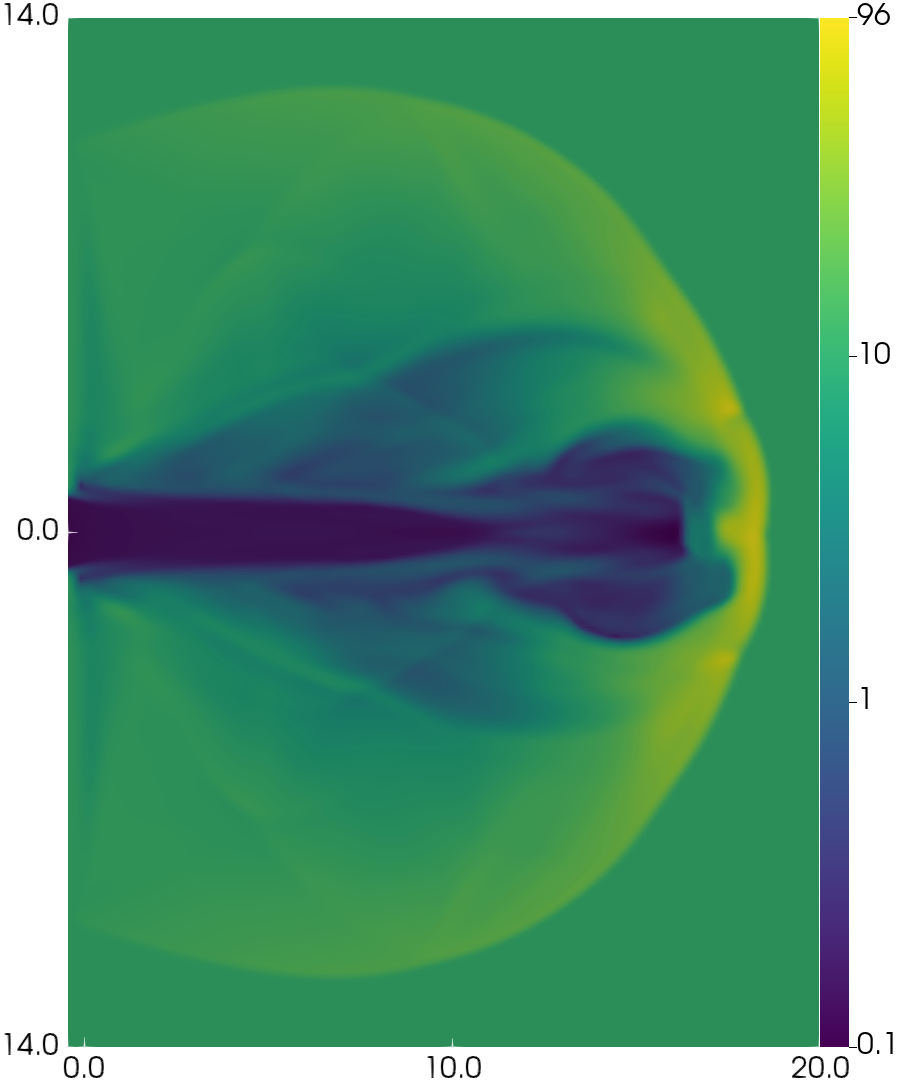}
    \\
    FD-10-6-2-2
  \end{minipage}
  \caption{Slab jet simulation $\rho$ at $t=27$ showing the results using
    different FD derivative orders and always using the PPAO9-5-2-1
    reconstruction method. We see significant symmetry breaking with FD-4 and
    FD-8, while all other schemes respect the symmetry quite well. Importantly,
    the FD-10-6-2-2 scheme is as robust as the FD-2 scheme while being very
    accurate.
  \label{fig:SlabJet}}
\end{figure}

We plot the rest mass density $\rho$ at the final time $t_f=27$ in
figure~\ref{fig:SlabJet} for simulations using the PPAO9-5-2-1 reconstruction
scheme with various FD derivative orders. The FD-4 and FD-8 schemes produce
significant deviations from symmetry about the $y=0$ plane, while all other
scheme preserve the symmetry quite well. In particular, we find that the FD
derivative order adaptation (FD-10-6-2-2) is as robust as the FD-2 scheme, and
can be run with larger CFL factors than the non-adaptive high-order derivatives.

\subsection{TOV star}\label{sec:TOV star}

To test the proposed method's ability to simulate fluids in curved spacetimes, we
evolve a Tolman-Oppenheimer-Volkoff (TOV) star~\cite{Tolman:1939jz,
  Oppenheimer:1939ne}. We perform simulations of both
magnetized and non-magnetized TOV stars. We adopt the same configuration as
in~\cite{Cipolletta:2019geh,Deppe:2021bhi,Deppe:2021ada}. Specifically, we use a
polytropic equation of state,
\begin{equation}
  \label{eq:polytropic EOS}
  p(\rho)=K \rho^\Gamma
\end{equation}
with $\Gamma=2$, $K=100$, and a central density
$\rho_c=1.28\times10^{-3}M_{\odot}^{-2}$. This choice of $K$ and $\Gamma$ mean
we use geometric units where $G=c=M_{\odot}=1$. For the magnetized case, we
choose a magnetic field given by a vector potential
\begin{equation}
  \label{eq:TOV vector potential}
  A_\phi=A_b\left(x^2+y^2\right)\max\left(p-p_{\mathrm{cut}},0\right)^{n_s},
\end{equation}
with $A_b=2500$, $p_{\mathrm{cut}}=0.04p_{\max}$, and $n_s=2$. This
configuration yields a magnetic field strength in CGS units of
$|B_{\mathrm{CGS}}|=1.03\times10^{16}\,\mathrm{G}$. The magnetic field is only a
perturbation to the dynamics of the star, since
$(p_{\mathrm{mag}}/p)(r=0)\sim5\times10^{-5}$. The magnetic field is given by
\begin{eqnarray}
  \label{eq:TOV magnetic field}
  B^x&=-\frac{1}{\sqrt{\gamma}}\frac{xz}{r}
       A_bn_s(p-p_{\mathrm{cut}})^{n_s-1}\partial_rp, \\
  B^y&=-\frac{1}{\sqrt{\gamma}}\frac{yz}{r}
       A_bn_s(p-p_{\mathrm{cut}})^{n_s-1}\partial_rp, \\
  B^z&=\frac{A_b}{\sqrt{\gamma}}\left[
       2(p-p_{\mathrm{cut}})^{n_s}
       +\frac{x^2+y^2}{r}
       n_s(p-p_{\mathrm{cut}})^{n_s-1}\partial_r p
       \right],
\end{eqnarray}
away from $r=0$ and by
\begin{eqnarray}
  \label{eq:TOV magnetic field origin}
  B^x&=0, \\
  B^y&=0, \\
  B^z&=\frac{A_b}{\sqrt{\gamma}}
       2(p-p_{\mathrm{cut}})^{n_s},
\end{eqnarray}
at $r=0$.

\begin{figure}[h]
  \raggedleft
  \begin{minipage}{0.45\columnwidth}
    \centering
    \includegraphics[width=1.0\textwidth]{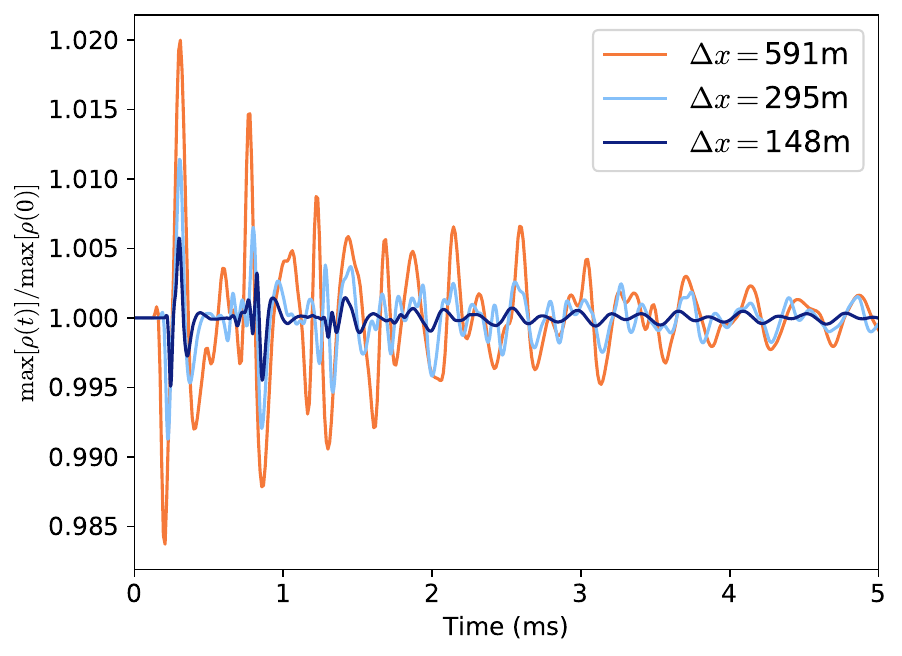}
    \\
    PPAO5-2-2 \& FD-4
  \end{minipage}
  \begin{minipage}{0.45\columnwidth}
    \centering
    \includegraphics[width=1.0\textwidth]{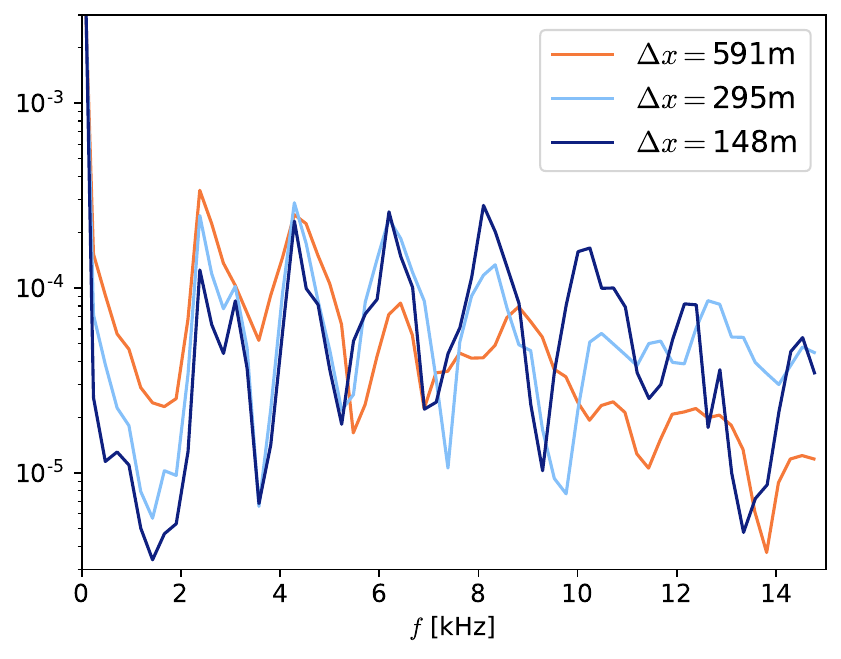}
    \\
    PPAO5-2-2 \& FD-4
  \end{minipage}

  \begin{minipage}{0.45\columnwidth}
    \centering
    \includegraphics[width=1.0\textwidth]{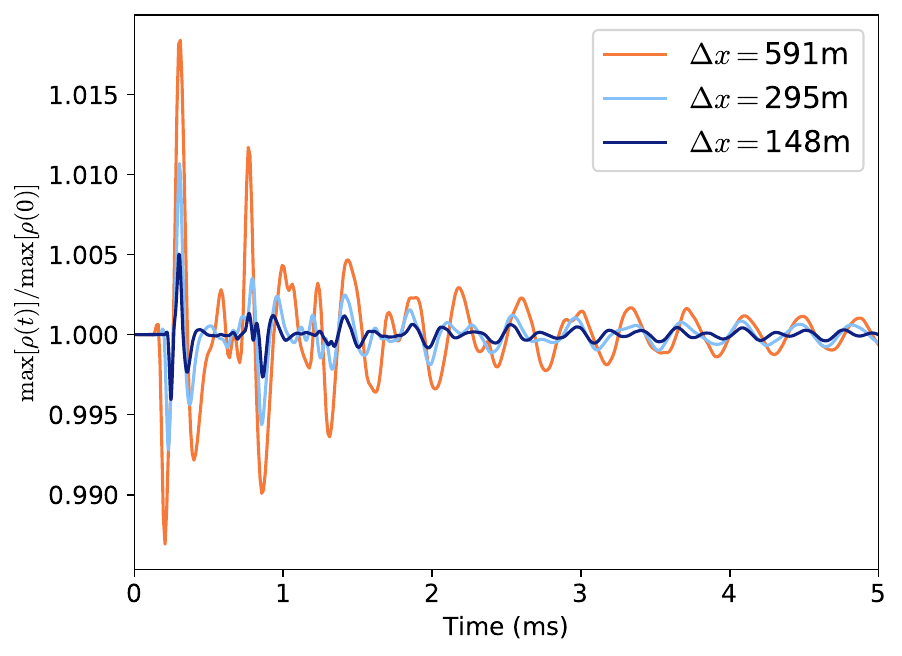}
    \\
    PPAO5-2-2 \& FD-6-2-2
  \end{minipage}
  \begin{minipage}{0.45\columnwidth}
    \centering
    \includegraphics[width=1.0\textwidth]{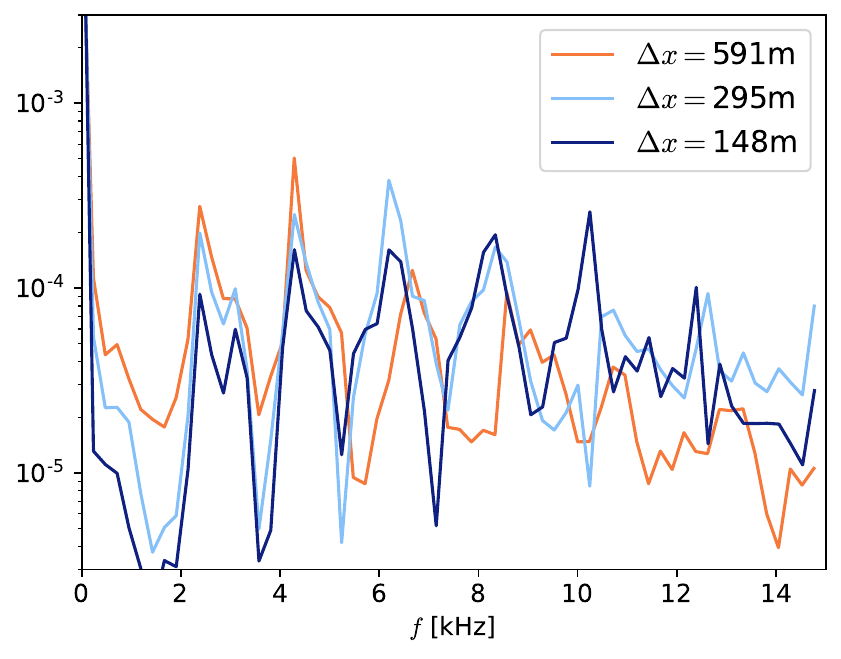}
    \\
    PPAO5-2-2 \& FD-6-2-2
  \end{minipage}

  \begin{minipage}{0.45\columnwidth}
    \centering
    \includegraphics[width=1.0\textwidth]{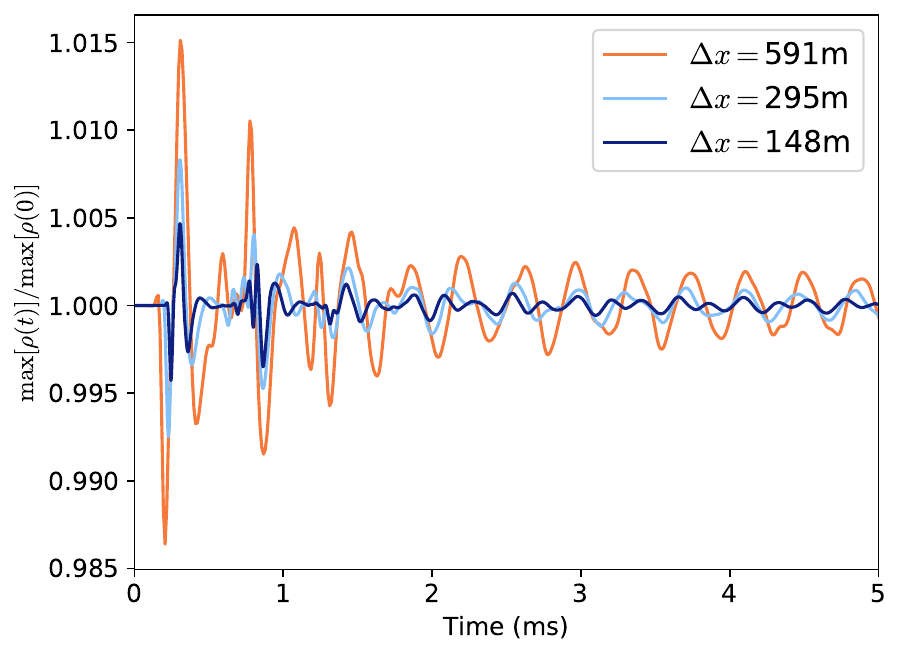}
    \\
    PPAO9-5-2-2 \& FD-10-6-2-2
  \end{minipage}
  \begin{minipage}{0.45\columnwidth}
    \centering
    \includegraphics[width=1.0\textwidth]{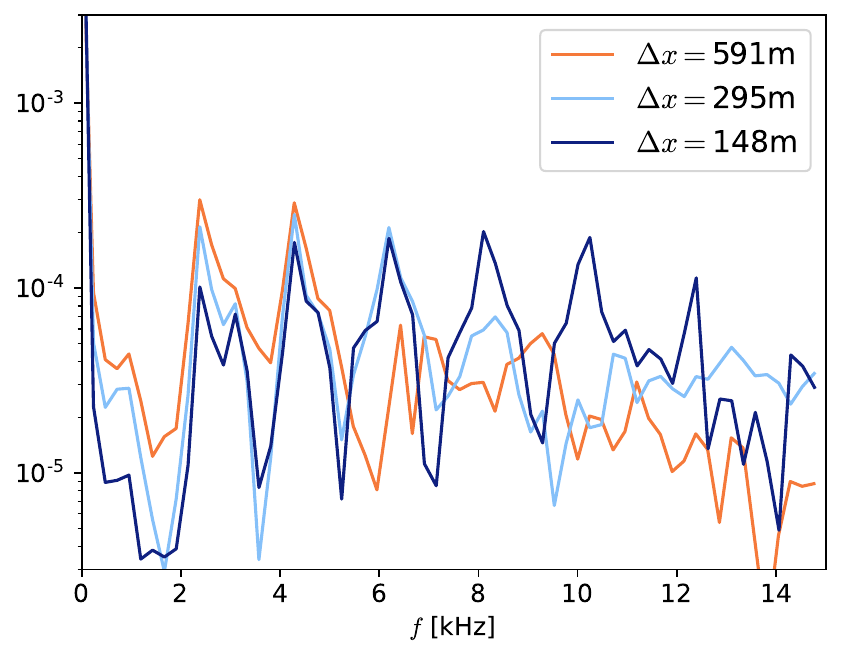}
    \\
    PPAO9-5-2-2 \& FD-10-6-2-2
  \end{minipage}
  \caption{Results from evolutions of a TOV star to a final time of 5ms using
    different PPAO reconstruction and FD order schemes. The left panels show the
    maximum rest mass density as a function of time divided by the maximum rest
    mass density at $t=0$, which oscillates about 1 with the amplitude decaying
    over time. The right panels show a power spectrum of the left panels with
    vertical dashed grey lines showing the analytically known radial oscillation
    frequencies~\cite{Font:1999wh, Stergioulas:1998id}. We see that the
    PPAO9-5-2-2 scheme does especially well at resolving higher frequency
    modes.
  \label{fig:TovStar}}
\end{figure}

\begin{figure}[h]
  \raggedleft
  \begin{minipage}{0.48\columnwidth}
    \centering
    \includegraphics[width=1.0\textwidth]{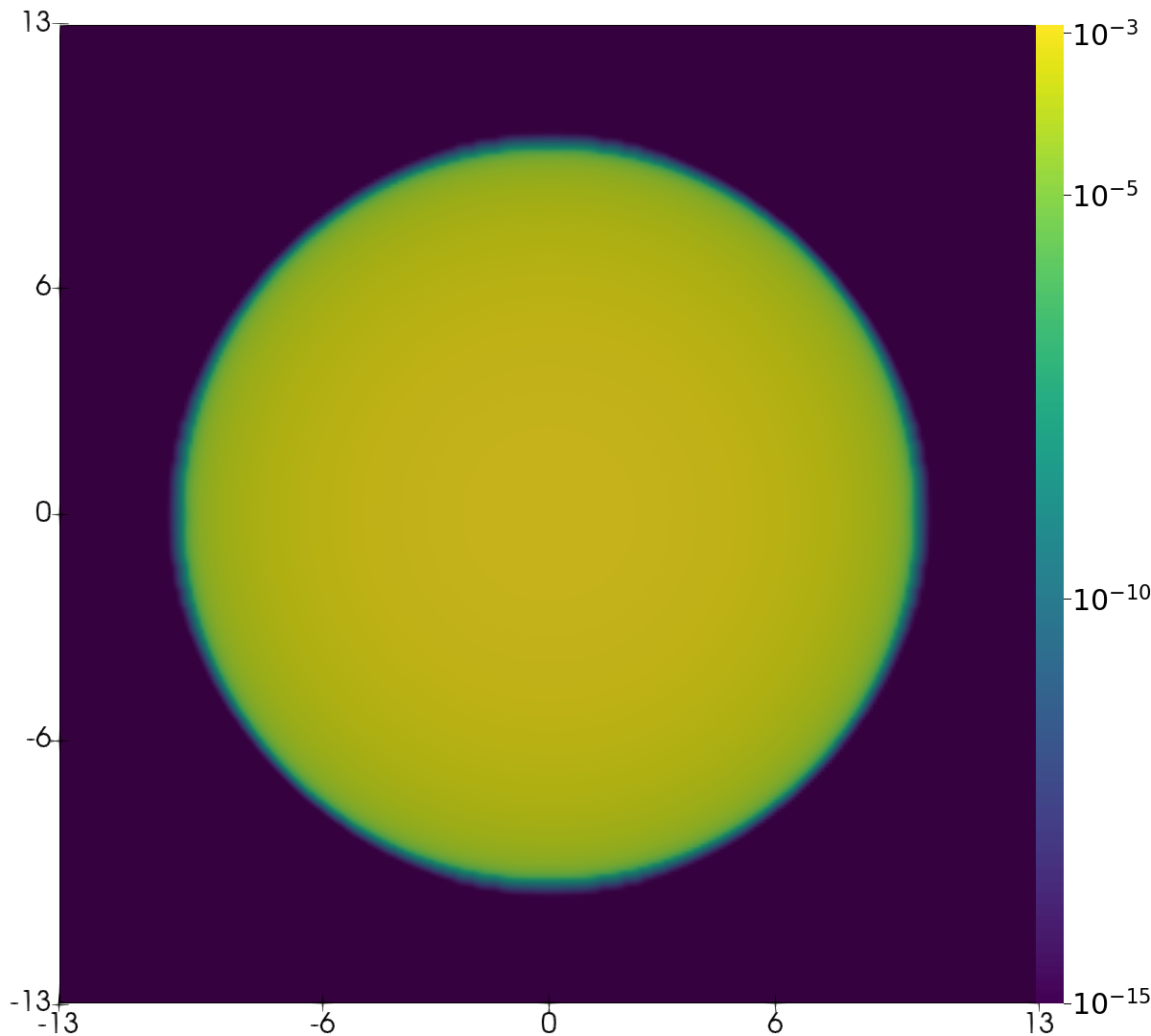}
    \\
    PPAO5-2-2 \& FD-4
  \end{minipage}
  \begin{minipage}{0.48\columnwidth}
    \centering
    \includegraphics[width=1.0\textwidth]{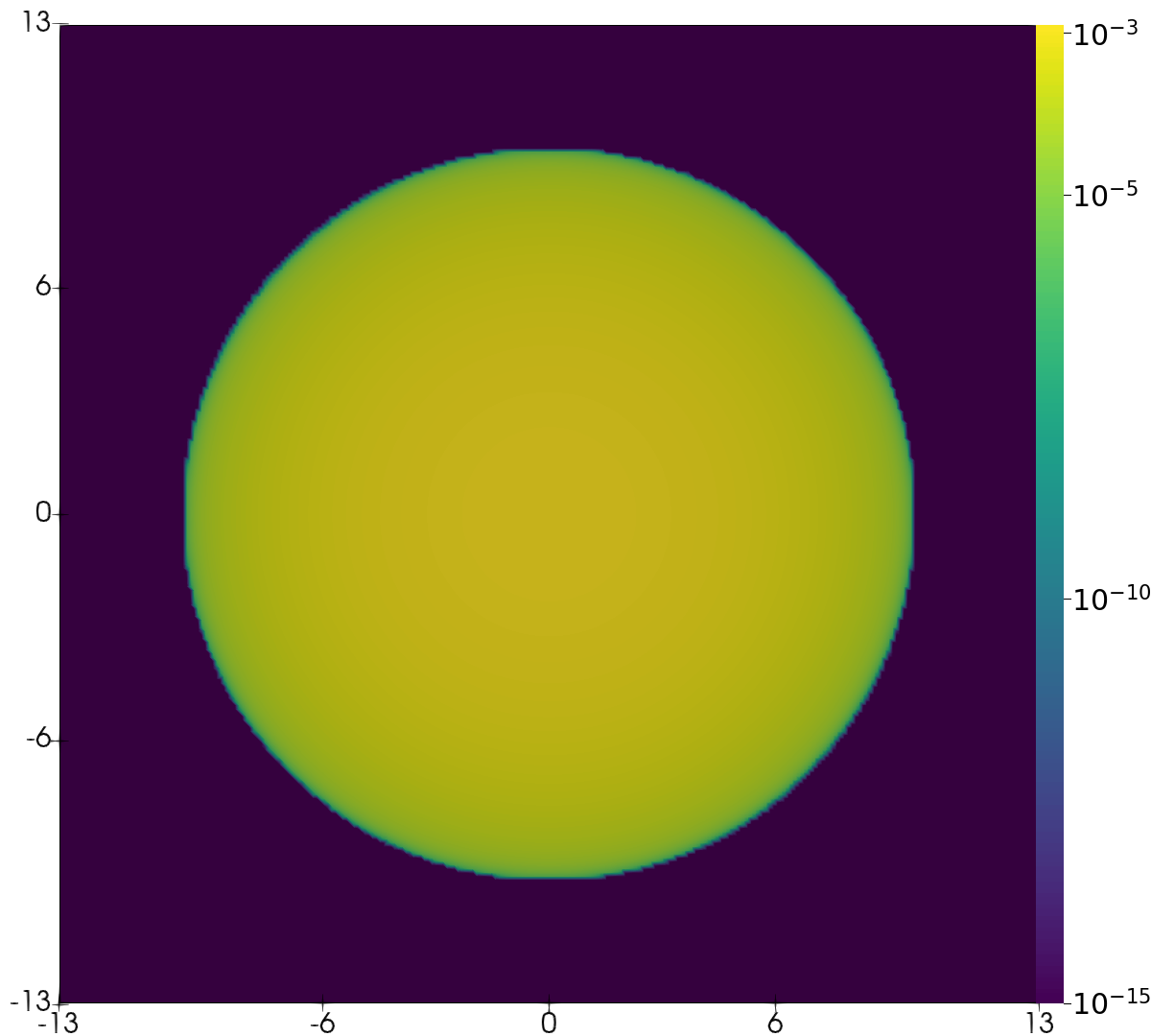}
    \\
    PPAO5-2-2 \& FD-4
  \end{minipage}

  \centering
  \includegraphics[width=0.48\textwidth]{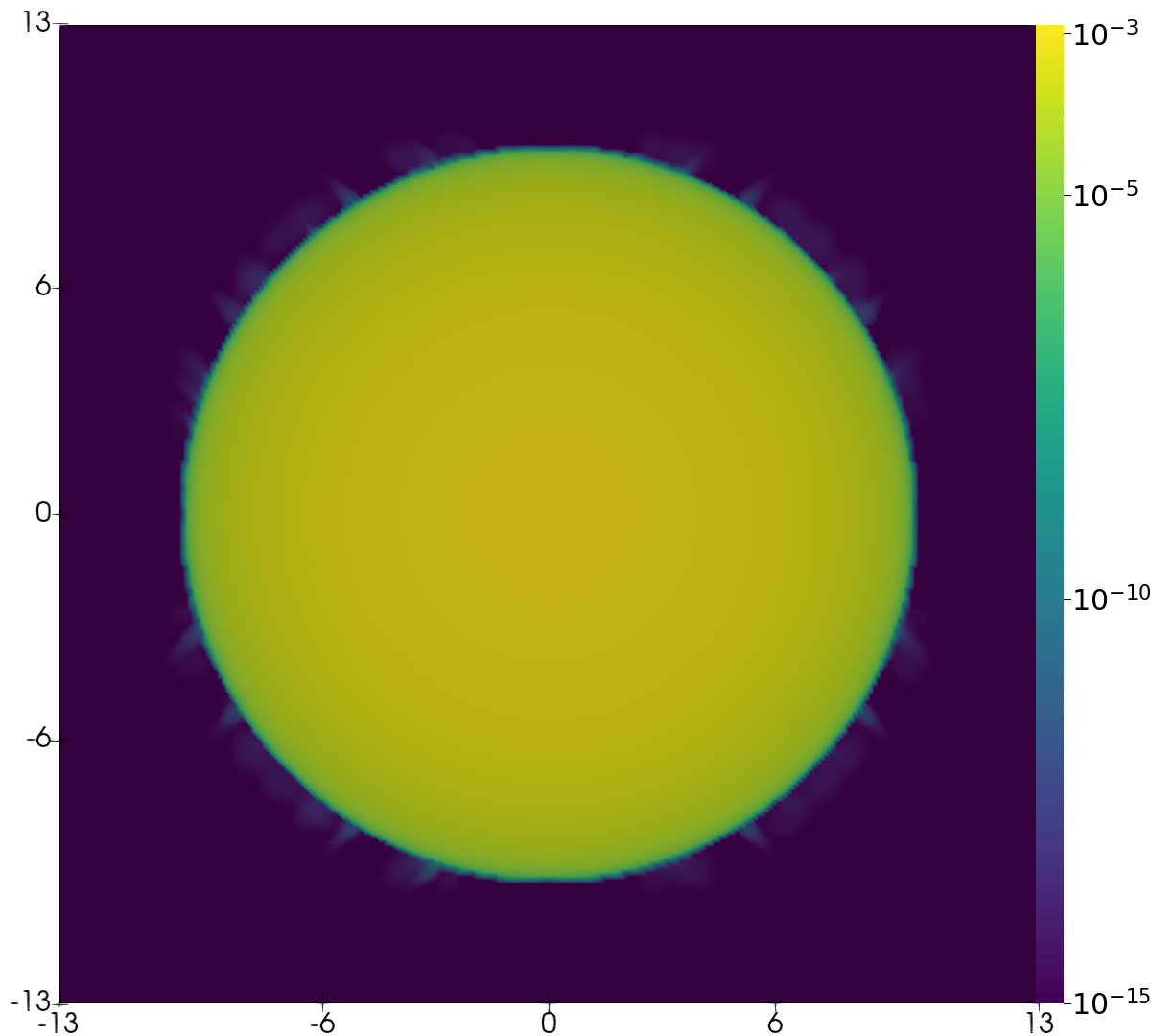}
  \\
  PPAO9-5-2-2 \& FD-10-6-2-2

  \caption{Visualization of the log of the rest mass density on the $z=-0.1$
    plane at $t=4.7$ms. The PPAO5-2-2+FD-4 scheme smears the surface more
    that the PPAO5-2-2+FD-6-2-2 scheme, while the PPAO9-5-2-2+FD-10-6-2-2 scheme
    produces spurious artifacts just outside the star. Dropping from ninth to
    fifth order earlier might remove the spurious artifacts, but we have not
    tested this. Ultimately, the adaptive order scheme is able to produce crisp
    surfaces while achieving high order in smooth regions.
  \label{fig:TovVolume}}
\end{figure}

All simulations are done in full 3d with no symmetry assumptions, but in the
Cowling approximation, i.e.~the spacetime is static. We use a cube $[-13,13]^3$
in geometric units, which is slightly higher resolution
than was used in~\cite{Cipolletta:2019geh,Deppe:2021bhi,Deppe:2021ada}. We
convert grid spacing to meters assuming a maximum mass of the neutron star of
$M_{\max}=2M_{\odot}$. We run simulations to a final time of $t_f=5\mathrm{ms}$
and monitor the maximum density over the domain as a function of time. In
figure~\ref{fig:TovStar} we plot the relative change of the maximum rest mass
density at three resolutions in the left panels and its power spectrum in the
right panels. We compare the oscillation frequencies to the known
frequencies~\cite{Font:1999wh, Stergioulas:1998id} and find good agreement in
all cases. In particular, the PPAO9-5-2-2+FD-10-6-2-2 scheme very nicely
resolves the seven frequencies plotted in figure~\ref{fig:TovStar}. The
PPAO5-2-2+FD-6-2-2 scheme starts to lose accuracy around the sixth peak, while
PPAO5-2-2+FD-4 resolves the first six frequencies and arguably the seventh, but
not as cleanly as the PPAO9-5-2-2+FD-10-6-2-2 scheme. Overall, our PPAO scheme
works very well for TOV simulations. We do not plot the magnetized case since
the behavior is effectively identical to the non-magnetized case, so much so
that the plots are indistinguishable by eye. In figure~\ref{fig:TovVolume} we
show volume renderings of the $z=-0.1$ plane for the three different methods
at the highest resolution. The PPAO5-2-2+FD-4 scheme smears the surface of the
star out more than the two adaptive schemes, while PPAO9-5-2-2+FD-10-6-2-2
has spurious artifacts just outside the star. Given that these artifacts are
absent in the PPAO5-2-2+FD-6-2-2 case, it might be possible to more aggressively
drop from ninth to fifth order to remove them, but we have not tested
this. Ultimately, this leads us to conclude that the adaptive scheme is able to
produce crisp surfaces while achieving high order in smooth regions.

\subsection{Rotating neutron star}\label{sec:rotating neutron star}

\begin{figure}[h]
  \raggedleft
  \begin{minipage}{0.45\columnwidth}
    \centering
    \includegraphics[width=1.0\textwidth]{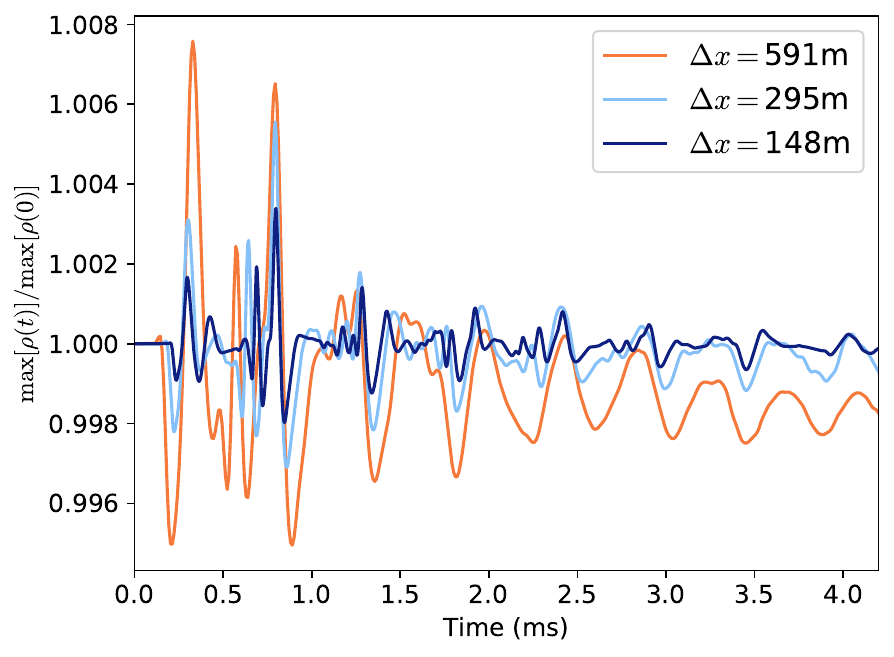}
    \\
    PPAO5-2-2 \& FD-4
  \end{minipage}
  \begin{minipage}{0.45\columnwidth}
    \centering
    \includegraphics[width=1.0\textwidth]{./SpectrumFd4.pdf}
    \\
    PPAO5-2-2 \& FD-4
  \end{minipage}

  \begin{minipage}{0.45\columnwidth}
    \centering
    \includegraphics[width=1.0\textwidth]{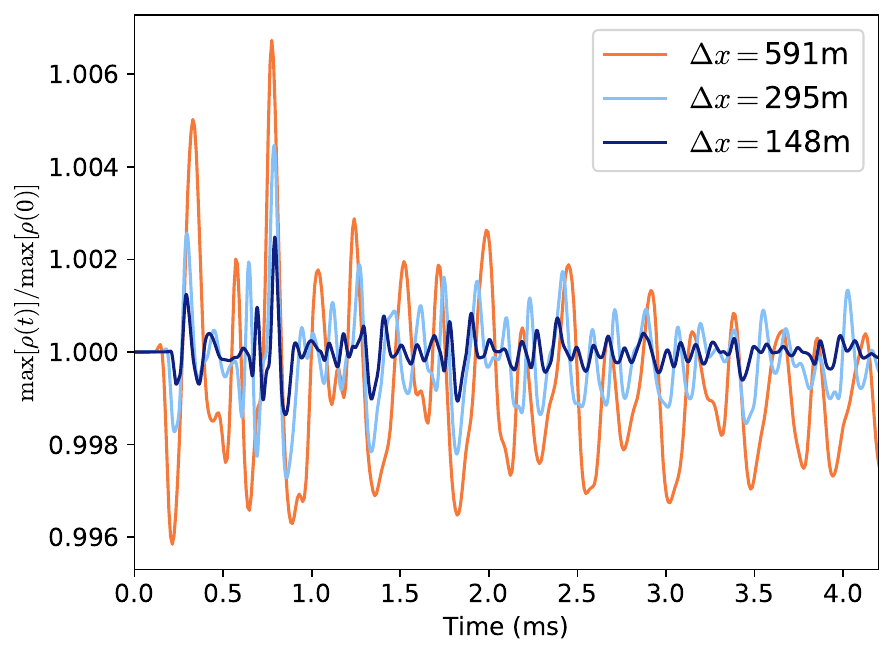}
    \\
    PPAO5-2-2 \& FD-6-2-2
  \end{minipage}
  \begin{minipage}{0.45\columnwidth}
    \centering
    \includegraphics[width=1.0\textwidth]{./SpectrumFd622.pdf}
    \\
    PPAO5-2-2 \& FD-6-2-2
  \end{minipage}

  \begin{minipage}{0.45\columnwidth}
    \centering
    \includegraphics[width=1.0\textwidth]{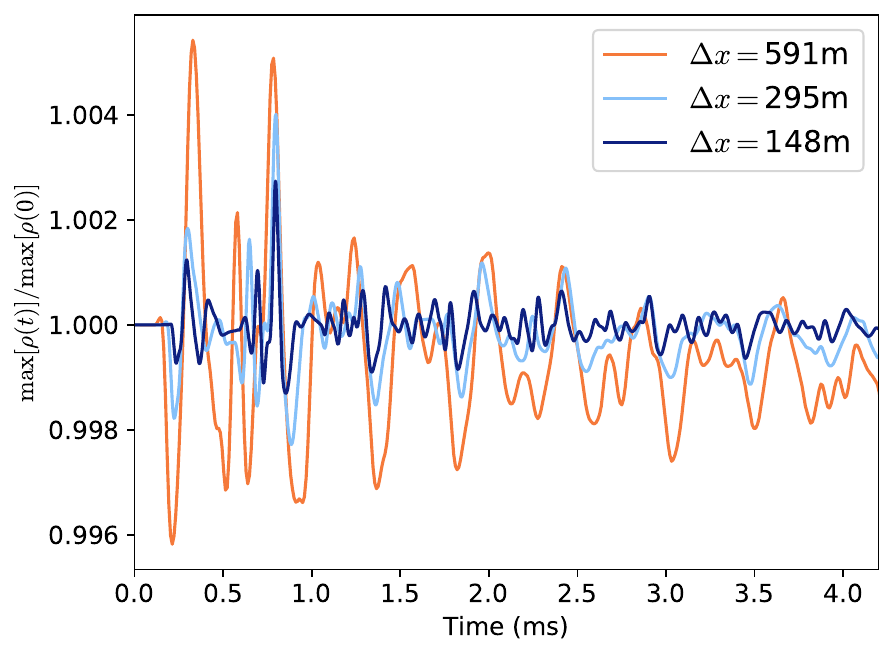}
    \\
    PPAO9-5-2-2 \& FD-10-6-2-2
  \end{minipage}
  \begin{minipage}{0.45\columnwidth}
    \centering
    \includegraphics[width=1.0\textwidth]{./SpectrumFd10622.pdf}
    \\
    PPAO9-5-2-2 \& FD-10-6-2-2
  \end{minipage}
  \caption{Results from evolutions of a rotating star with polar to equatorial
    radii ratio of 0.7 to a final time of 4.2ms using different PPAO
    reconstruction and FD order schemes. The left panels show the maximum rest
    mass density as a function of time divided by the maximum rest mass density
    at $t=0$, which oscillates about 1 with the amplitude decaying over
    time. The right panels show a power spectrum of the left panels.
  \label{fig:RotatingStar}}
\end{figure}

As a final challenging 3d test case, we simulate a uniformly rotating neutron
star with a ratio of polar to equatorial radii of 0.7, as
in~\cite{Deppe:2021bhi,Deppe:2021ada} and similar
to~\cite{2002PhRvD..65h4024F}. The initial data is constructed using the
\texttt{RotNS} code described in~\cite{1992ApJ...398..203C,
  1994ApJ...424..823C}. We again use a polytropic equation of state with
$\Gamma=2$ and $K=100$. The simulations are done on a cubical domain of size
$[-13,13]\times[-13,13]\times[-11,11]$ to give roughly the same number of grid
points across the star in all three dimensions. We show the ratio of the maximum
rest mass density as a function of time to the maximum rest mass density at
$t=0$ in the left panels of figure~\ref{fig:RotatingStar}, while the right
panels show the power spectrum of the rest mass density ratio. We show three
different numerical schemes, PPAO5-2-2+FD-4, PPAO5-2-2+FD-6-2-2, and
PPAO9-5-2-2+FD-10-6-2-2. We see that the adaptive-order derivative schemes are
less dissipative while also resolving high-frequency radial pulsations. This
demonstrates that our proposed adaptive-order schemes are able to produce stable
long-term simulations of interesting astrophysical systems, and marks the first
set of simulations that are ninth order in space.

\section{Conclusions}\label{sec:conclusions}

We presented a new positivity-preserving adaptive-order (PPAO) finite-difference
scheme that adjusts both the order of the unlimited cell-centered polynomial and
the order of the finite-difference derivative based on a new oscillation
indicator and physical admissibility criterion. The scheme reconstructs the
primitive variables, which makes satisfying physical realizability relatively
easy even for complicated systems such as general relativistic
magnetohydrodynamics. The scheme does not make any assumptions about what the
physical realizability conditions are and allows for combining an arbitrary
number of admissibility conditions when selecting the reconstruction
polynomial. We implemented the PPAO scheme in the publicly available code
\texttt{SpECTRE}~\cite{spectrecode}. To demonstrate the efficacy of the proposed
scheme, we perform a number of standard and difficult test problems in 1d, 2d,
and 3d general relativistic magnetohydrodynamics. The scheme was also used to
successfully simulate hybrid quark-hadron stars in~\cite{Legred:2023zet}. The
PPAO scheme is capable of evolving strongly magnetized and rotating neutron
stars, and adapting the order of the FD derivative proves to significantly
increase the robustness for challenging test problems. Adapting the FD
derivative order also allows simulations to remain stable with larger time step
sizes than when only high-order FD derivatives are used. Given the promising
results, we share the viewpoint of~\cite{BALSARA20127504} that physical
realizability of the solution is as important as conservation.

We plan on adopting the flux limiter of~\cite{HU2013169, Radice:2013xpa} to
arrive at a scheme that is positivity-preserving for both the reconstruction and
the time integration. Additionally, we have combined the PPAO scheme with our
discontinuous Galerkin-finite difference hybrid scheme~\cite{Deppe:2021ada} and
are working towards using the PPAO scheme to evolve the spacetime together with
the GRMHD equations. We are also adding support for moving and semi-unstructured
meshes, like cubed-sphere domains. The PPAO scheme naturally lends itself to
semi-unstructured meshes since the linear polynomial can be obtained from any
grid structure with our admissibility criterion remaining easily implemented.

\section*{Acknowledgments}
We are grateful to Michael Pajkos for feedback and informative discussions on
the paper, especially the visualizations.
\texttt{SpECTRE} uses
\texttt{Charm++}/\texttt{Converse}~\cite{laxmikant_kale_2021_5597907,
  kale1996charm++}, which was developed by the Parallel Programming Laboratory
in the Department of Computer Science at the University of Illinois at
Urbana-Champaign. \texttt{SpECTRE} uses \texttt{Blaze}~\cite{Blaze1,Blaze2},
\texttt{HDF5}~\cite{hdf5}, the GNU Scientific Library (\texttt{GSL})~\cite{gsl},
\texttt{yaml-cpp}~\cite{yamlcpp}, \texttt{pybind11}~\cite{pybind11},
\texttt{libsharp}~\cite{libsharp}, and \texttt{LIBXSMM}~\cite{libxsmm}.  The
figures in this article were produced with
\texttt{matplotlib}~\cite{Hunter:2007, thomas_a_caswell_2020_3948793},
\texttt{NumPy}~\cite{harris2020array}, and \texttt{ParaView}~\cite{paraview,
  paraview2}. Computations were performed with the Wheeler cluster at Caltech.
This work was supported in part by the Sherman Fairchild Foundation and by NSF
Grants PHY-2011961, PHY-2011968, and OAC-2209655 at Caltech, and NSF
Grants PHY-2207342 and OAC-2209655 at Cornell. This work was supported
in part by NSF grants PHY-1654359 and PHY-2208014, by the Dan Black Family
Trust, and by Nicholas and Lee Begovich.

\appendix
\section{Use with discontinuous Galerkin-finite difference hybrid
  methods}\label{sec:dg-fd}

We ultimately will use our PPAO scheme together with our discontinuous
Galerkin-FD (DG-FD) hybrid method~\cite{Deppe:2021ada}. We leave a detailed
discussion to future work, but briefly summarize our approach here. In practice
we expect to use PPAO5-2-1 with derivative orders 4-2-2 (or possibly 6-2-2) to
achieve formally fourth (fifth) order convergence on the FD grid, while the
discontinuous Galerkin (DG) method will be at least sixth order. One question is
whether to use the high-order flux $G$ or the second-order flux $G^{(2)}$ on
boundaries between DG and FD. Using $G^{(2)}$ significantly simplifies the code
but formally violates conservation. However, if the DG solver is being used, the
solution is smooth anyway, meaning that no discontinuities are nearby, and so
strict conservation is not necessary. We advocate for using $G^{(2)}$ given the
simplicity this offers. One way of guaranteeing the DG-FD interfaces are far
from discontinuities is to use a halo of FD cells around discontinuities.

\section*{References}
\bibliographystyle{unsrt}
\bibliography{refs}

\end{document}